\documentclass[11pt,a4paper]{article}
\usepackage{amssymb}

\usepackage{amsmath}

\newcounter{resultnum}[section]

\newcounter{conclusionnum}[section]

\newcounter{conditionnum}[section]

\newcounter{conjecturenum}[section]

\newcounter{examplenum}[section]

\newcounter{exercisenum}[section]

\newcounter{lemmanum}[section]

\newcounter{notationnum}[section]

\newcounter{theoremnum}[section]

\newcounter{definitionnum}[section]

\newcounter{corollarynum}[section]

\newcounter{remarknum}[section]

\newcounter{propositionnum}[section]

\newcounter{acknowledgementnum}[section]

\newcounter{algorithmnum}[section]

\newcounter{axiomnum}[section]

\newcounter{casenum}[section]

\newcounter{claimnum}[section]

\newcounter{summarynum}[section]

\newcounter{problemnum}[section]

\begin{document}

\title{Finsler--Lagrange Geometries and\\
Standard Theories in Physics:\ New Methods \\
in Einstein and String Gravity }
\date{January 31, 2008}
\author{ Sergiu I. Vacaru\thanks{%
sergiu$_{-}$vacaru@yahoo.com, svacaru@fields.utoronto.ca } \\
{\quad} \\
\textsl{The Fields Institute for Research in Mathematical Science} \\
\textsl{222 College Street, 2d Floor, } \textsl{Toronto \ M5T 3J1, Canada} }
\maketitle

\begin{abstract}
In this article, we review the current status of Finsler--Lagrange geometry
and generalizations. The goal is to aid non--experts on Finsler spaces, but
physicists and geometers skilled in general relativity and particle
theories, to understand the crucial importance of such geometric methods for
applications in modern physics. We also would like to orient mathematicians
working in generalized Finsler and K\"ahler geometry and geometric mechanics
how they could perform their results in order to be accepted by the
community of ''orthodox'' physicists.

Although the bulk of former models of Finsler--Lagrange spaces where
elaborated on tangent bundles, the surprising result advocated in our works
is that such locally anisotropic structures can be modelled equivalently on
Riemann--Cartan spaces, even as exact solutions in Einstein and/or string
gravity, if nonholonomic distributions and moving frames of references are
introduced into consideration.

We also propose a canonical scheme when geometrical objects on a (pseudo)
Riemannian space are nonholonomically deformed into generalized Lagrange, or
Finsler, configurations on the same manifold or on a corresponding tangent
bundle. Such canonical transforms are defined by the coefficients of a prime
metric (it can be a solution of the Einstein equations) and generate target
spaces as generalized Lagrange structures, their models of almost Hermitian/
K\"{a}hler, or nonholonomic Riemann spaces with constant curvature, for some
Finsler like connections. There are formulated the criteria when such
constructions can be redefined equivalently in terms of the Levi Civita
connection.

Finally, we consider some classes of exact solutions in string and Einstein
gravity modelling Lagrange--Finsler structures with solitonic pp--waves and
speculate on their physical meaning.

\vskip0.3cm \textbf{Keywords:}\ Nonholonomic manifolds, Einstein spaces,
string gravity, Finsler and Lagrange geometry, nonlinear connections, exact
solutions, Riemann--Cartan spaces.

\vskip5pt

MSC:\ 53B40, 53B50, 53C21, 53C55, 83C15, 83E99

PACS:\ 04.20.Jb, 04.40.-b, 04.50.+h, 04.90.+e, 02.40.-k
\end{abstract}

\tableofcontents

\section{ Introduction}

The main purpose of this survey is to present an introduction to
Finsler--Lagrange geometry and the anholonomic frame method in general
relativity and gravitation. We review and discuss possible applications in
modern physics and provide alternative constructions in the language of the
geometry of nonholonomic Riemannian manifolds (enabled with nonintegrable
distributions and preferred frame structures). It will be emphasized the
approach when Finsler like structures are modelled in general relativity and
gravity theories with metric compatible connections and, in general,
nontrivial torsion.

Usually, gravity and string theory physicists may remember that Finsler
geometry is a quite ''sophisticate'' spacetime generalization when
Riemannian metrics $g_{ij}(x^{k})$ are extended to Finsler metrics $%
g_{ij}(x^{k},y^{l})$ depending both on local coordinates $x^{k}$ on a
manifold $M$ and ''velocities'' $y^{l}$ on its tangent bundle $TM.$
\footnote{%
we emphasize that Finsler geometries can be alternatively modelled if $y^{l}$
are considered as certain nonholonomic, i. e. constrained, coordinates on a
general manifold $\mathbf{V}$, not only as "velocities" or "momenta", see
further constructions in this work} Perhaps, they will say additionally that
in order to describe local anisotropies depending only on directions given
by vectors $y^{l},$ the Finsler metrics should be defined in the form $%
g_{ij}\sim \frac{\partial F^{2}}{\partial y^{i}\partial y^{j}},$ where $%
F(x^{k},\zeta y^{l})$ $=|\zeta |\ F(x^{k},y^{l}),$ for any real $\zeta \neq
0,$ is a fundamental Finsler metric function. A number of authors analyzing
possible locally anisotropic physical effects omit a rigorous study of
nonlinear connections and do not reflect on the problem of compatibility of
metric and linear connection structures. If a Riemannian geometry is
completely stated by its metric, various models of Finsler spaces and
generalizations are defined by three independent geometric objects (metric
and linear and nonlinear connections) which in certain canonical cases are
induced by a fundamental Finsler function $F(x,y).$ For models with
different metric compatibility, or non--compatibility, conditions, this is a
point of additional geometric and physical considerations, new terminology
and mathematical conventions. Finally, a lot of physicists and
mathematicians have concluded that such geometries with generic local
anisotropy are characterized by various types of connections, torsions and
curvatures which do not seem to have physical meaning in modern particle
theories but (may be?) certain Finsler like analogs of mechanical systems
and continuous media can be constructed.

There were published a few rigorous studies on perspectives of Finsler like
geometries in standard theories of gravity and particle physics (see, for
instance, Refs. \cite{bek,will}) but they do not analyze any physical
effects of the nonlinear connection and adapted linear connection structures
and the possibility to model Finsler like spaces as exact solutions in
Einstein and sting gravity \cite{vsgg}). The results of such works, on
Finsler models with violations of local Lorentz symmetry and nonmetricity
fields, can be summarized in a very pessimistic form:\ both fundamental
theoretic consequences and experimental data restrict substantially the
importance for modern physics of locally anisotropic geometries elaborated
on (co) tangent bundles,\footnote{%
In result of such opinions, the Editors and referees of some top physical
journals almost stopped to accept for publication manuscripts on Finsler
gravity models. If other journals were more tolerant with such theoretical
works, they were considered to be related to certain alternative classes of
theories or to some mathematical physics problems with speculations on
geometric models and "nonstandard" physics, mechanics and some applications
to biology, sociology or seismology etc} see Introduction to monograph \cite%
{vsgg} and article \cite{vesnc} and reference therein for more detailed
reviews and discussions.

Why we should give a special attention to Finsler geometry and methods and
apply them in modern physics ?\ We list here a set of contr--arguments and
discus the main sources of "anti--Finsler" skepticism which (we hope) will
explain and re--move the existing unfair situation when spaces with generic
local anisotropy are not considered in standard theories of physics:

\begin{enumerate}
\item One should be emphasized that in the bulk the criticism on locally
anisotropic geometries and applications in standard physics was motivated
only for special classes of models on tangent bundles, with violation of
local Lorentz symmetry (even such works became very important in modern
physics, for instance, in relation to brane gravity \cite{groj} and quantum
theories \cite{kost}) and nonmetricity fields. Not all theories with
generalized Finsler metrics and connections were elaborated in this form (on
alternative approaches, see next points) and in many cases, like \cite%
{bek,will}, the analysis of physical consequences was performed not
following the nonlinear connection geometric formalism and a tensor calculus
adapted to nonholonomic structures which is crucial in Finsler geometry and
generalizations.

\item More recently, a group of mathematicians \cite{bcs,shen} developed
intensively some directions on Finsler geometry and applications following
the Chern's linear connection formalism proposed in 1948 (this connection is
with vanishing torsion but noncompatible with the metric structure). For
non--experts in geometry and physics, the works of this group, and other
authors working with generalized local Lorentz symmetries, created a false
opinion that Finsler geometry can be elaborated only on tangent bundles and
that the Chern connection is the "best" Finsler generalization of the Levi
Civita connection. A number of very important constructions with the
so--called metric compatible Cartan connection, or other canonical
connections, were moved on the second plan and forgotten. One should be
emphasized that the geometric constructions with the well known Chern or
Berwald connections can not be related to standard theories of physics
because they contain nonmetricity fields. The issue of nonmetricity was
studied in details in a number of works on metric--affine gravity, see
review \cite{mag} and Chapter I in the collection of works \cite{vsgg}, the
last one containing a series of papers on generalized Finsler--affine
spaces. Such results are not widely accepted by physicists because of
absence of experimental evidences and theoretical complexity of geometric
constructions. Here we note that it is a quite sophisticate task to
elaborate spinor versions, supersymmetric and noncommutative generalizations
of Finsler like geometries if we work with metric noncompatible connections.

\item A non--expert in special directions of differential geometry and
geometric mechanics, may not know that beginning E. Cartan (1935) \cite{cart}
various models of Finsler geometry were developed alternatively by using
metric compatible connections which resulted in generalizations to the
geometry of Lagrange and Hamilton mechanics and their higher order
extensions. Such works and monographs were published by prominent schools
and authors on Finsler geometry and generalizations from Romania and Japan %
\cite%
{ma1987,ma,mhl,mhf,mhss,mhh,kaw1,kaw2,ikeda,takano,watanik,mats,opr1,opr2,opp1,opp2,oppor,bej,bejf}
following approaches quite different from the geometry of sympletic
mechanics and generalizations \cite{dleon,lib,mard,krup}. As a matter of
principle, all geometric constructions with the Chern and/or sympletic
connections can de redefined equivalently for metric compatible geometries,
but the philosophy, aims, mathematical formalism and physical consequences
are very different for different approaches and the particle physics
researches usually are not familiar with such results.

\item It should be noted that for a number of scientists working in Western
Countries there are less known the results on the geometry of nonholonomic
manifolds published in a series of monographs and articles by G. Vr\v
anceanu (1926), Z. Horak (1927) and others \cite{vr1,vr2,hor}, see
historical remarks and bibliography in Refs. \cite{bejf,vsgg}. The
importance for modern physics of such works follows from the idea and
explicit proofs (in quite sophisticate component forms) that various types
of locally anisotropic geometries and physical interactions can be modelled
on usual Riemannian manifolds by considering nonholonomic distributions and
holonomic fibrations enabled with certain classes of special connections.

\item In our works (see, for instance, reviews and monographs \cite%
{vfs,vstrf,vncsup,vhs,vmon1,vstav,vtnut,vv,vesnc,vsgg}, and references
therein), we re--oriented the research on Finsler spaces and generalizations
in some directions connected to standard models of physics and gauge,
supersymmetric and noncommutative extensions of gravity. Our basic idea was
that both the Riemann--Cartan and generalized Finsler--Lagrange geometries
can be modelled in a unified manner by corresponding geometric structures on
nonholono\-mic manifolds. It was emphasized, that prescribing a preferred
nonholonomic frame structure (equivalently, a nonintegrabie distribution
with associated nonlinear connection) on a manifold, or on a vector bundle,
it is possible to work equivalently both with the Levi Civita and the
so--called canonical distinguished connection. We provided a number of
examples when Finsler like structures and geometries can be modelled as
exact solutions in Einstein and string gravity and proved that certain
geometric methods are very important, for instance, in constructing new
classes of exact solutions.
\end{enumerate}

This review work has also pedagogical scopes. We attempt to cover key
aspects and open issues of generalized Finsler--Lagrange geometry related to
a consistent incorporation of nonlinear connection formalism and moving/
deformation frame methods into the Einstein and string gravity and analogous
models of gravity, see also Refs. \cite{vesnc,vsgg,ma,bej,rund} for general
reviews, written in the same spirit as the present one but in a more
comprehensive, or inversely, with more special purposes forms. While the
article is essentially self--contained, the emphasis is on communicating the
underlying ideas and methods and the significance of results rather than on
presenting systematic derivations and detailed proofs (these can be found in
the listed literature).

The subject of Finsler geometry and applications can be approached in
different ways. We choose one of which is deeply rooted in the well
established gravity physics and also has sufficient mathematical precision
to ensure that a physicist familiar with standard textbooks and monographs
on gravity \cite{haw,mtw,wald,stw,sb} and string theory \cite%
{string1,string2,string3} will be able without much efforts to understand
recent results and methods of the geometry of nonholonomic manifolds and
generalized Finsler--Lagrange spaces. In other turn, in order to keep the
article to a reasonable size, and avoid overwhelming non--experts, we have
to leave out several interesting topics, results and viewpoints. We list the
most important alternative directions and comment references in Appendix in
order to orient experts in gravity and field theories in existing literature
and researches related to applications of Finsler geometry methods in modern
physics. This is meant that the work is an introduction into some subjects
and new geometric methods which seem to be very important in standard
physics rather then an exhaustive review of them.

We shall use the terms "standard" and "nonstandard" models in geometry and
physics. In connection to Finsler geometry, we shall consider a model to be
a standard one if it contains locally anisotropic structures defined by
certain nonholonomic distributions and adapted frames of reference on a
(pseudo) Riemannian or Riemann--Cartan space (for instance, in general
relativity, Kaluza--Klein theories and low energy string gravity models).
Such constructions preserve, in general, the local Lorentz symmetry and they
are performed with metric compatible connections. The term "nonstandard"
will be used for those approaches which are related to metric
non--compatible connections and/or local Lorentz violations in Finsler
spacetimes and generalizations. Sure, any standard or nonstandard model is
rigorously formulated following certain purposes in modern geometry and
physics, geometric mechanics, biophysics, locally anisotropic thermodynamics
and stochastic and kinetic processes and classical or quantum gravity
theories. Perhaps, it will be the case to distinguish the class of "almost
standard" physical models with locally anisotropic interactions when certain
geometric objects from a (pseudo) Riemannian or Riemann--Cartan manifolds
are lifted on a (co) tangent or vector bundles and/or their supersymmetric,
non--commutative, Lie algebroid, Clifford space, quantum group ...
generalizations. There are possible various effects with "nonstandard"
corrections, for instance, violations of the local Lorentz symmetry by
quantum effects but in some classical or quantum limits such theories are
constrained to correspond to certain standard ones.

This contribution is organized as follows:

In section 2, we outline an unified approach to the geometry of nonholonomic
distributions on Riemann manifolds and Finsler--Lagrange spaces. The basic
concepts on nonholonomic manifolds and associated nonlinear connection
structures are explained and the possibility of equivalent (non) holonomic
formulations of gravity theories is analyzed.

Section 3 is devoted to nonholonomic deformations of manifolds and vector
bundles. There are reviewed the basic constructions in the geometry of
(generalized) Lagrange and Finsler spaces. We show how effective algebroid
structures can be generated by nonholonomic transforms. A general ansatz for
constructing exact solutions, with effective (algebroid) Lagrange and
Finsler structures, in Einstein and string gravity, is analyzed.

In section 4, the Finsler--Lagrange geometry is formulated as a variant of
almost Hermitian and/or K\"aher geometry with additional Lie algebroid
structure. We show how the Einstein gravity can be equivalently reformulated
in terms of almost Hermitian geometry with preferred frame structure.

Section 5 is focused on explicit examples of exact solutions in Einstein and
string gravity when (generalized) Finsler--Lagrange structures are modelled
on (pseudo) Riemannian and Riemann--Cartan spaces. We analyze some classes
of Einstein metrics which can be deformed into new exact solutions
characterized additionally by Lagrange--Finsler configurations. For string
gravity, there are constructed explicit examples of locally anisotropic
configurations describing gravitational solitonic pp--waves and their
effective Lagrange spaces. We also analyze some exact solutions for
Finsler--solitonic pp--waves on Schwarzschild spaces.

Conclusions and further perspectives of Finsler geometry and new geometric
methods for modern gravity theories are considered in section 6.

We provide an Appendix containing historical and bibliographical comments on
(generalized) Finsler geometry and physics.

Finally, we should note that our list of references is minimalist, trying to
concentrate on reviews and monographs rather than on original articles. More
complete reference lists can be found in the books \cite%
{vsgg,vmon1,vstav,ma,mhss}. Various guides for learning, both for experts
and beginners on geometric methods and further applications in modern
physics, with references, can be found in \cite{vsgg,ma,mhss,bej,rund}.

\subsubsection*{Notational remarks:}

We shall consider geometric and physical objects on different spaces. There
were elaborated very sophisticate systems of denotations and terminology in
various approaches to general relativity, string theory and generalized
Finsler--Lagrange geometry. In this work, one follows the conventions from %
\cite{vsgg,vesnc}. We shall use ''boldface'' letters, $\mathbf{A},\mathbf{B}%
_{\ \beta }^{\alpha },...$ for geometric objects and spaces adapted to
(provided with) a nonlinear connection structure. In general, small Greek
indices are considered as abstract ones, which may split into horizontal (h)
and vertical (v) indices, for instance $\alpha =(i,a),\beta =(j,b),...$
where with respect to a coordinate basis they run values of type $%
i,j,...=1,2,...,n $ and $a,b,...=n+1,n+2,...n+m,$ for $n\geq 2$ and $m\geq
1. $ One shall be considered primed indices, $\alpha ^{\prime }=(i^{\prime
},a^{\prime }),\beta ^{\prime }=(j^{\prime },b^{\prime }),...,$ working with
respect to a nonholonomically transformed bases, or underlined indices, $%
\underline{\alpha }=(\underline{i},\underline{a}),\underline{\beta }=(%
\underline{j},\underline{b}),...,$ in order to emphasize that coefficients
of geometric objects are defined with respect to a coordinate basis.
Various types of left ''up'' and ''low'' labels of geometric objects will be
used, for instance, $\ ^{RC}V$ means that the manifold $V$ is a
Riemannian--Cartan one, the Levi Civita connection will be labelled $\
_{\shortmid }D=\nabla $ and the corresponding Riemannian and Ricci tensors
will be written $\ _{\shortmid }\mathcal{R}=\{\ _{\shortmid }R_{\ \beta
\gamma \tau }^{\alpha }\},$ and$\ _{\shortmid }Ric(\ _{\shortmid }D)=\{\
_{\shortmid }R_{\ \beta \gamma }\}.$ We shall omit labels and indices if
that will not result in ambiguities.  Finally, we note that we shall write
with boldface letters a new term if it is introduced for the first time in
the text.

\section{Nonholonomic Einstein Gravity and Finsler--La\-grange Spa\-ces}

In this section we present in a unified form the Riemann--Cartan and
Finsler--Lagrange geometry. The reader is supposed to be familiar with
well--known geometrical approaches to gravity theories \cite%
{haw,mtw,wald,stw,sb} but may not know the basic concepts on Finsler
geometry and nonholonomic manifolds. The constructions for locally
anisotropic spaces will be derived by special parametrizations of the frame,
metric and connection structures on usual manifolds, or vector bundle
spaces, as we proved in details in Refs. \cite{vsgg,vesnc}.

\subsection{Metric--affine, Riemann--Cartan and Einstein manifolds}

Let $V$ be a necessary smooth class manifold of dimension $\dim V=n+m,$ when
$n\geq 2$ and $m\geq 1,$ enabled with \textbf{metric}, $g=g_{\alpha \beta
}e^{\alpha }\otimes e^{\beta },$ and \textbf{linear connection}, $D=\{\Gamma
_{\ \beta \gamma }^{\alpha }\},$ structures. The coefficients of $g$ and $D$
can be computed with respect to any local \textbf{frame}, $e_{\alpha },$ and
\textbf{co--frame}, $e^{\beta },$ bases, for which $e_{\alpha }\rfloor
e^{\beta }=\delta _{\alpha }^{\beta },$ where $\rfloor $ denotes the
interior (scalar) product defined by $g$ and $\delta _{\alpha }^{\beta }$ is
the Kronecker symbol. A local system of coordinates on $V$ is denoted $%
u^{\alpha }=(x^{i},y^{a}),$ or (in brief) $u=(x,y),$ where indices run
correspondingly the values: $i,j,k...=1,2,...,n$ and $%
a,b,c,...=n+1,n+2,...n+m$ for any splitting $\alpha =(i,a),\beta =(j,b),...$
We shall also use primed, underlined, or other type indices: for instance, $%
e_{\alpha ^{\prime }}=(e_{i^{\prime }},e_{a^{\prime }})$ and $e^{\beta
^{\prime }}=(e^{j^{\prime }},e^{b^{\prime }}),$ for a different sets of
local (co) bases, or $\underline{e}_{\alpha }=e_{\underline{\alpha }%
}=\partial _{\underline{\alpha }}=\partial /\partial u^{\underline{\alpha }%
}, $ $\underline{e}_{i}=e_{\underline{i}}=\partial _{\underline{i}}=\partial
/\partial x^{\underline{i}}$ and $\underline{e}_{a}=e_{\underline{a}%
}=\partial _{\underline{a}}=\partial /\partial y^{\underline{a}}$ if we wont
to emphasize that the coefficients of geometric objects (tensors,
connections, ...) are defined with respect to a local \textbf{coordinate
basis}. For simplicity, we shall omit underlining or priming of indices and
symbols if that will not result in ambiguities. The Einstein's summation
rule on repeating ''up-low'' indices will be applied if the contrary will
not be stated.

\textbf{Frame transforms} of a local basis $e_{\alpha }$ and its dual basis $%
e^{\beta }$ are paramet\-riz\-ed in the form
\begin{equation}
e_{\alpha }=A_{\alpha }^{\ \alpha ^{\prime }}(u)e_{\alpha ^{\prime }}%
\mbox{\
and\  }e^{\beta }=A_{\ \beta ^{\prime }}^{\beta }(u)e^{\beta ^{\prime }},
\label{ft}
\end{equation}%
where the matrix $A_{\ \beta ^{\prime }}^{\beta }$ is inverse to $A_{\alpha
}^{\ \alpha ^{\prime }}.$ In general, local bases are \textbf{nonholonomic}
(equivalently, \textbf{anholonomic}, or \textbf{nonintegrable}) and satisfy
certain anholonomy conditions
\begin{equation}
e_{\alpha }e_{\beta }-e_{\beta }e_{\alpha }=W_{\alpha \beta }^{\gamma
}e_{\gamma }  \label{nhr}
\end{equation}%
with nontrivial \textbf{anholonomy coefficients} $W_{\alpha \beta }^{\gamma
}(u).$ We consider the \textbf{holonomic} frames to be defined by $W_{\alpha
\beta }^{\gamma }=0,$ which holds, for instance, if we fix a local
coordinate basis.

Let us denote the covariant derivative along a vector field $X=X^{\alpha
}e_{\alpha }$ as $D_{X}=X\rfloor D.$ One defines three fundamental geometric
objects on manifold $V:$ \textbf{nonmetricity} field,
\begin{equation}
\mathcal{Q}_{X}\doteqdot D_{X}g,  \label{nm}
\end{equation}%
\textbf{torsion},
\begin{equation}
\mathcal{T}(X,Y)\doteqdot D_{X}Y-D_{Y}X-[X,Y],  \label{ators}
\end{equation}%
and \textbf{curvature},
\begin{equation}
\mathcal{R}(X,Y)Z\doteqdot D_{X}D_{Y}Z-D_{Y}D_{X}Z-D_{[X,Y]}Z,  \label{acurv}
\end{equation}%
where the symbol ''$\doteqdot $'' states ''by definition'' and $%
[X,Y]\doteqdot XY-YX.$ With respect to fixed local bases $e_{\alpha }$ and $%
e^{\beta },$ the coefficients $\mathcal{Q}=\{Q_{\alpha \beta \gamma
}=D_{\alpha }g_{\beta \gamma }\},\mathcal{T}=\{T_{\ \beta \gamma }^{\alpha
}\}$ and $\mathcal{R}=\{R_{\ \beta \gamma \tau }^{\alpha }\}$ can be
computed by introducing $X\rightarrow e_{\alpha },Y\rightarrow e_{\beta
},Z\rightarrow e_{\gamma }$ into respective formulas (\ref{nm}), (\ref{ators}%
) and (\ref{acurv}).

In gravity theories, one uses three others important geometric objects: the
\textbf{Ricci tensor}, $Ric(D)=\{R_{\ \beta \gamma }\doteqdot R_{\ \beta
\gamma \alpha }^{\alpha }\},$ the \textbf{scalar curvature}, $R\doteqdot
g^{\alpha \beta }R_{\alpha \beta }$ ($g^{\alpha \beta }$ being the inverse
matrix to $g_{\alpha \beta }),$ and the \textbf{Einstein tensor}, $\mathcal{E%
}=\{E_{\alpha \beta }\doteqdot R_{\alpha \beta }-\frac{1}{2}g_{\alpha \beta
}R\}.$

A manifold $\ ^{ma}V$ is a \textbf{metric--affine space }if it is provided
with arbitrary two independent metric $g$ and linear connection $D$
structures and characterized by three nontrivial fundamental geometric
objects $\mathcal{Q},\mathcal{T}$ and $\mathcal{R}.$

If the metricity condition, $\mathcal{Q}=0,$ is satisfied for a given couple
$g$ and $D,$ such a manifold $\ ^{RC}V$ is called a \textbf{Riemann--Cartan
space }with nontrivial torsion $\mathcal{T}$ of $D.$

A \textbf{Riemann space} $\ ^{R}V$ is provided with a metric structure $g$
which defines a unique Levi Civita connection $\ _{\shortmid }D=\nabla ,$
which is both metric compatible, $\ _{\shortmid }\mathcal{Q}=\nabla g=0,$
and torsionless, $\ _{\shortmid }\mathcal{T}=0.$ Such a space is pseudo-
(semi-) Riemannian if locally the metric has any mixed signature $(\pm 1,
\pm 1, ..., \pm 1).$\footnote{%
mathematicians usually use the term semi--Riemannian but physicists are more
familiar with pseudo--Riemannian; we shall apply both terms on convenience}
In brief, we shall call all such spaces to be Riemannian (with necessary
signature) and denote the main geometric objects in the form $\ _{\shortmid }%
\mathcal{R}=\{\ _{\shortmid }R_{\ \beta \gamma \tau }^{\alpha }\},$ $\
_{\shortmid }Ric(\ _{\shortmid }D)=\{\ _{\shortmid }R_{\ \beta \gamma }\},\
_{\shortmid }R$ and $\ _{\shortmid }\mathcal{E}=\{\ _{\shortmid }E_{\alpha
\beta }\}.$

The \textbf{Einstein gravity theory} is constructed canonically for $\dim
^{R}V=4 $ and Minkowski signature, for instance, $(-1,+1,+1,+1).$ Various
generalizations in modern \textbf{string and/or gauge gravity} consider
Riemann, Riemann--Cartan and metric--affine spaces of higher dimensions.

The \textbf{Einstein equations }are postulated in the form
\begin{equation}
\mathcal{E}(D)\doteqdot Ric(D)-\frac{1}{2}\ g~Sc(D)=\Upsilon ,  \label{einst}
\end{equation}%
where the source $\Upsilon $ contains contributions of matter fields and
corrections from, for instance, string/brane theories of gravity. In a
physical model, the equations (\ref{einst}) have to be completed with
equations for the matter fields and torsion (for instance, in the\ \textbf{%
Einstein--Cartan theory} \cite{mag}, one considers algebraic equations for
the torsion and its source). It should be noted here that because of
possible nonholonomic structures on a manifold $V$ (we shall call such
spaces to be locally anisotropic), see next section, the tensor $Ric(D)$ is
not symmetric and $D\left[ \mathcal{E}(D)\right] \neq 0.$ This imposes a
more sophisticate form of conservation laws on spaces with generic ''local
anisotropy'', see discussion in \cite{vsgg} (a similar situation arises in
Lagrange mechanics \cite{dleon,lib,mard,krup,ma} when nonholonomic
constraints modify the definition of conservation laws).

For \textbf{general relativity}, $\dim V=4$ and $D=\nabla, $ the field
equations can be written in the well--known component form%
\begin{equation}
\ _{\shortmid }E_{\alpha \beta }=\ _{\shortmid }R_{\ \beta \gamma }-\frac{1}{%
2}\ _{\shortmid }R=\Upsilon _{\alpha \beta }  \label{einstgr}
\end{equation}%
when $\nabla (\ _{\shortmid }E_{\alpha \beta })=\nabla (\Upsilon _{\alpha
\beta })=0.$ The coefficients in equations (\ref{einstgr}) are defined with
respect to arbitrary nonholomomic frame (\ref{ft}).

\subsection{Nonholonomic manifolds and adapted frame structures}

A \textbf{nonholonomic manifold} $(M,\mathcal{D})$ is a manifold $M$ of
necessary smooth class enabled with a nonholonomic distribution $\mathcal{D}%
, $ see details in Refs. \cite{bejf,vsgg}. Let us consider a $(n+m)$%
--dimensional manifold $\mathbf{V,}$ with $n\geq 2$ and $m\geq 1$ (for a
number of physical applications, it will be considered to model a physical
or geometric space). In a particular case, $\mathbf{V=}TM,$ with $n=m$ (i.e.
a tangent bundle), or $\mathbf{V=E}=(E,M),$ $\dim M=n,$ is a vector bundle
on $M,$ with total space $E$ (we shall use such spaces for traditional
definitions of Finsler and Lagrange spaces \cite%
{ma1987,ma,mats,bej,rund,bcs,shen}). In a general case, a manifold $\mathbf{V%
}$ is provided with a local fibred structure into conventional
''horizontal'' and ''vertical'' directions defined by a nonholonomic
(nonintegrable) distribution with associated nonlinear connection
(equivalently, nonholonomic frame) structure. Such nonholonomic manifolds
will be used for modelling locally anisortropic structures in Einstein
gravity and generalizations \cite{vhs,vtnut,vv,vesnc,vsgg}.

\subsubsection{Nonlinear connections and N--adapted frames}

We denote by $\pi ^{\top }:T\mathbf{V}\rightarrow TM$ the differential of a
map $\pi :\mathbf{V}\rightarrow V$ defined by fiber preserving morphisms of
the tangent bundles $T\mathbf{V}$ and $TM.$ The kernel of $\pi ^{\top }$ is
just the vertical subspace $v\mathbf{V}$ with a related inclusion mapping $%
i:v\mathbf{V}\rightarrow T\mathbf{V}.$

A \textbf{nonlinear connection (N--connection)} $\mathbf{N}$ on a manifold $%
\mathbf{V}$ is defined by the splitting on the left of an exact sequence
\begin{equation*}
0\rightarrow v\mathbf{V}\overset{i}{\rightarrow }T\mathbf{V}\rightarrow T%
\mathbf{V}/v\mathbf{V}\rightarrow 0,
\end{equation*}%
i. e. by a morphism of submanifolds $\mathbf{N:\ \ }T\mathbf{V}\rightarrow v%
\mathbf{V}$ such that $\mathbf{N\circ i}$ is the unity in $v\mathbf{V}.$

Locally, a N--connection is defined by its coefficients $N_{i}^{a}(u),$%
\begin{equation}
\mathbf{N}=N_{i}^{a}(u)dx^{i}\otimes \frac{\partial }{\partial y^{a}}.
\label{coeffnc}
\end{equation}%
In an equivalent form, we can say that any N--connection is defined by a
\textbf{Whitney sum} of conventional horizontal (h) space, $\left( h\mathbf{V%
}\right) ,$ and vertical (v) space, $\left( v\mathbf{V}\right) ,$
\begin{equation}
T\mathbf{V}=h\mathbf{V}\oplus v\mathbf{V}.  \label{whitney}
\end{equation}%
The sum (\ref{whitney}) states on $T\mathbf{V}$ a nonholonomic
(equivalently, anholonomic, or nonintegrable) distribution of h- and
v--space. The well known class of linear connections consists on a
particular subclass with the coefficients being linear on $y^{a},$ i.e.
\begin{equation}
N_{i}^{a}(u)=\Gamma _{bj}^{a}(x)y^{b}.  \label{lincon}
\end{equation}

The geometric objects on $\mathbf{V}$ can be defined in a form adapted to a
N--connection structure, following decompositions which are invariant under
parallel transports preserving the splitting (\ref{whitney}). In this case,
we call them to be distinguished (by the N--connection structure), i.e.
\textbf{d--objects.} For instance, a vector field $\mathbf{X}\in T\mathbf{V}$
\ is expressed
\begin{equation*}
\mathbf{X}=(hX,\ vX),\mbox{ \ or \ }\mathbf{X}=X^{\alpha }\mathbf{e}_{\alpha
}=X^{i}\mathbf{e}_{i}+X^{a}e_{a},
\end{equation*}%
where $hX=X^{i}\mathbf{e}_{i}$ and $vX=X^{a}e_{a}$ state, respectively, the
adapted to the N--connection structure horizontal (h) and vertical (v)
components of the vector. In brief, $\mathbf{X}$ is called a distinguished
vectors, \textbf{d--vector}.\footnote{%
We shall use always ''boldface'' symbols if it would be necessary to
emphasize that certain spaces and/or geometrical objects are
provided/adapted to a\ N--connection structure, or with the coefficients
computed with respect to N--adapted frames.} In a similar fashion, the
geometric objects on $\mathbf{V}$ like tensors, spinors, connections, ...
are called respectively \textbf{d--tensors, d--spinors, d--connections} if
they are adapted to the N--connection splitting (\ref{whitney}).

The \textbf{N--connection curvature} is defined as the \textbf{Neijenhuis
tensor}%
\begin{equation}
\mathbf{\Omega }(\mathbf{X,Y})\doteqdot \lbrack vX,vY]+\ v[\mathbf{X,Y}]-v[vX%
\mathbf{,Y}]-v[\mathbf{X,}vY].  \label{njht}
\end{equation}%
In local form, we have for (\ref{njht})
\begin{equation*}
\mathbf{\Omega }=\frac{1}{2}\Omega _{ij}^{a}\ d^{i}\wedge d^{j}\otimes
\partial _{a},
\end{equation*}%
with coefficients%
\begin{equation}
\Omega _{ij}^{a}=\frac{\partial N_{i}^{a}}{\partial x^{j}}-\frac{\partial
N_{j}^{a}}{\partial x^{i}}+N_{i}^{b}\frac{\partial N_{j}^{a}}{\partial y^{b}}%
-N_{j}^{b}\frac{\partial N_{i}^{a}}{\partial y^{b}}.  \label{ncurv}
\end{equation}

Any N--connection $\mathbf{N}$ may be characterized by an associated frame
(vielbein) structure $\mathbf{e}_{\nu }=(\mathbf{e}_{i},e_{a}),$ where
\begin{equation}
\mathbf{e}_{i}=\frac{\partial }{\partial x^{i}}-N_{i}^{a}(u)\frac{\partial }{%
\partial y^{a}}\mbox{ and
}e_{a}=\frac{\partial }{\partial y^{a}},  \label{dder}
\end{equation}%
and the dual frame (coframe) structure $\mathbf{e}^{\mu }=(e^{i},\mathbf{e}%
^{a}),$ where
\begin{equation}
e^{i}=dx^{i}\mbox{ and }\mathbf{e}^{a}=dy^{a}+N_{i}^{a}(u)dx^{i},
\label{ddif}
\end{equation}%
see formulas (\ref{ft}). These vielbeins are called respectively \textbf{%
N--adapted fra\-mes and coframes.} In order to preserve a relation with the
previous denotations \cite{vesnc,vsgg}, we emphasize that $\mathbf{e}_{\nu
}=(\mathbf{e}_{i},e_{a})$ and $\mathbf{e}^{\mu }=(e^{i},\mathbf{e}^{a})$ are
correspondingly the former ''N--elongated'' partial derivatives $\delta
_{\nu }=\delta /\partial u^{\nu }=(\delta _{i},\partial _{a})$ and
N--elongated differentials $\delta ^{\mu }=\delta u^{\mu }=(d^{i},\delta
^{a}).$ This emphasizes that the operators (\ref{dder}) and (\ref{ddif})
define certain ``N--elongated'' partial derivatives and differentials which
are more convenient for tensor and integral calculations on such
nonholonomic manifolds. The vielbeins (\ref{ddif}) satisfy the nonholo\-nomy
relations
\begin{equation}
\lbrack \mathbf{e}_{\alpha },\mathbf{e}_{\beta }]=\mathbf{e}_{\alpha }%
\mathbf{e}_{\beta }-\mathbf{e}_{\beta }\mathbf{e}_{\alpha }=W_{\alpha \beta
}^{\gamma }\mathbf{e}_{\gamma }  \label{anhrel}
\end{equation}%
with (antisymmetric) nontrivial anholonomy coefficients $W_{ia}^{b}=\partial
_{a}N_{i}^{b}$ and $W_{ji}^{a}=\Omega _{ij}^{a}$ defining a proper
parametrization (for a $n+m$ splitting by a N--connection $N_{i}^{a})$ of (%
\ref{ddif}).

\subsubsection{N--anholonomic manifolds and d--metrics}

For simplicity, we shall work with a particular class of nonholonomic
manifolds: A manifold $\mathbf{V}$ is \textbf{N--anholonomic} if its tangent
space $T\mathbf{V}$ is enabled with a N--connection structure (\ref{whitney}%
).\footnote{%
In a similar manner, we can consider different types of (super) spaces and
low energy string limits \cite{vncsup,vmon1,vcv,vstrf}, Riemann or
Riemann--Cartan manifolds \cite{vsgg}, noncommutative bundles, or
superbundles and gauge models \cite{vggr,vdgrg,dvgrg,vgonch,vesnc},
Clifford--Dirac spinor bundles and algebroids \cite%
{vclalg,vfs,vhs,vstav,vv,vtnut}, Lagrange--Fedosov manifolds \cite{esv}...
provided with nonholonomc (super) distributions (\ref{whitney}) and
preferred systems of reference (supervielbeins).}

A \textbf{distinguished metric} (in brief, \textbf{d--metric}) on a
N--anholo\-nom\-ic manifold $\mathbf{V}$ is a usual second rank metric
tensor $\mathbf{g}$ which with respect to a N--adapted basis (\ref{ddif})
can be written in the form%
\begin{equation}
\mathbf{g}=\ g_{ij}(x,y)\ e^{i}\otimes e^{j}+\ h_{ab}(x,y)\ \mathbf{e}%
^{a}\otimes \mathbf{e}^{b}  \label{m1}
\end{equation}%
defining a N--adapted decomposition $\mathbf{g=}hg\mathbf{\oplus _{N}}%
vg=[hg,vg].$

A \textbf{metric structure } $\ \breve{g}$ on a N--anholonomic manifold $%
\mathbf{V}$ is a symmetric covariant second rank tensor field which is not
degenerated and of constant signature in any point $\mathbf{u\in V.}$ Any
metric on $\mathbf{V,}$ with respect to a local coordinate basis $du^{\alpha
}=\left( dx^{i},dy^{a}\right) ,$ can be parametrized in the form
\begin{equation}
\ \breve{g}=\underline{g}_{\alpha \beta }\left( u\right) du^{\alpha }\otimes
du^{\beta }  \label{metr}
\end{equation}%
where%
\begin{equation}
\underline{g}_{\alpha \beta }=\left[
\begin{array}{cc}
g_{ij}+N_{i}^{a}N_{j}^{b}h_{ab} & N_{j}^{e}h_{ae} \\
N_{i}^{e}h_{be} & h_{ab}%
\end{array}%
\right] .  \label{ansatz}
\end{equation}%
Such a metric (\ref{ansatz})\ is generic off--diagonal, i.e. it can not be
diagonalized by coordinate transforms if $N_{i}^{a}(u)$ are any general
functions.

In general, a metric structure is not adapted to a N--connection structure,
but we can transform it into a d--metric
\begin{equation}
\mathbf{g}=hg(hX,hY)+vg(vX,vY)  \label{dmetra}
\end{equation}%
adapted to a N--connection structure defined by coefficients $N_{i}^{a}.$ We
introduce denotations $h\breve{g}(hX,hY)=hg(hX,hY)$ and $v\breve{g}(vX,$ $%
vY)=vg(vX,vY)$ and try to find a N--connection when
\begin{equation}
\breve{g}(hX,vY)=0  \label{algn01}
\end{equation}%
for any d--vectors $\mathbf{X,Y.}$ In local form, for $hX\rightarrow e_{i}$
and $\ vY\rightarrow e_{a},$\ the equation (\ref{algn01}) is an algebraic
equation for the N--connection coefficients $N_{i}^{a},$
\begin{equation}
\breve{g}(e_{i},e_{a})=0, \mbox{ equivalently, } \underline{g}%
_{ia}-N_{i}^{b}h_{ab}=0,  \label{aux1a}
\end{equation}%
where $\underline{g}_{ia}$ $\doteqdot g(\partial /\partial x^{i},\partial
/\partial y^{a}),$ which allows us to define in a unique form the
coefficients $N_{i}^{b}=h^{ab}\underline{g}_{ia}$ where $h^{ab}$ is inverse
to $h_{ab}.$ We can write the metric $\breve{g}$ with ansatz (\ref{ansatz})\
in equivalent form, as a d--metric (\ref{m1}) adapted to a N--connection
structure, if we define $g_{ij}\doteqdot \mathbf{g}\left( e_{i},e_{j}\right)
$ and $h_{ab}\doteqdot \mathbf{g}\left( e_{a},e_{b}\right) $ \ and consider
the vielbeins $\mathbf{e}_{\alpha }$ and $\mathbf{e}^{\alpha }$ to be
respectively of type (\ref{dder}) and (\ref{ddif}).

A metric $\ \breve{g}$ (\ref{metr}) can be equivalently transformed into a
d--metric (\ref{m1}) by performing a frame (vielbein) transform
\begin{equation}
\mathbf{e}_{\alpha }=\mathbf{e}_{\alpha }^{\ \underline{\alpha }}\partial _{%
\underline{\alpha }}\mbox{ and }\mathbf{e}_{\ }^{\beta }=\mathbf{e}_{\
\underline{\beta }}^{\beta }du^{\underline{\beta }},  \label{ftnc}
\end{equation}%
with coefficients
\begin{eqnarray}
\mathbf{e}_{\alpha }^{\ \underline{\alpha }}(u) &=&\left[
\begin{array}{cc}
e_{i}^{\ \underline{i}}(u) & N_{i}^{b}(u)e_{b}^{\ \underline{a}}(u) \\
0 & e_{a}^{\ \underline{a}}(u)%
\end{array}%
\right] ,  \label{vt1} \\
\mathbf{e}_{\ \underline{\beta }}^{\beta }(u) &=&\left[
\begin{array}{cc}
e_{\ \underline{i}}^{i\ }(u) & -N_{k}^{b}(u)e_{\ \underline{i}}^{k\ }(u) \\
0 & e_{\ \underline{a}}^{a\ }(u)%
\end{array}%
\right] ,  \label{vt2}
\end{eqnarray}%
being linear on $N_{i}^{a}.$

It should be noted here that parametrizations of metrics of type (\ref%
{ansatz}) have been introduced in Kaluza--Klein gravity \cite{ow} for the
case of linear connections (\ref{lincon}) and compactified extra dimensions $%
y^{a}.$ For the five (or higher) dimensions, the coefficients $\Gamma _{\
bi}^{a}(x)$ were considered as Abelian or non--Abelian gauge fields. In our
approach, the coefficients $N_{i}^{b}(x,y)$ are general ones, not obligatory
linearized and/or compactified on $y^{a}.$ For some models of Finsler
gravity, the values $N_{i}^{a}$ were treated as certain generalized
nonlinear gauge fields (see Appendix to Ref. \cite{ma1987}), or as certain
objects defining (semi) spray configurations in generalized Finsler and
Lagrange gravity \cite{ma1987,ma,aim}.

The N--connection coefficients can be associated to certain off--diagonal
metric coefficients $N_{i}^{b}$ in (\ref{ansatz}) when a $\left( n+m\right) $%
--splitting is prescribed for a manifold $V$ (such a manifold may be a
Riemannian or an Einstein space). We can also say that such a splitting and
corresponding coefficients $N_{i}^{a}$ induce preferred (in general,
nonholonomic) frame and/or coframe structures, respectively (\ref{dder})
and/or (\ref{ddif}). In general, this does not violate the frame and
coordinate diffeomorphisms invariance because any formulas can be written in
any system of references, or coordinates. Nevertheless, there is a class of
frame transforms (\ref{ftnc}) with coefficients (\ref{vt1}) and (\ref{vt2})
preserving the prescribed nonintegrable $\left( n+m\right) $--splitting (\ref%
{whitney}). This is like on a Schwarzschild space when we prefer the
spherical symmetry but all formulas can be written in arbitrary coordinates
(frames), for instance, in Cartezian coordinates.

Formal $\left( n+m\right) $--splitting exist naturally on vector/ tangent
bundles when $x^{i}$ label the base space coordinates and $y^{a}$ label the
fiber coordinates. If such splitting are defined by nonintegrable
distributions, we also get N--connection structures. In order to give to the
N--connections a gauge like interpretation, we can say that they broke
nonholonomically the spacetime symmetry and define certain type of nonlinear
gauge fields. In this work we shall not consider gauge like models of
locally anisotropic gravity (see \cite{vggr,vdgrg,dvgrg,vgonch,vesnc,vsgg}).

On N--anholonomic manifolds, we can say that the coordinates $x^{i}$ are
holonomic and the coordinates $y^{a}$ are nonholonomic (on N--anholonomic
vector bundles, such coordinates are called respectively to be the
horizontal and vertical ones). We conclude that a N--anholonomic manifold $%
\mathbf{V}$ provided with a metric structure $\breve{g}$ (\ref{metr})
(equivalently, with a d--metric (\ref{m1})) is a usual manifold (in
particular, a pseudo--Riemannian one) with a prescribed nonholonomic $n+m$
splitting into conventional ``horizontal'' and ``vertical'' subspaces (\ref%
{whitney}) induced by the ``off--diagonal'' terms $N_{i}^{b}(u)$ and the
corresponding preferred nonholonomic frame structure (\ref{anhrel}).

\subsubsection{d--torsions and d--curvatures}

From the general class of linear connections which can be defined on a
manifold $V,$ and any its N--anholonomic versions $\mathbf{V},$ we
distinguish those which are adapted to a N--connection structure $\mathbf{N.}
$

A \textbf{distinguished connection (d--connection) }$\mathbf{D}$ on a
N--anho\-lo\-no\-mic manifold $\mathbf{V}$ is a linear connection conserving
under parallelism the Whitney sum (\ref{whitney}). For any d--vector $%
\mathbf{X,}$ there is a decomposition of $\mathbf{D}$ into h-- and
v--covariant derivatives,%
\begin{equation}
\mathbf{D}_{\mathbf{X}}\mathbf{\doteqdot X}\rfloor \mathbf{D=}\ hX\rfloor
\mathbf{D+}\ vX\rfloor \mathbf{D=}Dh_{X}+D_{vX}=hD_{X}+vD_{X}.
\label{dconcov}
\end{equation}%
The symbol ''$\rfloor "$ in (\ref{dconcov}) denotes the interior product
defined by a metric (\ref{metr}) (equivalently, by a d--metric (\ref{m1})).
The N--adapted components $\mathbf{\Gamma }_{\ \beta \gamma }^{\alpha }$ of
a d--connection $\mathbf{D}_{\alpha }=(\mathbf{e}_{\alpha }\rfloor \mathbf{D}%
)$ are defined by the equations
\begin{equation}
\mathbf{D}_{\alpha }\mathbf{e}_{\beta }=\mathbf{\Gamma }_{\ \alpha \beta
}^{\gamma }\mathbf{e}_{\gamma },\mbox{\ or \ }\mathbf{\Gamma }_{\ \alpha
\beta }^{\gamma }\left( u\right) =\left( \mathbf{D}_{\alpha }\mathbf{e}%
_{\beta }\right) \rfloor \mathbf{e}^{\gamma }.  \label{dcon1}
\end{equation}%
The N--adapted splitting into h-- and v--covariant derivatives is stated by
\begin{equation*}
h\mathbf{D}=\{\mathbf{D}_{k}=\left( L_{jk}^{i},L_{bk\;}^{a}\right) \},%
\mbox{
and }\ v\mathbf{D}=\{\mathbf{D}_{c}=\left( C_{jc}^{i},C_{bc}^{a}\right) \},
\end{equation*}%
where
\begin{equation*}
L_{jk}^{i}=\left( \mathbf{D}_{k}\mathbf{e}_{j}\right) \rfloor e^{i},\quad
L_{bk}^{a}=\left( \mathbf{D}_{k}e_{b}\right) \rfloor \mathbf{e}%
^{a},~C_{jc}^{i}=\left( \mathbf{D}_{c}\mathbf{e}_{j}\right) \rfloor
e^{i},\quad C_{bc}^{a}=\left( \mathbf{D}_{c}e_{b}\right) \rfloor \mathbf{e}%
^{a}.
\end{equation*}%
The components $\mathbf{\Gamma }_{\ \alpha \beta }^{\gamma }=\left(
L_{jk}^{i},L_{bk}^{a},C_{jc}^{i},C_{bc}^{a}\right) $ completely define a
d--connec\-ti\-on $\mathbf{D}$ on a N--anholonomic manifold $\mathbf{V}.$ We
shall write conventionally that $\mathbf{D=}(hD,\ vD),$ or $\mathbf{D}%
_{\alpha }=(D_{i},D_{a}),$ with $hD=(L_{jk}^{i},L_{bk}^{a})$ and $%
vD=(C_{jc}^{i},$ $C_{bc}^{a}),$ see (\ref{dcon1}).

The \textbf{torsion and curvature} of a d--connection $\mathbf{D=}(hD,\ vD),$
\textbf{d--torsions and d--curvatures},\textbf{\ }are defined similarly to
formulas (\ref{ators}) and (\ref{acurv}) with further h-- and
v--decompositions. \ The simplest way to perform computations with
d--connections is to use \textbf{N--adapted differential forms} like
\begin{equation}
\mathbf{\Gamma }_{\ \beta }^{\alpha }=\mathbf{\Gamma }_{\ \beta \gamma
}^{\alpha }\mathbf{e}^{\gamma }  \label{dconf}
\end{equation}%
with the coefficients defined with respect to (\ref{ddif}) and (\ref{dder}).
For instance, torsion can be computed in the form
\begin{equation}
\mathcal{T}^{\alpha }\doteqdot \mathbf{De}^{\alpha }=d\mathbf{e}^{\alpha
}+\Gamma _{\ \beta }^{\alpha }\wedge \mathbf{e}^{\beta }.  \label{tors}
\end{equation}%
Locally it is characterized by (N--adapted) d--torsion coefficients
\begin{eqnarray}
T_{\ jk}^{i} &=&L_{\ jk}^{i}-L_{\ kj}^{i},\ T_{\ ja}^{i}=-T_{\ aj}^{i}=C_{\
ja}^{i},\ T_{\ ji}^{a}=\Omega _{\ ji}^{a},\   \notag \\
T_{\ bi}^{a} &=&-T_{\ ib}^{a}=\frac{\partial N_{i}^{a}}{\partial y^{b}}-L_{\
bi}^{a},\ T_{\ bc}^{a}=C_{\ bc}^{a}-C_{\ cb}^{a}.  \label{dtors}
\end{eqnarray}%
By a straightforward d--form calculus, we can compute the N--adapted
components $\mathbf{R=\{\mathbf{R}_{\ \beta \gamma \delta }^{\alpha }\}}$ of
the curvature
\begin{equation}
\mathcal{R}_{~\beta }^{\alpha }\doteqdot \mathbf{D\Gamma }_{\ \beta
}^{\alpha }=d\mathbf{\Gamma }_{\ \beta }^{\alpha }-\mathbf{\Gamma }_{\ \beta
}^{\gamma }\wedge \mathbf{\Gamma }_{\ \gamma }^{\alpha }=\mathbf{R}_{\ \beta
\gamma \delta }^{\alpha }\mathbf{e}^{\gamma }\wedge \mathbf{e}^{\delta },
\label{curv}
\end{equation}%
of a d--connection $\mathbf{D},$
\begin{eqnarray}
R_{\ hjk}^{i} &=&e_{k}L_{\ hj}^{i}-e_{j}L_{\ hk}^{i}+L_{\ hj}^{m}L_{\
mk}^{i}-L_{\ hk}^{m}L_{\ mj}^{i}-C_{\ ha}^{i}\Omega _{\ kj}^{a},  \notag \\
R_{\ bjk}^{a} &=&e_{k}L_{\ bj}^{a}-e_{j}L_{\ bk}^{a}+L_{\ bj}^{c}L_{\
ck}^{a}-L_{\ bk}^{c}L_{\ cj}^{a}-C_{\ bc}^{a}\Omega _{\ kj}^{c},  \notag \\
R_{\ jka}^{i} &=&e_{a}L_{\ jk}^{i}-D_{k}C_{\ ja}^{i}+C_{\ jb}^{i}T_{\
ka}^{b},  \label{dcurv} \\
R_{\ bka}^{c} &=&e_{a}L_{\ bk}^{c}-D_{k}C_{\ ba}^{c}+C_{\ bd}^{c}T_{\
ka}^{c},  \notag \\
R_{\ jbc}^{i} &=&e_{c}C_{\ jb}^{i}-e_{b}C_{\ jc}^{i}+C_{\ jb}^{h}C_{\
hc}^{i}-C_{\ jc}^{h}C_{\ hb}^{i},  \notag \\
R_{\ bcd}^{a} &=&e_{d}C_{\ bc}^{a}-e_{c}C_{\ bd}^{a}+C_{\ bc}^{e}C_{\
ed}^{a}-C_{\ bd}^{e}C_{\ ec}^{a}.  \notag
\end{eqnarray}

Contracting respectively the components of (\ref{dcurv}), one proves that
the Ricci tensor $\mathbf{R}_{\alpha \beta }\doteqdot \mathbf{R}_{\ \alpha
\beta \tau }^{\tau }$ is characterized by h- v--components, i.e. d--tensors,%
\begin{equation}
R_{ij}\doteqdot R_{\ ijk}^{k},\ \ R_{ia}\doteqdot -R_{\ ika}^{k},\
R_{ai}\doteqdot R_{\ aib}^{b},\ R_{ab}\doteqdot R_{\ abc}^{c}.
\label{dricci}
\end{equation}%
It should be noted that this tensor is not symmetric for arbitrary
d--connecti\-ons $\mathbf{D,}$ i.e. $\mathbf{R}_{\alpha \beta }\neq \mathbf{R%
}_{\beta \alpha }.$

The \textbf{scalar curvature} of a d--connection is
\begin{equation}
\ ^{s}\mathbf{R}\doteqdot \mathbf{g}^{\alpha \beta }\mathbf{R}_{\alpha \beta
}=g^{ij}R_{ij}+h^{ab}R_{ab},  \label{sdccurv}
\end{equation}%
defined by a sum the h-- and v--components of (\ref{dricci}) and d--metric (%
\ref{m1}).

The Einstein d--tensor is defined and computed similarly to (\ref{einstgr}),
but for d--connections,
\begin{equation}
\mathbf{E}_{\alpha \beta }=\mathbf{R}_{\alpha \beta }-\frac{1}{2}\mathbf{g}%
_{\alpha \beta }\ ^{s}\mathbf{R}  \label{enstdt}
\end{equation}%
This \ d--tensor defines an alternative to $\ _{\shortmid }E_{\alpha \beta }$
(nonholonomic) Einstein configuration if its d--connection is defined in a
unique form for an off--diagonal metric (\ref{ansatz}).

\subsubsection{Some classes of distinguished or non--adapted linear
connections}

From the class of arbitrary d--connections $\mathbf{D}$ on $\mathbf{V,}$ one
distinguishes those which are \textbf{metric compatible (metrical
d--connections)} satisfying the condition%
\begin{equation}
\mathbf{Dg=0}  \label{metcomp}
\end{equation}%
including all h- and v-projections
\begin{equation*}
D_{j}g_{kl}=0,D_{a}g_{kl}=0,D_{j}h_{ab}=0,D_{a}h_{bc}=0.
\end{equation*}%
Different approaches to Finsler--Lagrange geometry modelled on $\mathbf{TM}$
(or on the dual tangent bundle $\mathbf{T}^{\ast }\mathbf{M,}$ in the case
of Cartan--Hamilton geometry) were elaborated for different d--metric
structures which are metric compatible \cite%
{cart,ma1987,ma,mhf,mhss,mhh,vncsup,vmon1,vstav} or not metric compatible %
\cite{bcs}.

For any d--metric $\mathbf{g}=[hg,vg]$ on a N--anholonomic manifold $\mathbf{%
V,}$ there is a unique metric canonical d--connection $\widehat{\mathbf{D}}$
satisfying the conditions $\widehat{\mathbf{D}}\mathbf{g=}0$ and with
vanishing $h(hh)$--torsion, $v(vv)$--torsion, i. e. $h\widehat{T}(hX,hY)=0$
and $\mathbf{\ }v\widehat{T}(vX,\mathbf{\ }vY)=0.$ By straightforward
calculations, we can verify that $\widehat{\mathbf{\Gamma }}_{\ \alpha \beta
}^{\gamma }=\left( \widehat{L}_{jk}^{i},\widehat{L}_{bk}^{a},\widehat{C}%
_{jc}^{i},\widehat{C}_{bc}^{a}\right) ,$ when
\begin{eqnarray}
\widehat{L}_{jk}^{i} &=&\frac{1}{2}g^{ir}\left(
e_{k}g_{jr}+e_{j}g_{kr}-e_{r}g_{jk}\right) ,  \label{candcon} \\
\widehat{L}_{bk}^{a} &=&e_{b}(N_{k}^{a})+\frac{1}{2}h^{ac}\left(
e_{k}h_{bc}-h_{dc}\ e_{b}N_{k}^{d}-h_{db}\ e_{c}N_{k}^{d}\right) ,  \notag \\
\widehat{C}_{jc}^{i} &=&\frac{1}{2}g^{ik}e_{c}g_{jk},\ \widehat{C}_{bc}^{a}=%
\frac{1}{2}h^{ad}\left( e_{c}h_{bd}+e_{c}h_{cd}-e_{d}h_{bc}\right)  \notag
\end{eqnarray}%
result in $\widehat{T}_{\ jk}^{i}=0$ and $\widehat{T}_{\ bc}^{a}=0$ but $%
\widehat{T}_{\ ja}^{i},\widehat{T}_{\ ji}^{a}$ and $\widehat{T}_{\ bi}^{a}$
are not zero, see formulas (\ref{dtors}) written for this canonical
d--connection.

For any metric structure $\mathbf{g}$ on a manifold $\mathbf{V,}$ there is a
unique metric compatible and torsionless \textbf{Levi Civita connection} $%
\bigtriangledown =\{\ _{\shortmid }\Gamma _{\beta \gamma }^{\alpha }\}$ for
which $~\ _{\shortmid }\mathcal{T}=0$ and $\bigtriangledown g=0\mathbf{.}$
This is not a d--connection because it does not preserve under parallelism
the N--connection splitting (\ref{whitney}) (it is not adapted to the
N--connection structure). \ Let us parametrize its coefficients in the form
\begin{equation*}
_{\shortmid }\Gamma _{\beta \gamma }^{\alpha }=\left( _{\shortmid
}L_{jk}^{i},_{\shortmid }L_{jk}^{a},_{\shortmid }L_{bk}^{i},\ _{\shortmid
}L_{bk}^{a},_{\shortmid }C_{jb}^{i},_{\shortmid }C_{jb}^{a},_{\shortmid
}C_{bc}^{i},_{\shortmid }C_{bc}^{a}\right) ,
\end{equation*}%
where
\begin{eqnarray*}
\bigtriangledown _{\mathbf{e}_{k}}(\mathbf{e}_{j}) &=&\ _{\shortmid
}L_{jk}^{i}\mathbf{e}_{i}+\ _{\shortmid }L_{jk}^{a}e_{a},\ \bigtriangledown
_{\mathbf{e}_{k}}(e_{b})=\ _{\shortmid }L_{bk}^{i}\mathbf{e}_{i}+\
_{\shortmid }L_{bk}^{a}e_{a}, \\
\bigtriangledown _{e_{b}}(\mathbf{e}_{j}) &=&\ _{\shortmid }C_{jb}^{i}%
\mathbf{e}_{i}+\ _{\shortmid }C_{jb}^{a}e_{a},\ \bigtriangledown
_{e_{c}}(e_{b})=\ _{\shortmid }C_{bc}^{i}\mathbf{e}_{i}+\ _{\shortmid
}C_{bc}^{a}e_{a}.
\end{eqnarray*}%
A straightforward calculus\footnote{%
Such results were originally considered by R. Miron and M. Anastasiei for
vector bundles provided with N--connection and metric structures, see Ref. %
\cite{ma}. Similar proofs hold true for any nonholonomic manifold provided
with a prescribed N--connection structure \cite{vsgg}.} shows that the
coefficients of the Levi--Civita connection can be expressed in the form%
\begin{eqnarray}
\ _{\shortmid }L_{jk}^{i} &=&L_{jk}^{i},\ _{\shortmid
}L_{jk}^{a}=-C_{jb}^{i}g_{ik}h^{ab}-\frac{1}{2}\Omega _{jk}^{a},
\label{lccon} \\
\ _{\shortmid }L_{bk}^{i} &=&\frac{1}{2}\Omega _{jk}^{c}h_{cb}g^{ji}-\frac{1%
}{2}(\delta _{j}^{i}\delta _{k}^{h}-g_{jk}g^{ih})C_{hb}^{j},  \notag \\
\ _{\shortmid }L_{bk}^{a} &=&L_{bk}^{a}+\frac{1}{2}(\delta _{c}^{a}\delta
_{d}^{b}+h_{cd}h^{ab})\left[ L_{bk}^{c}-e_{b}(N_{k}^{c})\right] ,  \notag \\
\ _{\shortmid }C_{kb}^{i} &=&C_{kb}^{i}+\frac{1}{2}\Omega
_{jk}^{a}h_{cb}g^{ji}+\frac{1}{2}(\delta _{j}^{i}\delta
_{k}^{h}-g_{jk}g^{ih})C_{hb}^{j},  \notag \\
\ _{\shortmid }C_{jb}^{a} &=&-\frac{1}{2}(\delta _{c}^{a}\delta
_{b}^{d}-h_{cb}h^{ad})\left[ L_{dj}^{c}-e_{d}(N_{j}^{c})\right] ,\
_{\shortmid }C_{bc}^{a}=C_{bc}^{a},  \notag \\
\ _{\shortmid }C_{ab}^{i} &=&-\frac{g^{ij}}{2}\left\{ \left[
L_{aj}^{c}-e_{a}(N_{j}^{c})\right] h_{cb}+\left[ L_{bj}^{c}-e_{b}(N_{j}^{c})%
\right] h_{ca}\right\} ,  \notag
\end{eqnarray}%
where $\Omega _{jk}^{a}$ are computed as in formula (\ref{ncurv}). For
certain considerations, it is convenient to express
\begin{equation}
\ _{\shortmid }\Gamma _{\ \alpha \beta }^{\gamma }=\widehat{\mathbf{\Gamma }}%
_{\ \alpha \beta }^{\gamma }+\ _{\shortmid }Z_{\ \alpha \beta }^{\gamma }
\label{cdeft}
\end{equation}%
where the explicit components of \textbf{distorsion tensor} $\ _{\shortmid
}Z_{\ \alpha \beta }^{\gamma }$ can be defined by comparing the formulas (%
\ref{lccon}) and (\ref{candcon}):%
\begin{eqnarray}
\ _{\shortmid }Z_{jk}^{i} &=&0,\ _{\shortmid
}Z_{jk}^{a}=-C_{jb}^{i}g_{ik}h^{ab}-\frac{1}{2}\Omega _{jk}^{a},  \notag \\
\ _{\shortmid }Z_{bk}^{i} &=&\frac{1}{2}\Omega _{jk}^{c}h_{cb}g^{ji}-\frac{1%
}{2}(\delta _{j}^{i}\delta _{k}^{h}-g_{jk}g^{ih})C_{hb}^{j},  \notag \\
\ _{\shortmid }Z_{bk}^{a} &=&\frac{1}{2}(\delta _{c}^{a}\delta
_{d}^{b}+h_{cd}h^{ab})\left[ L_{bk}^{c}-e_{b}(N_{k}^{c})\right] ,  \notag \\
\ _{\shortmid }Z_{kb}^{i} &=&\frac{1}{2}\Omega _{jk}^{a}h_{cb}g^{ji}+\frac{1%
}{2}(\delta _{j}^{i}\delta _{k}^{h}-g_{jk}g^{ih})C_{hb}^{j},  \notag \\
\ _{\shortmid }Z_{jb}^{a} &=&-\frac{1}{2}(\delta _{c}^{a}\delta
_{b}^{d}-h_{cb}h^{ad})\left[ L_{dj}^{c}-e_{d}(N_{j}^{c})\right] ,\
_{\shortmid }Z_{bc}^{a}=0,  \label{cdeftc} \\
\ _{\shortmid }Z_{ab}^{i} &=&-\frac{g^{ij}}{2}\left\{ \left[
L_{aj}^{c}-e_{a}(N_{j}^{c})\right] h_{cb}+\left[ L_{bj}^{c}-e_{b}(N_{j}^{c})%
\right] h_{ca}\right\} .  \notag
\end{eqnarray}%
It should be emphasized that all components of $\ _{\shortmid }\Gamma _{\
\alpha \beta }^{\gamma },\widehat{\mathbf{\Gamma }}_{\ \alpha \beta
}^{\gamma }$ and$\ _{\shortmid }Z_{\ \alpha \beta }^{\gamma }$ are uniquely
defined by the coefficients of d--metric (\ref{m1}) and N--connection (\ref%
{coeffnc}), or equivalently by the coefficients of the corresponding generic
off--diagonal metric\ (\ref{ansatz}).

\subsection{On equivalent (non)holonomic formulations of gra\-vity theories}

A \textbf{N--anholonomic Riemann--Cartan manifold} $\ ^{RC}\mathbf{V}$ is
defined by a d--metric $\mathbf{g}$ and a metric d--connection $\mathbf{D}$
structures. We can say that a space$\ ^{R}\widehat{\mathbf{V}}$ is a
canonical N--anholonomic Riemann manifold if its d--connecti\-on structure
is canonical, i.e. $\mathbf{D=}\widehat{\mathbf{D}}.$ The d--metric
structure $\mathbf{g}$ on$\ ^{RC}\mathbf{V}$ is of type (\ref{m1}) and
satisfies the metricity conditions (\ref{metcomp}). With respect to a local
coordinate basis, the metric $\mathbf{g}$ is parametrized by a generic
off--diagonal metric ansatz (\ref{ansatz}). For a particular case, we can
treat the torsion $\widehat{\mathbf{T}}$ as a nonholonomic frame effect
induced by a nonintegrable N--splitting. We conclude that a manifold $^{R}%
\widehat{\mathbf{V}}$ is enabled with a nontrivial torsion (\ref{dtors})
(uniquely defined by the coefficients of N--connection (\ref{coeffnc}), and
d--metric (\ref{m1}) and canonical d--connection (\ref{candcon})
structures). Nevertheless, such manifolds can be described alternatively,
equivalently, as a usual (holonomic) Riemann manifold \ with the usual Levi
Civita for the metric (\ref{metr}) with coefficients (\ref{ansatz}). We do
not distinguish the existing nonholonomic structure for such geometric
constructions.

Having prescribed a nonholonomic $n+m$ splitting on a manifold $V,$ we can
define two canonical linear connections $\nabla $ and $\widehat{\mathbf{D}}.$
Correspondingly, these connections are characterized by two curvature
tensors, $_{\shortmid }R_{~\beta \gamma \delta }^{\alpha }(\nabla )$
(computed by introducing $_{\shortmid }\Gamma _{\beta \gamma }^{\alpha }$
into (\ref{dconf}) and (\ref{curv})) and $\mathbf{R}_{~\beta \gamma \delta
}^{\alpha }(\widehat{\mathbf{D}})$ (with the N--adapted coefficients
computed following formulas (\ref{dcurv})). Contracting indices, we can
commute the Ricci tensor $Ric(\nabla )$ and the Ricci d--tensor $\mathbf{Ric}%
(\widehat{\mathbf{D}})$ following formulas (\ref{dricci}), correspondingly
written for $\nabla $ and $\widehat{\mathbf{D}}.$ Finally, using the inverse
d--tensor $\mathbf{g}^{\alpha \beta }$ for both cases, we compute the
corresponding scalar curvatures $\ ^{s}R(\nabla )$ and $\ ^{s}\mathbf{R(%
\widehat{\mathbf{D}}),}$ see formulas (\ref{sdccurv}) by contracting,
respectively, with the Ricci tensor and Ricci d--tensor.

The standard formulation of the Einstein gravity is for the connection $%
\nabla ,$ when the field equations are written in the form (\ref{einstgr}).
But it can be equivalently reformulated by using the canonical
d--connection, or other connections uniquely defined by the metric
structure. If a metric (\ref{ansatz}) $\underline{g}_{\alpha \beta }$ is a
solution of the Einstein equations$\ _{\shortmid }E_{\alpha \beta }=\Upsilon
_{\alpha \beta },$ having prescribed a $\left( n+m\right) $--decomposition,
we can define algebraically the coefficients of a N--connection, $N_{i}^{a},$
N--adapted frames $e_{\alpha }$ (\ref{dder}) and $e^{\beta }$ (\ref{ddif}),
and d--metric $\mathbf{g}_{\alpha \beta }=[g_{ij},h_{ab}]$ (\ref{m1}). The
next steps are to compute $\widehat{\mathbf{\Gamma }}_{\ \alpha \beta
}^{\gamma }, $ following formulas (\ref{candcon}), and then using (\ref%
{dcurv}), (\ref{dricci}) and (\ref{sdccurv}) for $\widehat{\mathbf{D}},$ to
define $\widehat{\mathbf{E}}_{\alpha \beta }$ (\ref{enstdt}). The Einstein
equations with matter sources, written in equivalent form by using the
canonical d--connection, are
\begin{equation}
\widehat{\mathbf{E}}_{\alpha \beta }=\mathbf{\Upsilon }_{\alpha \beta }+\
^{Z}\mathbf{\Upsilon }_{\alpha \beta },  \label{einstgrdef}
\end{equation}%
where the effective source $\ ^{Z}\mathbf{\Upsilon }_{\alpha \beta }$ is
just the deformation tensor of the Einstein tensor computed by introducing
deformation (\ref{cdeft}) into the left part of (\ref{einstgr}); all
decompositions being performed with respect to the N--adapted co--frame (\ref%
{ddif}), when $_{\shortmid }E_{\alpha \beta }=\widehat{\mathbf{E}}_{\alpha
\beta }-\ ^{Z}\mathbf{\Upsilon }_{\alpha \beta }.$ For certain matter field/
string gravity configurations, the solutions of (\ref{einstgrdef}) also
solve the equations (\ref{einstgr}). Nevertheless, because of generic
nonlinear character of gravity and gravity--matter field interactions and
functions defining nonholonomic distributions, one could be certain special
conditions when even vacuum configurations contain a different physical
information if to compare with usual holonomic ones. We analyze some
examples:

In our works \cite{vesnc,vsgg,vclalg}, we investigated a series of exact
solutions defining N--anholonomic Einstein spaces related to generic
off--diagonal solutions in general relativity by such nonholonomic
constraints when $\mathbf{Ric}(\widehat{\mathbf{D}})=Ric(\mathbf{\nabla }),$
even $\widehat{\mathbf{D}}\neq \nabla .$\footnote{%
One should be emphasized here that different type of connections on
N--anholonomic manifolds have different coordinate and frame transform
properties. It is possible, for instance, to get equalities of coefficients
for some systems of coordinates even the connections are very different. The
transformation laws of tensors and d--tensors are also different if some
objects are adapted and other are not adapted to a prescribed N--connection
structure.} In this case, for instance, the solutions of the Einstein
equations with cosmological constant $\lambda ,$%
\begin{equation}
\widehat{\mathbf{R}}_{\alpha \beta }=\lambda \mathbf{g}_{\alpha \beta }
\label{nhesp}
\end{equation}%
can be transformed into metrics for usual Einstein spaces with Levi Civita
connection $\nabla .$ The idea is that for certain general metric ansatz,
see section \ref{exsolans}, the equations (\ref{nhesp}) can be integrated in
general form just for the connection $\widehat{\mathbf{D}}$ but not for $%
\nabla .$ The nontrivial torsion components
\begin{equation}
\ \widehat{T}_{\ ja}^{i}==-\widehat{T}_{\ aj}^{i}=\widehat{C}_{\ ja}^{i},\
\widehat{T}_{\ ji}^{a}=\ \widehat{T}_{\ ij}^{a}=\Omega _{\ ji}^{a},\
\widehat{T}_{\ bi}^{a}=-\widehat{T}_{\ ib}^{a}=\frac{\partial N_{i}^{a}}{%
\partial y^{b}}-\widehat{L}_{\ bi}^{a},  \label{dtorsc}
\end{equation}%
see (\ref{dtors}), for some configurations, may be associated with an
absolute antisymmetric $H$--fields in string gravity \cite{string1,string2},
but nonholonomically transformed to N--adapted bases, see details in \cite%
{vesnc,vsgg}.

For more restricted configurations, we can search solutions with metric
ansatz defining Einstein foliated spaces, when
\begin{equation}
\Omega _{jk}^{c}=0,\ \widehat{L}_{bk}^{c}=e_{b}(N_{k}^{c}),\ \widehat{C}%
_{jb}^{i}=0,  \label{intcond}
\end{equation}%
and the d--torsion components (\ref{dtorsc}) vanish, but the N--adapted
frame structure has, in general, nontrivial anholonomy coefficients, see (%
\ref{anhrel}). One present a special interest a less constrained
configurations with $\ \widehat{T}_{\ jk}^{c}=\Omega _{jk}^{c}\neq 0$ when $%
\mathbf{Ric}(\widehat{\mathbf{D}})=Ric(\mathbf{\nabla })$ and $\widehat{T}%
_{jk}^{i}=\widehat{T}_{bc}^{a}=0,$ for certain general ansatz $\widehat{T}%
_{\ ja}^{i}=0$ and $\widehat{T}_{\ bi}^{a}=0,$ but $\widehat{\mathbf{\mathbf{%
R}}}\mathbf{_{\ \beta \gamma \delta }^{\alpha }\neq \ }_{\shortmid }R\mathbf{%
_{\ \beta \gamma \delta }^{\alpha }.}$ In such cases, we constrain the
integral varieties of equations (\ref{nhesp}) in such a manner that we
generate integrable or nonintegrable distributions on a usual Einstein space
defined by $\nabla .$ This is possible because if the conditions (\ref%
{intcond}) are satisfied, the deformation tensor $\ _{\shortmid }Z_{\ \alpha
\beta }^{\gamma }=0,$ see (\ref{cdeftc}). For $\lambda =0,$ if $n+m=4,$ for
corresponding signature, we get foliated vacuum configurations in general
relativity.

For N--anholonomic manifolds $\mathbf{V}^{n+n}$ of odd dimensions, when $%
m=n, $ and if $g_{ij}=h_{ij}$ (we identify correspondingly, the h- and
v--indices), we can consider a canonical d--connection$\ \widehat{\mathbf{D}}%
=(h\widehat{D},v\widehat{D})$ with the nontrivial coefficients with respect
to $\mathbf{e}_{\nu }$ and $\mathbf{e}^{\mu }$ paramet\-riz\-ed respectively
$\widehat{\mathbf{\Gamma }}_{\ \beta \gamma }^{\alpha }=(\widehat{L}_{\
jk}^{i}=\widehat{L}_{\ bk}^{a},\widehat{C}_{jc}^{i}=\widehat{C}_{bc}^{a}),$%
\footnote{%
the equalities of indices ''$i=a"$ are considered in the form $"i=1=a=n+1,$ $%
i=2=a=n+2,$ ... $i=n=a=n+n"$} for
\begin{eqnarray}
\widehat{L}_{\ jk}^{i} &=&\frac{1}{2}g^{ih}(\mathbf{e}_{k}g_{jh}+\mathbf{e}%
_{j}g_{kh}-\mathbf{e}_{h}g_{jk}),  \label{3cdctb} \\
\widehat{C}_{\ bc}^{a} &=&\frac{1}{2}%
g^{ae}(e_{b}g_{ec}+e_{c}g_{eb}-e_{e}g_{bc}),  \notag
\end{eqnarray}%
defining the generalized Christoffel symbols. Such nonholonomic
configurations can be used for modelling generalized Finsler--Lagrange, and
particular cases, defined in Refs. \cite{ma1987,ma} for $\mathbf{V}^{n+n}=%
\mathbf{TM,}$ see below section \ref{sfls}. \ There are only three classes
of d--curvatures for the d--connection (\ref{3cdctb}),%
\begin{eqnarray}
\widehat{R}_{\ hjk}^{i} &=&\mathbf{e}_{k}\widehat{L}_{\ hj}^{i}-\mathbf{e}%
_{j}\widehat{L}_{\ hk}^{i}+\widehat{L}_{\ hj}^{m}\widehat{L}_{\ mk}^{i}-%
\widehat{L}_{\ hk}^{m}\widehat{L}_{\ mj}^{i}-\widehat{C}_{\ ha}^{i}\Omega
_{\ kj}^{a},  \label{3dcurvtb} \\
\widehat{P}_{\ jka}^{i} &=&e_{a}\widehat{L}_{\ jk}^{i}-\widehat{\mathbf{D}}%
_{k}\widehat{C}_{\ ja}^{i},\ \widehat{S}_{\ bcd}^{a}=e_{d}\widehat{C}_{\
bc}^{a}-e_{c}\widehat{C}_{\ bd}^{a}+\widehat{C}_{\ bc}^{e}\widehat{C}_{\
ed}^{a}-\widehat{C}_{\ bd}^{e}\widehat{C}_{\ ec}^{a},  \notag
\end{eqnarray}%
where all indices $a,b,...,i,j,...$ run the same values and, for instance, $%
C_{\ bc}^{e}\rightarrow $ $C_{\ jk}^{i},...$ Such locally anisotropic
configurations are not integrable if $\Omega _{\ kj}^{a}\neq 0,$ even the
d--torsion components $\widehat{T}_{\ jk}^{i}=0$ and $\widehat{T}_{\
bc}^{a}=0.$ We note that for geometric models on $\mathbf{V}^{n+n},$ or on $%
\mathbf{TM,}$ with $g_{ij}=h_{ij},$ one writes, in brief, $\widehat{\mathbf{%
\Gamma }}_{\ \beta \gamma }^{\alpha }=\left( \widehat{L}_{\ jk}^{i},\widehat{%
C}_{\ bc}^{a}\right) ,$ or, for more general d--connections, $\mathbf{\Gamma
}_{\ \beta \gamma }^{\alpha }=\left( L_{\ jk}^{i},C_{\ bc}^{a}\right) ,$ see
below section \ref{sfls}, on Lagrange and Finsler spaces.

\section{Nonholonomic Deformations of Manifolds and Vector Bundles}

This section will deal mostly with nonholonomic distributions on manifolds
and vector/ tangent bundles and their nonholonomic deformations modelling,
on Riemann and Riemann--Cartan manifolds, different types of generalized
Finsler--Lagrange geometries.

\subsection{Finsler--Lagrange spaces and generalizations}

\label{sfls}The notion of Lagrange space was introduced by J. Kern \cite%
{kern} and elaborated in details by R. Miron's school, see Refs. \cite%
{ma1987,ma,mhl,mhf,mhss,mhh}, as a natural extension of Finsler geometry %
\cite{cart,rund,mats,bej} (see also Refs. \cite{vncsup,vmon1,vcv,vesnc,vsgg}%
, on Lagrange--Finsler super/noncommutative geometry). Originally, such
geometries were constructed on tangent bundles, but they also can be
modelled on N--anholonomic manifolds, for instance, as models for certain
gravitational interactions with prescribed nonholonomic constraints deformed
symmetries.

\subsubsection{Lagrange spaces}

A \textbf{differentiable Lagrangian} $L(x,y),$ i.e. a fundamental Lagrange
function, is defined by a map $L:(x,y)\in TM\rightarrow L(x,y)\in \mathbb{R}$
of class $\mathcal{C}^{\infty }$ on $\widetilde{TM}=TM\backslash \{0\}$ and
continuous on the null section $0:\ M\rightarrow TM$ of $\pi .$ A regular
Lagrangian has non-degenerate \textbf{Hessian}
\begin{equation}
\ ^{L}g_{ij}(x,y)=\frac{1}{2}\frac{\partial ^{2}L(x,y)}{\partial
y^{i}\partial y^{j}},  \label{lqf}
\end{equation}%
when $rank\left| g_{ij}\right| =n$ and $^{L}g^{ij}$ is the inverse matrix. A
\textbf{Lagrange space }is a pair $L^{n}=\left[ M,L(x,y)\right] $ with $\
^{L}g_{ij}$ being of fixed signature over $\mathbf{V}=\widetilde{TM}.$

One holds the results:\ The \textbf{Euler--Lagrange equations}%
\begin{equation*}
\frac{d}{d\tau }\left( \frac{\partial L}{\partial y^{i}}\right) -\frac{%
\partial L}{\partial x^{i}}=0
\end{equation*}%
where $y^{i}=\frac{dx^{i}}{d\tau }$ for $x^{i}(\tau )$ depending on
parameter $\tau ,$ are equivalent to the \textbf{``nonlinear'' geodesic
equations}
\begin{equation*}
\frac{d^{2}x^{a}}{d\tau ^{2}}+2G^{a}(x^{k},\frac{dx^{b}}{d\tau })=0
\end{equation*}%
defining paths of a canonical \textbf{semispray}%
\begin{equation*}
S=y^{i}\frac{\partial }{\partial x^{i}}-2G^{a}(x,y)\frac{\partial }{\partial
y^{a}}
\end{equation*}%
where
\begin{equation*}
2G^{i}(x,y)=\frac{1}{2}\ ^{L}g^{ij}\left( \frac{\partial ^{2}L}{\partial
y^{i}\partial x^{k}}y^{k}-\frac{\partial L}{\partial x^{i}}\right) .
\end{equation*}%
There exists on $\mathbf{V\simeq }$ $\widetilde{TM}$ a canonical
N--connection $\ $%
\begin{equation}
\ ^{L}N_{j}^{a}=\frac{\partial G^{a}(x,y)}{\partial y^{j}}  \label{cncl}
\end{equation}%
defined by the fundamental Lagrange function $L(x,y),$ which prescribes
nonholonomic frame structures of type (\ref{dder}) and (\ref{ddif}), $^{L}%
\mathbf{e}_{\nu }=(\ ^{L}\mathbf{e}_{i},e_{a})$ and $^{L}\mathbf{e}^{\mu
}=(e^{i},\ ^{L}\mathbf{e}^{a}).$ One defines the canonical metric structure%
\begin{equation}
\ ^{L}\mathbf{g}=\ ^{L}g_{ij}(x,y)\ e^{i}\otimes e^{j}+\ ^{L}g_{ij}(x,y)\
^{L}\mathbf{e}^{i}\otimes \ ^{L}\mathbf{e}^{j}  \label{slm}
\end{equation}%
constructed as a Sasaki type lift from $M$ for $\ ^{L}g_{ij}(x,y),$ see
details in \cite{yano,ma1987,ma}.

There is a unique canonical d--connection$\ ^{L}\widehat{\mathbf{D}}=(h\ ^{L}%
\widehat{D},v\ ^{L}\widehat{D})$ with the coefficients $\ ^{L}\widehat{%
\Gamma }_{\ \beta \gamma }^{\alpha }=(\ ^{L}\widehat{L}_{\ jk}^{i},\ ^{L}%
\widehat{C}_{bc}^{a})$ computed by formulas (\ref{3cdctb}) for the d--metric
(\ref{slm}) with respect to $^{L}\mathbf{e}_{\nu }$ and $^{L}\mathbf{e}^{\mu
}.$ All such geometric objects, including the corresponding to $^{L}\widehat{%
\Gamma }_{\ \beta \gamma }^{\alpha },\ ^{L}\mathbf{g}$ and $^{L}N_{j}^{a}$
d--curvatures\newline
$\ ^{L}\widehat{\mathbf{\mathbf{R}}}\mathbf{_{\ \beta \gamma \delta
}^{\alpha }=}\left( \ ^{L}\widehat{R}_{\ hjk}^{i},\ ^{L}\widehat{P}_{\
jka}^{i},\ ^{L}\widehat{S}_{\ bcd}^{a}\right) ,$ see (\ref{3dcurvtb}), are
completely defined by a Lagrange fundamental function $L(x,y)$ for a
nondegerate $^{L}g_{ij}.$

Let us consider how a Lagrange mechanics can be modelled on nonholonomic
Riemann, usual Riemann, or Riemann--Cartan manifolds. We take a manifold $%
V^{n+n},$ $\dim V=n+n,$ and consider a metric structure (\ref{ansatz}) for a
particular case when the values $g_{ij}$ and $N_{i}^{a}$ are respectively of
type (\ref{lqf}) and (\ref{cncl}) for a function $L(x,y)$ on $V,$ with
nondegenerate $\ ^{L}g_{ij}.$ The preferred frame structure on $V$ is
defined by introducing $^{L}N_{j}^{a}$ in the class of vierbein transforms (%
\ref{ftnc}) with coefficients (\ref{vt1}) and (\ref{vt2}). All data can be
redefined for a d-metric (\ref{m1}) but generated by $L$ in a form
equivalent to $\ ^{L}\mathbf{g} $ (\ref{slm}), with that difference that the
first geometric object is defined on a N--anholonomic manifold but the
second one is considered on a $\mathbf{TM}.$

The next step of modelling is to decide what kind of linear connection we
chose. There are two canonical, equivalent, possibilities. If we take (\ref%
{candcon}), on $V^{n+n},$ we model a Riemann--Cartan manifold with induced
torsion (\ref{dtorsc}), in this case, completely defined by $L$ and
respective $\ ^{L}g_{ij}$ and $^{L}N_{j}^{a}.$ We can simplify the
constructions for a normal canonical d--connection (\ref{3cdctb}) and
generate a nonholonomic Riemann manifold with nonintegrable structure $\
^{L}\Omega _{\ kj}^{a}.$ Finally, we note that all constructions can be
re--defined for the Levi Civita connection if we consider $\ _{\shortmid
}^{L}\Gamma _{\ \alpha \beta }^{\gamma }=\ ^{L}\widehat{\mathbf{\Gamma }}_{\
\alpha \beta }^{\gamma }+\ _{\shortmid }^{L}Z_{\ \alpha \beta }^{\gamma }$
of type (\ref{lccon}), where the values are computed following formulas (\ref%
{cdeft}) and (\ref{cdeftc}) (also completely defined by $L$ and respective $%
\ ^{L}g_{ij}$ and $^{L}N_{j}^{a}).$ Such constructions are not adapted to
the N--connection structure: we work with arbitrary frame and coordinate
transforms and hidden Lagrange structure which appear in explicit form only
with respect to certain preferred, N--adapted, frames of reference.

We conclude that any regular Lagrange mechanics can be geometrized as a
nonholonomic Riemann manifold $^{L}\mathbf{V}$ equipped with the canonical
N--connecti\-on $^{L}N_{j}^{a}$ (\ref{cncl}). This geometrization was
performed in such a way that the N--connecti\-on is induced canonically by
the semispray configurations subjected the condition that the generalized
nonlinear geodesic equations are equivalent to the Euler--Lagrange equations
for $L.$ Such mechanical models and semispray configurations can be used for
a study of certain classes of nonholonomic effective analogous of
gravitational interactions. The approach can be extended for more general
classes of effective metrics, then those parametrized by (\ref{slm}), see
next sections. After Kern and Miron and Anastasiei works, it was elaborated
the so--called ''analogous gravity'' approach \cite{blv} with similar ideas
modelling related to continuous mechanics, condensed media.... It should be
noted here, that the constructions for higher order generalized Lagrange and
Hamilton spaces \cite{mhl,mhf,mhss,mhh} provided a comprehensive geometric
formalism for analogous models in gravity, geometric mechanics, continuous
media, nonhomogeneous optics etc etc.

\subsubsection{Finsler spaces}

Following the ideas of the Romanian school on Finsler--Lagrange geometry and
generalizations, any Finsler space defined by a \textbf{fundamental Finsler
function} $F(x,y),$ being homogeneous of type $F(x,\lambda y)=|\lambda |\
F(x,y),$ for nonzero $\lambda \in \mathbb{R},$ may be considered as a
particular case of Lagrange space when $L=F^{2}$ (on different rigorous
mathematical definitions of Finsler spaces, see \cite%
{rund,mats,ma1987,ma,bej,bcs}; in our approach with applications to physics,
we shall not constrain ourself with special signatures, smooth class
conditions and special types of connections). Historically, the bulk of
mathematicians worked in an inverse direction, by generalizing the
constructions from the Cartan's approach to Finsler geometry in order to
include into consideration regular Lagrange mechanical systems, or to define
Finsler geometries with another type of nonlinear and linear connection
structures. The Finsler geometry, in terms of the normal canonical
d--connection (\ref{3cdctb}), derived for respective $\ ^{F}g_{ij}$ and $%
^{F}N_{j}^{a},$ can be modelled as for the case of Lagrange spaces
consudered in the previous section: we have to change formally all labels $%
L\rightarrow F$ $\ $and take into consideration possible conditions of
homogeneity (or $TM,$ see the monographs \cite{ma1987,ma}).

For generalized Finsler spaces, a N--connection can be stated by a general
set of coeficients $N_{j}^{a}$ subjected to certain nonholonomy conditions.
Of course, working with homogeneous functions on a manifold $V^{n+n},$ we
can model a Finsler geometry both on holonomic and nonholonomic Riemannian
manifolds, or on certain types of Riemann--Cartan manifolds enabled with
preferred frame structures $^{F}\mathbf{e}_{\nu }=(\ ^{F}\mathbf{e}%
_{i},e_{a})$ and $^{F}\mathbf{e}^{\mu }=(e^{i},\ ^{F}\mathbf{e}^{a}).$
Bellow, in the section \ref{exsolans}, we shall discuss how certain type
Finsler configurations can be derived as exact solutions in Einstein
gravity. Such constructions allow us to argue that Finsler geometry is also
very important in standard physics and that it was a big confusion to treat
it only as a ''sophisticated'' generalization of Riemann geometry, on
tangent bundles, with not much perspectives for modern physics.

In a number of works (see monographs \cite{ma1987,ma,bcs}), it is emphasized
that the first example of Finsler metric was considered in the famous
inauguration thesis of B. Riemann \cite{riemann}, long time before P.\
Finsler \cite{fg}. Perhaps, this is a reason, for many authors, to use the
term Riemann--Finsler geometry. Nevertheless, we would like to emphasize
that a Finsler space is not completely defined only by a metric structure of
type
\begin{equation}
\ ^{F}g_{ij}=\frac{1}{2}\frac{\partial ^{2}F}{\partial y^{i}\partial y^{j}}
\label{fhes}
\end{equation}%
originally considered on the vertical fibers of a tangent bundle. There are
necessary additional conventions about metrics on a total Finsler space,
N--connections and linear connections. This is the source for different
approaches, definitions, constructions and ambiguities related to Finsler
spaces and applications. Roughly speaking, different famous mathematicians,
and their schools, elaborated their versions of Finsler geometries following
some special purposes in geometry, mechanics and physics.

The first complete model of Finsler geometry exists due to E. Cartan \cite%
{cart} who in the 20-30th years of previous century elaborated the concepts
of vector bundles, Rieman--Cartan spaces with torsion, moving frames,
developed the theory of spinors, Pfaff forms ... and (in coordinate form)
operated with nonlinear connection coefficients. The Cartan's constructions
were performed with metric compatible linear connections which is very
important for applications to standard models in physics.

Latter, there were proposed different models of Finsler spaces with metric
not compatible linear connections. The most notable connections were those
by L. Berwald, S. -S. Chern (re--discovered by H. Rund), H. Shimada and
others (see details, discussions and bibliography in monographs \cite%
{ma1987,ma,bcs,rund}). For d--connections of type (\ref{3cdctb}), there are
distinguished three cases of metric compatibility (compare with h- and
v-projections of formula (\ref{metcomp})):\ A Finsler connection $^{F}%
\mathbf{D}_{\alpha }=(\ ^{F}D_{k},\ ^{F}D_{a})$ is called h--metric if $%
^{F}D_{i}^{F}g_{ij}=0;$ it is called v--metric if $^{F}D_{a}\ ^{F}g_{ij}=0$
and it is metrical if both conditions are satisfied.

Here, we note four of the most important Finsler d--connections having their
special geometric and (possible) physical merits:

\begin{enumerate}
\item The \textbf{canonical Finsler connection} $\ ^{F}\widehat{\mathbf{D}}$
is defined by formulas (\ref{3cdctb}), but for $^{F}g_{ij},$ i.e. as $\ ^{F}%
\widehat{\mathbf{\Gamma }}_{\ \beta \gamma }^{\alpha }=\left( \ ^{F}\widehat{%
L}_{\ jk}^{i},\ ^{F}\widehat{C}_{\ bc}^{a}\right) .$ This d--connection is
metrical. For a special class of N--connections $%
^{C}N_{j}^{a}(x^{k},y^{b})=y^{k}\ ^{C}L_{\ kj}^{i},$ we get the famous
\textbf{Cartan connection} for Finsler spaces, $\ ^{C}\mathbf{\Gamma }_{\
\beta \gamma }^{\alpha }=\left( \ ^{C}L_{\ jk}^{i},\ ^{C}C_{\ bc}^{a}\right)
,$ with
\begin{eqnarray}
\ ^{C}L_{\ jk}^{i} &=&\frac{1}{2}\ ^{F}g^{ih}(\ ^{C}\mathbf{e}_{k}\
^{F}g_{jh}+\ ^{C}\mathbf{e}_{j}\ ^{F}g_{kh}-\ ^{C}\mathbf{e}_{h}\
^{F}g_{jk}),  \label{cartan} \\
\ ^{C}C_{\ bc}^{a} &=&\frac{1}{2}\ ^{F}g^{ae}(e_{b}\ ^{F}g_{ec}+e_{c}\
^{F}g_{eb}-e_{e}\ ^{F}g_{bc}),  \notag
\end{eqnarray}%
where
\begin{equation*}
\ ^{C}\mathbf{e}_{k}=\frac{\partial }{\partial x^{k}}-\ ^{C}N_{j}^{a}\frac{%
\partial }{\partial y^{a}}\mbox{\ and \ }e_{b}=\frac{\partial }{\partial
y^{b}},
\end{equation*}%
which can be defined in a unique axiomatic form \cite{mats}. Such canonical
and Cartan--Finsler connections, being metric compatible, for nonholonomic
geometric models with local anisotropy on Riemann or Riemann--Cartan
manifolds, are more suitable with the paradigm of modern standard physics.

\item The \textbf{Berwald connection} $\ ^{B}\mathbf{D}$ was introduced in
the form $\ ^{B}\mathbf{\Gamma }_{\ \beta \gamma }^{\alpha }=\left( \frac{%
\partial \ ^{C}N_{j}^{b}}{\partial y^{a}},0\right)$ \cite{bw}. This
d--connection is defined completely by the N--connec\-tion structure but it
is not metric compatible, both not h--metric and not v--metric.

\item The \textbf{Chern connection} $^{Ch}\mathbf{D}$ was considered as a
minimal Finsler extension of the Levi Civita connection, $\ ^{Ch}\mathbf{%
\Gamma }_{\ \beta \gamma }^{\alpha }=\left( \ ^{C}L_{\ jk}^{i},0\right) ,$
with $\ ^{C}L_{\ jk}^{i}$ defined as in (\ref{cartan}), preserving the
torsionless condition, being h--metric but not v--metric. It is an
interesting case of nonholonomic geometries when torsion is completely
transformed into nonmetricity which for physicists presented a substantial
interest in connection to the Weyl nonmetricity introduced as a method of
preserving conformal symmetry of certain scalar field constructions in
general relativity, see discussion in \cite{mag}. Nevertheless, it should be
noted that the constructions with the Chern connection, in general, are not
metric compatible and can not be applied in direct form to standard models
of physics.

\item There is also the \textbf{Hashiguchi connection} $^{H}\mathbf{\Gamma }%
_{\ \beta \gamma }^{\alpha }=\left( \frac{\partial \ ^{C}N_{j}^{b}}{\partial
y^{a}},\ ^{C}C_{\ bc}^{a}\right) ,$ with $\ \ ^{C}C_{\ bc}^{a}$ defined as
in (\ref{cartan}), which is v--metrical but not h--metrical.
\end{enumerate}

It should be noted that all mentioned types of d--connections are uniquely
defined by the coefficients of Finsler type d--metric and N--connection
structure (equivalently, by the coefficients of corresponding generic
off--diagonal metric of type (\ref{ansatz})) following well defined
geometric conditions. From such d--connections, we can always 'extract' the
Levi Civita connection, using formulas of type (\ref{lccon}), (\ref{cdeft})
and (\ref{cdeftc}), and work in 'non--adapted' (to N--connection) form. From
geometric point of view, we can work with all types of Finsler connections
and elaborate equivalent approaches even different connections have
different merits in some directions of physics. For instance, in \cite%
{ma1987,ma}, there are considered the Kawaguchi metrization procedure and
the Miron's method of computing all metric compatible Finsler connections
starting with a canonical one. It was analyzed also the problem of
transforming one Finsler connection into different ones on tangent bundles
and the formalism of mutual transforms of connections was reconsidered for
nonholonomic manifolds, see details in \cite{vsgg}.

Different models of Finsler spaces can be elaborated in explicit form for
different types of d--metrics, N--connections and d--connections. For
instance, for a Finsler Hessian (\ref{fhes}) defining a particular case of
d--metrics (\ref{slm}), or (\ref{m1}), denoted $\ ^{F}\mathbf{g,}$ for any
type of connection (for instance, canonical d--connection, Cartan--Finsler,
Berwald, Chern, Hashigushi etc), we can compute the curvatures by using
formulas (\ref{3dcurvtb}) when ''hat'' labels are changed into the
corresponding ones ''$C,B,Ch,H...".$ This way, we model Finsler geometries
on tangent bundles, like it is considered in the bulk of monographs \cite%
{cart,mats,rund,ma1987,ma,bej,bcs}, or on nonholonomic manifolds \cite%
{vr1,vr2,hor,bejf,vsgg}.

With the aim to develop new applications in standard models of physics, let
say in classical general relativity, when Finsler like structures are
modelled on a (pseudo) Riemannian manifold (we shall consider explicit
examples in the next sections), it is positively sure that the canonical
Finsler and Cartan connections, and their variants of canonical
d--connection on vector bundles and nonholonomic manifolds, should be
preferred for constructing new classes of Einstein spaces and defining
certain low energy limits to locally anisotropic string gravity models. Here
we note that it is a very difficult problem to define Finsler--Clifford
spaces with Finsler spinors, noncommutative generalizations to
supersymmetric/ noncommutative Finsler geometry if we work with nonmetric
d--connections, see discussions in \cite{vsgg,vstav,vmon1}.

We cite a proof \cite{bcs} that any Lagrange fundamental function $L$ can be
modelled as a singular case for a certain class of Finsler geometries of
extra dimension (perhaps, the authors were oriented to prove from a
mathematical point of view that it is not necessary to develop Finsler
geometry as a new theory for Lagrange spaces, or their dual constructions
for the Hamilton spaces). This idea, together with the method of
Kawaguchi--Miron transforms of connections, can be related to the H.
Poincare philosophical concepts about conventionality of the geometric space
and field interaction theories \cite{poinc1,poinc2}. According to the
Poincare's geometry--physics dualism, the procedure of choosing a geometric
arena for a physical theory is a question of convenience for researches to
classify scientific data in an economical way, but not an action to be
verified in physical experiments: as a matter of principe, any physical
theory can be equivalently described on various types of geometric spaces by
using more or less "simple" geometric objects and transforms.

Nevertheless, the modern physics paradigm is based on the ideas of objective
reality of physical laws and their experimental and theoretical
verifications, at least in indirect form. The concept of Lagrangian is a
very important geometrical and physical one and we shall distinguish the
cases when we model a Lagrange or a Finsler geometry. A physical or
mechanical model with a Lagrangian is not only a ''singular'' case for a
Finsler geometry but reflects a proper set of \ concepts, fundamental
physical laws and symmetries describing real physical effects. We use the
terms Finsler and Lagrange spaces in order to emphasize that they are
different both from geometric and physical points of view. Certain geometric
concepts and methods (like the N--connection geometry and nonholonomic frame
transforms ...) are very important for both types of geometries, modelled on
tangent bundles or on nonholonomic manifolds. This will be noted when we use
the term Finsler--Lagrange geometry (structures, configurations, spaces).

One should be emphasized that the author of this review should not be
considered as a physicist who does not accept nonmetric geometric
constructions in modern physics. For instance, the Part I in monograph \cite%
{vsgg} is devoted to a deep study of the problem when generalized
Finsler--Lagrange structures can be modelled on metric--affine spaces, even
as exact solutions in gravity with nonmetricity \cite{mag}, and, inversely,
the Lagrange--affine and Finsler--affine spaces are classified by
nonholonomic structures on metric--affine spaces. It is a question of
convention on the type of physical theories one models by geometric methods.
The standard theories of physics are formulated for metric compatible
geometries, but further developments in quantum gravity may request certain
type of nonmetric Finsler like geometries, or more general constructions.
This is a topic for further investigations.

\subsubsection{Generalized Lagrange spaces}

There are various application in optics of nonhomogeneous media and gravity
(see, for instance, Refs. \cite{ma,vsgg,esv,vesnc}) considering metrics of
type $g_{ij}\sim e^{\lambda (x,y)}\ ^{L}g_{ij}(x,y)\ $\ which can not be
derived directly from a mechanical Lagrangian. The ideas and methods to work
with arbitrary symmetric and nondegenerated tensor fields $g_{ij}(x,y)$ were
concluded in geometric and physical models for generalized Lagrange spaces,
denoted $GL^{n}=(M,g_{ij}(x,y)),$ on $\widetilde{TM},$ see \cite{ma1987,ma},
where $g_{ij}(x,y)$ is called the \textbf{fundamental tensor field}. Of
course, the geometric constructions will be equivalent if we shall work on
N--anholonomic manifolds $\mathbf{V}^{n+n}$ with nonholonomic coordinates $%
y. $ If we prescribe an arbitrary N--connection $N_{i}^{a}(x,y)$ and
consider that a metric $g_{ij}$ defines both the h-- and -v--components of a
d--metric (\ref{m1}), we can introduce the canonical d--connection (\ref%
{3cdctb}) and compute the components of d--curvature (\ref{3dcurvtb}),
define Ricci and Einstein tensors, elaborate generalized Lagrange models of
gravity.

If we work with a general fundamental tensor field $g_{ij}$ which can not be
transformed into $\ ^{L}g_{ij},$ we can consider an effective Lagrange
function \footnote{%
in \cite{ma1987,ma}, it is called the absolute energy of a $GL^{n}$--space,
but for further applications in modern gravity the term ''energy'' may
result in certain type ambiguities},
\begin{equation*}
\mathcal{L}(x,y)\doteqdot g_{ab}(x,y)y^{a}y^{b}
\end{equation*}%
and use
\begin{equation}
\ ^{\mathcal{L}}g_{ab}\doteqdot \frac{1}{2}\frac{\partial ^{2}\mathcal{L}}{%
\partial y^{a}\partial y^{b}}  \label{glhf}
\end{equation}%
as a Lagrange Hessian (\ref{slm}). A space $GL^{n}=(M,g_{ij}(x,y))$ is said
to be with a weakly regular metric if $L^{n}=\left[ M,L=\sqrt{|\mathcal{L}|})%
\right] $ is a Lagrange space. For such spaces, we can define a canonical
nonlinear connection structure%
\begin{equation}
\ ^{\mathcal{L}}N_{j}^{a}(x,y)\doteqdot \frac{\partial \ ^{\mathcal{L}}G^{a}%
}{\partial y^{j}},  \label{glnc}
\end{equation}%
for
\begin{eqnarray*}
\ ^{\mathcal{L}}G^{a} &=&\frac{1}{4}\ ^{\mathcal{L}}g^{ab}\left( y^{k}\frac{%
\partial \mathcal{L}}{\partial y^{b}\partial x^{k}}-\frac{\partial \mathcal{L%
}}{\partial x^{a}}\right) \\
&=&\frac{1}{4}\ ^{\mathcal{L}}g^{ab}\left( \frac{\partial g_{bc}}{\partial
y^{d}}+\frac{\partial g_{bd}}{\partial y^{c}}-\frac{\partial g_{cd}}{%
\partial y^{b}}\right) y^{c}y^{d},
\end{eqnarray*}%
which allows us to write $\ ^{\mathcal{L}}N_{j}^{a}$ is terms of the
fundamental tensor field $g_{ij}(x,y).$ The geometry of such generalized
Lagrange spaces is completely similar to that of usual Lagrange one, with
that difference that we start not with a Lagrangian but with a fundamental
tensor field.

In our papers \cite{vsh1,vsh2}, we proposed to see also nonholonomic
transforms of a metric $\ ^{\mathcal{L}}g_{a^{\prime }b^{\prime }}(x,y)$
\begin{equation}
g_{ab}(x,y)=e_{a}^{\ a^{\prime }}(x,y)e_{b}^{\ b^{\prime }}(x,y)\ ^{\mathcal{%
L}}g_{a^{\prime }b^{\prime }}(x,y)  \label{ftgfs}
\end{equation}%
when
\begin{equation*}
\ ^{\mathcal{L}}g_{a^{\prime }b^{\prime }}\doteqdot \frac{1}{2}\left(
e_{a^{\prime }}e_{b^{\prime }}\mathcal{L}+e_{b^{\prime }}e_{a^{\prime }}%
\mathcal{L}\right) =\ ^{0}g_{a^{\prime }b^{\prime }},
\end{equation*}%
for $e_{a^{\prime }}=e_{\ a^{\prime }}^{a}(x,y)\frac{\partial }{\partial
y^{a}},$ where $\ ^{0}g_{a^{\prime }b^{\prime }}$ are constant coefficients
(or in a more general case, they should result in a constant matrix for the
d--curvatures (\ref{dcurv}) of a canonical d--connection (\ref{candcon})).
Such constructions allowed to derive proper solitonic hierarchies and
bi--Hamilton structures for any (pseudo) Riemannian or generalized
Finsler--Lagrange metric. The point was to work not with the Levi Civita
connection (for which the solitonic equations became very cumbersome) but
with a correspondingly defined canonical d--connection allowing to apply
well defined methods from the geometry of nonlinear connections. Having
encoded the ''gravity and geometric mechanics'' information into solitonic
hierarchies and convenient d--connections, the constructions were shown to
hold true if they are ''inverted'' to those with usual Levi Civita
connections.

\subsection{Effective Finsler--Lagrange (algebroid) structures}

All valuable physical solutions in general gravity and generalizations are
characterized by corresponding symmetries of spacetime metrics, see reviews
of such constructions in Refs. \cite{kramer,bic} and, for nonholonomic
solutions, \cite{vsgg}. For instance, a special importance is given to
spacetimes with spherical, cylindrical, or toroidal symmetries, and, in
general, to gravitational distributions characterized by Killing symmetries,
or by metrics conformal to the flat Minkowski metric. In this section, we
show that, as a matter of principle, any (pseudo) Riemannian manifold can be
deformed by nonholonomic trasforms into some classes of N--anholonomic
manifolds and/or gravitational Lie algebroid configurations \cite{vclalg}.

We note that applying any general frame or coordinate transforms (\ref{ft}),
when the metric transforms are of type $g_{\alpha \beta }=A_{\alpha }^{\
\alpha ^{\prime }}(u)A_{\beta }^{\ \beta ^{\prime }}(u)g_{\alpha ^{\prime
}\beta ^{\prime }},$ a Lagrange, or Finsler, structure, characterized by a
''prime'' d--metrics of type (\ref{slm}), with coefficients (\ref{lqf}), or (%
\ref{fhes}), became ''hidden'' into some general formulas for metrics and
linear connections. The bulk of solutions in Einstein gravity and
generalizations can be associated to certain models of generalized Lagrange
(or Finsler) spaces not in explicit form but via some special types of
nonholonomic deformations (transforms) of the frame/ N--connection, metric
and linear connection structures.

The aim of this section is to examine some important examples of
nonholonomic transforms preserving the N--connection splitting (\ref{whitney}%
).

\subsubsection{N--adapted nonholonomic transforms}

\label{ssnft}We use the term ''transforms/ transformations'' of geometric
objects if we work with usual transforms of the local frames. The spacetime
geometry and geometrical objects are not changed under such transforms.
Fixing a frame structure (holonomic or nonholonomic one), we can consider
''deformations'' of geometric objects induced by deformations of the metric,
or linear connection structure.

The spacetime geometry and geometric objects are changed under holonomic of
nonholonomic deformations. For instance, fixing a co-frame $e^{\alpha
^{\prime }}=du^{\alpha ^{\prime }},$ we define conformal maps of metrics as
local re--scaling of metric coefficients,%
\begin{equation*}
g=g_{\alpha ^{\prime }\beta ^{\prime }}\ e^{\alpha ^{\prime }}\otimes
e^{\beta ^{\prime }}\rightarrow \ ^{\omega }g=\omega ^{2}g_{\alpha \beta }\
e^{\alpha }\otimes e^{\beta },
\end{equation*}%
i.e. $\ ^{\omega }g_{\alpha \beta }=\omega ^{2}(u)g_{\alpha \beta }(u),$
when $e^{\alpha }=\delta _{\alpha ^{\prime }}^{\alpha }e^{\alpha ^{\prime }}$
which mean that we deform the metric structure and the spacetime geometry is
changed under such maps (we get another types of connections and
curvatures). In an alternative form, we can say that a conformal transform
of metric, $\ ^{\omega }g,$ is generated from $g$ by an active frame
transform $e^{\alpha ^{\prime }}\rightarrow e^{\alpha }=\omega \delta
_{\alpha ^{\prime }}^{\alpha }e^{\alpha ^{\prime }}.$ Such (active) frame
transforms preserve the spacetime geometry. Similar properties exist for
more general classes of transforms (deformations) on N--anholonomic
manifolds.

In the simplest case, for a fixed trivial N--connection structure in (\ref%
{vt1}), when $N_{i}^{a}=0$ and $e_{i}^{\ \underline{i}}(u)=\omega (u)\
\delta _{i}^{\ \underline{i}}$ and $e_{a}^{\ \underline{a}}(u)=\omega (u)\
\delta _{a}^{\ \underline{a}},$ we model conformal transforms $\ ^{\omega
}g_{\alpha \beta }=\omega ^{2}(u)\underline{g}_{\alpha \beta },$ where $%
\underline{g}_{\alpha \beta }$ is a flat metric on a manifold $V^{n+m}.$ For
nontrivial N--connection structures, it is convenient to distinguish four
general classes of nonholonomic frame transforms (deformations):

\paragraph{General frame transforms on N--anholonomic manifolds:}

{\ }\newline
Any N--anholonomic structure can be induced by a series of chains of one,
two, three,..., $k,$ ... frame transforms (\ref{ft}):%
\begin{eqnarray}
\mathbf{e}_{\alpha } &=&\mathbf{e}_{\alpha }^{\ \underline{\alpha }}(u)e_{%
\underline{\alpha }},  \label{cgft} \\
\ ^{2}\mathbf{e}_{\alpha } &=&\ ^{2}\mathbf{e}_{\alpha }^{\ \underline{%
\alpha }}(u)e_{\underline{\alpha }}=A_{\alpha }^{\ \alpha ^{\prime
}}(u)A_{\alpha ^{\prime }}^{\ \underline{\alpha }}(u)e_{\underline{\alpha }},
\notag \\
\ ^{3}\mathbf{e}_{\alpha } &=&\ ^{3}\mathbf{e}_{\alpha }^{\ \underline{%
\alpha }}(u)e_{\underline{\alpha }}=A_{\alpha }^{\ \alpha ^{\prime \prime
}}(u)A_{\alpha ^{\prime \prime }}^{\ \alpha ^{\prime }}(u)A_{\alpha ^{\prime
}}^{\ \underline{\alpha }}(u)e_{\underline{\alpha }},  \notag \\
&&...  \notag \\
\ ^{k}\mathbf{e}_{\alpha } &=&\ ^{k}\mathbf{e}_{\alpha }^{\ \underline{%
\alpha }}(u)e_{\underline{\alpha }}=A_{\alpha }^{\ \alpha ^{\prime \prime
\prime ...k}}(u)...A_{\alpha }^{\ \alpha ^{\prime \prime }}(u)A_{\alpha
^{\prime \prime }}^{\ \alpha ^{\prime }}(u)A_{\alpha ^{\prime }}^{\
\underline{\alpha }}(u)e_{\underline{\alpha }},  \notag \\
&&...  \notag
\end{eqnarray}%
where the left--up index label the number of transforms in a chain and we
can chose a coordinate base $e_{\underline{\alpha }}=\partial _{\underline{%
\alpha }}.$ The $(n+m)\times (n+m)$ dimensional matrices $A_{\alpha }^{\
\alpha ^{\prime \prime \prime ...k}},...,A_{\alpha }^{\ \alpha ^{\prime
\prime }},A_{\alpha ^{\prime \prime }}^{\ \alpha ^{\prime }},A_{\alpha
^{\prime }}^{\ \underline{\alpha }}(u)$ parametrize arbitrary frame
transforms but subjected to the condition that their product $\ ^{k}\mathbf{e%
}_{\alpha }^{\ \underline{\alpha }}$ results is a triangle matrix of type (%
\ref{vt1}), which induces at the final step of transforms, with fixed $%
\left( n+m\right) $--splitting, a N--connection structure (\ref{dder}).

Having generated a N--anholonomic frame structure $\mathbf{e}_{\overline{%
\alpha }},$ applying superpositions of nonholonomic transforms, we get, in
general, hidden N--anholonomic frame structures of type
\begin{equation}
e_{\alpha }=\ ^{k}A_{\alpha }^{\ \overline{\alpha }}(u)\mathbf{e}_{\overline{%
\alpha }}=A_{\alpha }^{\ \alpha ^{\prime \prime \prime ...k}}(u)...A_{\alpha
}^{\ \alpha ^{\prime \prime }}(u)A_{\alpha ^{\prime \prime }}^{\ \alpha
^{\prime }}(u)A_{\alpha ^{\prime }}^{\ \overline{\alpha }}(u)\mathbf{e}_{%
\overline{\alpha }}.  \label{cgfti}
\end{equation}%
We conclude that if a $\left( n+m\right) $--splitting is prescribed by a
N--connection $N_{i}^{a}$ on a nonholonomic manifold $\mathbf{V}$, we always
can model this structure by certain chains of nonholonomic frames even the
elements of the chains may not result in explicit forms of N--anholonomic
frames. If the N--anholonomic structure of $\mathbf{V}^{n+n}$ is, for
instance, of Lagrange type with canonical N--connection (\ref{dder}), by
chains of transforms (\ref{cgfti}), we can hide the Lagrange structure (and,
inversely, we can extract the Lagrange structure by chains of frame
transforms (\ref{cgft}) from certain special Riemannian or Riemann--Cartan
configutations). In a more general context, we can work with generalized
Lagrange structures $\ ^{\mathcal{L}}g_{ab}(u)$ and $\ ^{\mathcal{L}%
}N_{j}^{a}(u),$ see respectively (\ref{glhf}) and (\ref{glnc}), and (\ref%
{ftgfs}), hidden or embedded in explicit form into a nonholonomic Riemannian
space. The aim of such transforms is to relate a (pseudo) Riemannian metric
structure (it can be a solution of the gravitational field equations) to
certain generalized Lagrange, or Finsler, geometries which allows to apply
new geometric methods and define additional symmetries and conservation
laws, for instance, associated to bi--Hamilton structures and solitonic
hierarchies.

\paragraph{ N--adapted frame transforms:}

{\ }\newline
For the class of general nonholonomic transforms considered above, we can
can associate subclasses of matrices of type (\ref{vt1}) for any element of
the chains,
\begin{eqnarray}
\mathbf{e}_{\alpha } &=&\mathbf{e}_{\alpha }^{\ \underline{\alpha }}(u)e_{%
\underline{\alpha }},  \label{cngft} \\
\ ^{2}\mathbf{e}_{\alpha } &=&\ ^{2}\mathbf{e}_{\alpha }^{\ \underline{%
\alpha }}(u)e_{\underline{\alpha }}=\mathbf{e}_{\alpha }^{\ \alpha ^{\prime
}}(u)\mathbf{e}_{\alpha ^{\prime }}^{\ \underline{\alpha }}(u)e_{\underline{%
\alpha }},  \notag \\
\ ^{3}\mathbf{e}_{\alpha } &=&\ ^{3}\mathbf{e}_{\alpha }^{\ \underline{%
\alpha }}(u)e_{\underline{\alpha }}=\mathbf{e}_{\alpha }^{\ \alpha ^{\prime
\prime }}(u)\mathbf{e}_{\alpha ^{\prime \prime }}^{\ \alpha ^{\prime }}(u)%
\mathbf{e}_{\alpha ^{\prime }}^{\ \underline{\alpha }}(u)e_{\underline{%
\alpha }},  \notag \\
&&...  \notag \\
\ ^{k}\mathbf{e}_{\alpha } &=&\ ^{k}\mathbf{e}_{\alpha }^{\ \underline{%
\alpha }}(u)e_{\underline{\alpha }}=\mathbf{e}_{\alpha }^{\ \alpha ^{\prime
\prime \prime ...k}}(u)...\mathbf{e}_{\alpha }^{\ \alpha ^{\prime \prime
}}(u)\mathbf{e}_{\alpha ^{\prime \prime }}^{\ \alpha ^{\prime }}(u)\mathbf{e}%
_{\alpha ^{\prime }}^{\ \underline{\alpha }}(u)e_{\underline{\alpha }},
\notag \\
&&...  \notag
\end{eqnarray}%
which, for every step, transform an N--adapted base into another one, in
general, with different N--connection coefficients, but with the same $n$
and $m$ for the $\left( n+m\right) $--splitting. The reason to introduce
such transforms is to relate a d--metric and N--connection structure to a
special type ansatz for which the Einstein equations became integrable in
general form, see section \ref{exsolans}. There are possible chains of
N--anholonomic frame transforms when some Lagrange spaces are
nonholonomically related to generalized Lagrange spaces and then to some
classes of exact solutions of the Einstein equations. The superpositions of
nonholonomic frame transforms may be defined to depend on some classes of
parameters, see details in ref. \cite{vparsol}. All types of nonholonomic
deformations may be considered to change the signature of metrics if such
constructions are necessary.

\paragraph{ N--connection transforms with fixed h-- and v--metrics:}

{\ }\newline
For some purposes, for instance, in constructing exact solutions, or
defining analogous models of gravity, it is useful to work with more special
classes of nonholonomic deformations. The transforms of vielbeins (\ref{vt1}%
) of type%
\begin{equation*}
\mathbf{e}_{\alpha }^{\ \underline{\alpha }}=\left[
\begin{array}{cc}
e_{i}^{\ \underline{i}}(u) & N_{i}^{b}(u)e_{b}^{\ \underline{a}}(u) \\
0 & e_{a}^{\ \underline{a}}(u)%
\end{array}%
\right] \rightarrow \ ^{\eta }\mathbf{e}_{\alpha }^{\ \underline{\alpha }}=%
\left[
\begin{array}{cc}
e_{i}^{\ \underline{i}}(u) & \eta _{i}^{b}(u)N_{i}^{b}(u)e_{b}^{\ \underline{%
a}}(u) \\
0 & e_{a}^{\ \underline{a}}(u)%
\end{array}%
\right] ,
\end{equation*}%
preserve the h-- and v--components of a d--metric (\ref{m1}), $\ ^{\eta
}g_{ij}=g_{ij}=e_{i}^{\ \underline{i}}e_{j}^{\ \underline{j}}g_{\underline{i}%
\underline{j}}$ and $\ ^{\eta }g_{ab}=g_{ab}=e_{a}^{\ \underline{a}}e_{b}^{\
\underline{b}}g_{\underline{a}\underline{b}},$ for some prescribed values $%
g_{\underline{i}\underline{j}}$ and $g_{\underline{a}\underline{b}},$ but
transform the N--connection,
\begin{equation}
N_{j}^{i}\rightarrow \ ^{\eta }N_{j}^{i}=\eta _{i}^{b}N_{i}^{b},
\label{pfnc}
\end{equation}%
where we do not consider summation on repeating both up/low indices.
Usually, it is supposed that such deformations of the N--connection
structure preserve the $n$-- and $m$--dimensions of splitting (\ref{whitney}%
).

\paragraph{ N--anholonomic transforms of h-- and v--metrics:}

{\ }\newline
We can fix a N--connection $N_{i}^{b}(u)$ and consider frame tranforms of
type
\begin{equation*}
\mathbf{e}_{\alpha }^{\ \underline{\alpha }}=\left[
\begin{array}{cc}
e_{i}^{\ \underline{i}} & N_{i}^{b}e_{b}^{\ \underline{a}} \\
0 & e_{a}^{\ \underline{a}}%
\end{array}%
\right] \rightarrow \ ^{\eta }\mathbf{e}_{\alpha }^{\ \underline{\alpha }}=%
\left[
\begin{array}{cc}
\eta _{i}^{\ \underline{i}}\ e_{i}^{\ \underline{i}} & N_{i}^{b}e_{b}^{\
\underline{a}} \\
0 & \eta _{a}^{\ \underline{a}}\ e_{a}^{\ \underline{a}}%
\end{array}%
\right] ,
\end{equation*}%
which result in deformations of the h-- and v--metrics, respectively,
\begin{eqnarray}
g_{ij}(u) &=&e_{i}^{\ \underline{i}}(u)e_{j}^{\ \underline{j}}(u)g_{%
\underline{i}\underline{j}}  \label{htransm} \\
&\rightarrow &\ ^{N}g_{ij}(u)=\eta _{i}^{\ \underline{i}}(u)\eta _{j}^{\
\underline{j}}(u)e_{i}^{\ \underline{i}}(u)e_{j}^{\ \underline{j}}(u)g_{%
\underline{i}\underline{j}}  \notag
\end{eqnarray}%
and
\begin{eqnarray}
g_{ab}(u) &=&e_{a}^{\ \underline{a}}(u)e_{b}^{\ \underline{b}}(u)g_{%
\underline{a}\underline{b}}(u)  \label{vtransm} \\
&\rightarrow &\ ^{N}g_{ab}(u)=\eta _{a}^{\ \underline{a}}(u)\eta _{b}^{\
\underline{b}}(u)e_{a}^{\ \underline{a}}(u)e_{b}^{\ \underline{b}}(u)g_{%
\underline{a}\underline{b}},  \notag
\end{eqnarray}%
where we do not consider summation on indices for $\eta _{i}^{\ \underline{i}%
}\ e_{i}^{\ \underline{i}}$ but the Einstein summation rule is applied, for
instance to ''up--low'' repeating indices like on underlined ones in $%
e_{a}^{\ \underline{a}}e_{b}^{\ \underline{b}}g_{\underline{a}\underline{b}%
}. $

For $n+m=2+2,$ for simplicity, we can work only with diagonalized matrices
for the h-- and v--components of d--metrics of type (\ref{m1}), when $%
g_{ij}=diag[g_{1}(u),g_{2}(u)]$ and $h_{ab}=diag[h_{3}(u),h_{4}(u)]$ (such
diagonalizations can be performed by coordinate transforms for matrices of
dimension $3\times 3$ and $2\times 2$). We can write in effectively
diagonalized forms the deformations of h-- and v--metrics, respectively, (%
\ref{htransm}) and (\ref{vtransm}),%
\begin{equation*}
g_{\alpha }=[g_{i}(u),h_{a}(u)]\rightarrow \ ^{\eta }g_{\alpha }=\left[ \
^{\eta }g_{i}=\eta _{i}(u)g_{i}(u),\ ^{\eta }h_{i}=\eta _{a}(u)h_{a}(u)%
\right] ,
\end{equation*}%
where $\eta _{\alpha }(u)=\left[ \eta _{i}(u),\eta _{a}(u)\right] $ are
called polarization functions, see exact solutions with such polarization
functions in Ref. \cite{vsgg,vesnc}. The reason to introduce such
polarizations was that for small polarizations $\eta _{a}\sim 1+\varepsilon
_{a}(u),$ where $|\varepsilon _{a}|\ll 1,$ it was possible to generate small
nonholonomic deformations of solutions of the Einstein equations, belonging
in general to a class of exact solutions, but for small deformations
possessing, for instance, black ellipsoid properties.

\subsubsection{Lie algebroids and N--connections}

A \textbf{Lie algebroid} $\mathcal{A}\doteqdot (E,[\cdot ,\cdot ],\rho )$ is
defined as a vector bundle $\mathcal{E}=(E,\pi ,$ $M),$ with a surjective
projection $\pi :E\rightarrow M,\dim M=n$ and $\dim E=n+m,$ provided with
\textbf{algebroid structure} $([\cdot ,\cdot ],\rho ),$ where $[\cdot ,\cdot
]$ is a \textbf{Lie bracket} on the $C^{\infty }(M)$--module of sections of $%
E,$ denoted $Sec(E),$ and the '\textbf{anchor}' $\rho $ is defined as a
bundle map $\rho :E\rightarrow M$ such that
\begin{equation*}
\left[ X,fY\right] =f[X,Y]+\rho (X)(f)Y
\end{equation*}%
for $X,Y\in Sec(E)$ and $f\in C^{\infty }(M),$ see \cite{liberm,mcz} for
general results and some applications of Lie algebroid geometry. In local
form, the Lie algebroid structure is defined by its \textbf{structure
functions} $\rho _{a}^{i}(x)$ and $C_{ab}^{d}(x)$ on $M,$ determined by the
relations%
\begin{eqnarray}
\rho (e_{a}) &=&\rho _{a}^{i}(x)e_{i},  \label{anch} \\
\lbrack e_{a},e_{b}] &=&C_{ab}^{d}(x)e_{d}  \label{liea}
\end{eqnarray}%
and subjected to the structure equations%
\begin{equation}
\rho _{a}^{j}\frac{\partial \rho _{b}^{i}}{\partial x^{j}}-\rho _{b}^{j}%
\frac{\partial \rho _{b}^{i}}{\partial x^{j}}=\rho _{d}^{j}C_{ab}^{d}%
\mbox{\
and \ }\sum\limits_{cyclic(a,b,c)}\left( \rho _{a}^{j}\frac{\partial
C_{bc}^{d}}{\partial x^{j}}+C_{af}^{d}C_{bc}^{f}\right) =0,  \label{lasa}
\end{equation}%
for any local basis $e_{a}=(e_{i},e_{a})$ on $\mathcal{E}.$

We extended the Lie algebroid constructions for nonholonomic manifolds and
vector bundles provided with N--connection structure $\mathbf{N}%
=\{N_{i}^{a}(x,y)\}$, respectively, on $\mathbf{V}$ and $\mathbf{E},$ and
introduced the concept of N--anholonomic manifold $\ ^{N}\mathcal{A}%
\doteqdot (\mathbf{V},[\cdot ,\cdot ],\check{\rho}),$ see details and
references in \cite{vclalg,valg} (we note that in this work we use $\check{%
\rho}$ instead of $\widehat{\rho }$).

The Lie algebroid and N--connection structures prescribe a subclass of local
both N--adapted and $[\cdot ,\cdot ]$--adapted frames
\begin{equation}
\mathbf{\check{e}}_{\alpha }=(\mathbf{e}_{i}=\frac{\partial }{\partial x^{i}}%
-\check{N}_{\ i}^{a}\check{e}_{a},\check{e}_{b})  \label{ddera}
\end{equation}%
\ and dual coframes
\begin{equation}
\mathbf{\check{e}}^{\alpha }=(e^{i},\ \mathbf{\check{e}}^{b}=\check{e}^{b}+%
\check{N}_{\ i}^{b}dx^{i}),  \label{ddifa}
\end{equation}%
for some
\begin{equation}
\check{e}_{b}=\check{e}_{b}^{\ \underline{b}}\partial _{\underline{b}},
\label{vnhb}
\end{equation}%
when $\check{e}_{c}\mathbf{\rfloor }\check{e}^{b}=\delta _{c}^{b}.$ The \
N--connection coefficients may be redefined as%
\begin{equation*}
\mathbf{N}=N_{\ \underline{i}}^{\underline{a}}(u)dx^{\underline{i}}\otimes
\frac{\partial }{\partial y^{\underline{a}}}=\check{N}_{\
i}^{b}(u)e^{i}\otimes \check{e}_{b},
\end{equation*}%
where $\check{N}_{\ i}^{b}=\check{e}_{\ \underline{a}}^{b\ }$ $N_{\
\underline{i}}^{\underline{a}}$ and there are underlined the indices
defining the coefficients with respect to a local coordinate basis.

Any Lie algebroid structure can be adapted to a prescribed N--connection and
resulting frame structures (\ref{ddera}) \ and (\ref{ddifa}). This can be
done following the procedure: Let us re--define the coefficients of the
anchor and structure functions with respect to the $\mathbf{e}_{\alpha }$
and $\mathbf{e}^{\alpha },$ when
\begin{eqnarray*}
\rho _{\underline{b}}^{\underline{i}}(x) &\rightarrow &\ \mathbf{\check{\rho}%
}_{b}^{i}(x,y)=\mathbf{e}_{\ \underline{i}}^{i}(x,y)\ \mathbf{e}_{b}^{\
\underline{b}}(x,y)\rho _{\underline{b}}^{\underline{i}}(x), \\
C_{\underline{d}\underline{b}}^{\underline{f}}(x) &\rightarrow &\mathbf{C}%
_{db}^{f}(x,y)=\mathbf{e}_{\ \underline{f}}^{f}(x,y)\ \mathbf{e}_{d}^{\
\underline{d}}(x,y)\ \mathbf{e}_{b}^{\ \underline{b}}(x,y)C_{\underline{d}%
\underline{b}}^{\underline{f}}(x),
\end{eqnarray*}%
where the transform $\mathbf{e}$--matrices are linear on coefficients $%
N_{i}^{a}$ as can be obtained from (\ref{vt1}). In terms of N--adapted
anchor $\ \ \mathbf{\check{\rho}}_{b}^{i}(x,y)$ and structure functions $%
\mathbf{C}_{db}^{f}(x,y)$ (which depend also on variables $y^{a}$), the
structure equations of the Lie algebroids (\ref{anch}),\ (\ref{liea}) and (%
\ref{lasa}) transform respectively into
\begin{eqnarray}
{}\ \mathbf{\check{\rho}}(\check{e}_{b}) &=&{}\ \mathbf{\check{\rho}}%
_{b}^{i}(x,y)\ \mathbf{e}_{i},  \label{anch1d} \\
\lbrack \check{e}_{d},\check{e}_{b}] &=&\mathbf{C}_{db}^{f}(x,y)\ \check{e}%
_{f}  \label{lie1d}
\end{eqnarray}%
and
\begin{eqnarray}
{}\ \mathbf{\check{\rho}}_{a}^{j}\mathbf{e}_{j}({}\widehat{\mathbf{\rho }}%
_{b}^{i})-{}\ \mathbf{\check{\rho}}_{b}^{j}\mathbf{e}_{j}({}\widehat{\mathbf{%
\rho }}_{a}^{i}) &=&{}\ \mathbf{\check{\rho}}_{e}^{j}\mathbf{C}_{ab}^{e},
\label{lased} \\
\sum\limits_{cyclic(a,b,e)}\left( {}\ \mathbf{\check{\rho}}_{a}^{j}\mathbf{e}%
_{j}(\mathbf{C}_{be}^{f})+\mathbf{C}_{ag}^{f}\mathbf{C}_{be}^{g}-\mathbf{C}%
_{b^{\prime }e^{\prime }}^{f^{\prime }}{}\ \mathbf{\check{\rho}}_{a}^{j}%
\mathbf{Q}_{f^{\prime }bej}^{fb^{\prime }e^{\prime }}\right) &=&0,  \notag
\end{eqnarray}%
for $\mathbf{Q}_{f^{\prime }bej}^{fb^{\prime }e^{\prime }}=\mathbf{e}_{\
\underline{b}}^{b^{\prime }}\mathbf{e}_{\ \underline{e}}^{e^{\prime }}%
\mathbf{e}_{f^{\prime }}^{\ \underline{f}}~\mathbf{e}_{j}(\mathbf{e}_{b}^{\
\underline{b}}\mathbf{e}_{e}^{\ \underline{e}}\mathbf{e}_{\ \underline{f}%
}^{f})$ computed for the values $\mathbf{e}_{\ \underline{b}}^{b^{\prime }}$
and $\mathbf{e}_{f^{\prime }}^{\ \underline{f}}$ taken \ from (\ref{vt1})
and (\ref{vt2}).

Using N--anholonomic Lie algebroid structures, we can apply certain methods
of Finsler and Lagrange geometry to spacetimes provided with arbitrary
d--metrics (\ref{m1}) when $g_{ij}\neq h_{ab},$ see details and examples of
exact solutions in Refs. \cite{valg}. In the simplest case, for a
N--anholomic manifold $\mathbf{V}^{n+n},$ we can work with a trivial anchor
structure ${}\ \mathbf{\check{\rho}}_{a}^{j}=0,$ when the conditions (\ref%
{anch1d}) and (\ref{lased}) are satisfied for certain nonholonomic frame
configurations, but with the bracket $[\cdot ,\cdot ]$ induced by
decomposition%
\begin{equation*}
h_{ab}(u)=\check{e}_{\ a}^{a^{\prime }}(u)\check{e}_{\ b}^{b^{\prime
}}(u)g_{a^{\prime }b^{\prime }}(u)
\end{equation*}%
with
\begin{eqnarray}
\mathbf{\check{g}} &=&g_{ij}(x,y)\left( dx^{i}\otimes dx^{j}+\check{e}_{\
a}^{i}\check{e}_{\ b}^{i}dy^{a}\otimes dy^{b}\right)  \notag \\
&=&g_{ij}(x,y)\left( dx^{i}\otimes dx^{j}+\check{e}^{i}\otimes \check{e}%
^{j}\right)  \label{adlegdm}
\end{eqnarray}%
where $\check{e}^{i}=\check{e}_{\ a}^{i}dy^{a}$ states the coefficients $%
\mathbf{C}_{db}^{f}(x,y)$ for (\ref{lie1d}).\footnote{%
see Refs. \cite{vclalg,valg} for constructions with nontrivial $\ \mathbf{%
\check{\rho};}$ algebroid models with nontrivial anchor maps are useful in
extra dimension gravity, for instance, with nonholonomic splitting of
dimensions of type $n+m\geq 5,$ when $n\geq 2$ and $m\geq n;$ for
simplicity, we omit such constructions in this work} We can say that if $%
g_{ij}\neq h_{ab},$ on $\mathbf{V}^{n+n},$ or $\mathbf{TM},$ we work
similarly with the generalized Lagrange spaces but with modified prescribed
frame structures (\ref{ddera}) and (\ref{ddifa}) when nonholonomy
coefficients are nontrivial both for the h--part and v--part. In a
particular case, we can consider that $g_{ij}(x,y)$ from (\ref{adlegdm}) is
of type $\ ^{L}g_{ij}(x,y),$ see (\ref{lqf}) and (\ref{slm}) (in certain
more general or particular cases we can take (\ref{ftgfs}), (\ref{glhf}) or (%
\ref{fhes})), which models a Lagrange N--algebroid structure (respectively,
generalized Lagrange, or Finsler, N--algebroid structure). To work on
generalized Lagrange N--algebroids with d--metrics of type (\ref{adlegdm})
is convenient if we apply some methods from almost Hermitian/ K\"{ }ahler
geometry, see section \ref{sakeg}. The Einstein equations for the d--metric (%
\ref{adlegdm}) and canonical d--connection defined with respect to
N--adapted bases (\ref{ddera}) and (\ref{ddifa}) are equivalent to equations
(\ref{einstgrdef}) for the canonical d--connection (\ref{candcon}),
redefined with respect to N--algebroid bases.

\subsection{An ansatz for constructing exact solutions}

\label{exsolans}We consider a four dimensional (4D) manifold $\mathbf{V}$ of
necessary smooth class and conventional splitting of dimensions $\dim
\mathbf{V=}$ $n+m$ for $n=2$ and $m=2.$ The local coordinates are labelled
in the form $u^{\alpha }=(x^{i},y^{a})=(x^{1},x^{2},y^{3}=v,y^{4}),$ for $%
i=1,2$ and $a,b,...=3,4.$

The ansatz of type (\ref{m1}) is parametrized in the form
\begin{eqnarray}
\mathbf{g} &=&g_{1}(x^{i}){dx^{1}}\otimes {dx^{1}}+g_{2}\left( x^{i}\right) {%
dx^{2}}\otimes {dx^{2}}  \notag \\
&& + h_{3}\left( x^{k},v\right) \ {\delta v}\otimes {\delta v}+h_{4}\left(
x^{k},v\right) \ {\delta y}\otimes {\delta y},  \notag \\
\delta v &=&dv+w_{i}\left( x^{k},v\right) dx^{i},\ \delta y=dy+n_{i}\left(
x^{k},v\right) dx^{i}  \label{ans5d}
\end{eqnarray}%
with the coefficients defined by some necessary smooth class functions of
type
\begin{equation}
g_{1,2}=g_{1,2}(x^{1},x^{2}),h_{3,4}=h_{3,4}(x^{i},v),w_{i}=w_{i}(x^{k},v),n_{i}=n_{i}(x^{k},v).
\notag
\end{equation}%
The off--diagonal terms of this metric, written with respect to the
coordinate dual frame $du^{\alpha }=(dx^{i},dy^{a}),$ can be redefined to
state a N--connection structure $\mathbf{N}=[N_{i}^{3}=w_{i}(x^{k},v),$$%
N_{i}^{4}=n_{i}(x^{k},v)]$ with a N--elongated co--frame (\ref{ddif})
parametrized as
\begin{equation}
e^{1}=dx^{1},\ e^{2}=dx^{2},\mathbf{e}^{3}=\delta v=dv+w_{i}dx^{i},\ \mathbf{%
e}^{4}=\delta y=dy+n_{i}dx^{i}.  \label{ddif5}
\end{equation}%
This vielbein is dual to the local basis%
\begin{equation}
\mathbf{e}_{i}=\frac{\partial }{\partial x^{i}}-w_{i}\left( x^{k},v\right)
\frac{\partial }{\partial v}-n_{i}\left( x^{k},v\right) \frac{\partial }{%
\partial y^{5}},e_{3}=\frac{\partial }{\partial v},e_{4}=\frac{\partial }{%
\partial y^{5}},  \label{dder5}
\end{equation}%
which is a particular case of the N--adapted frame (\ref{dder}). The metric (%
\ref{ans5d}) does not depend on variable $y^{4},$ i.e. it possesses a
Killing vector $e_{4}=\partial /\partial y^{4},$ and distinguish the
dependence on the so--called ''anisotropic'' variable $y^{3}=v.$

Computing the components of the Ricci and Einstein tensors for the metric (%
\ref{ans5d}) and canonical d--connection (see details on tensors components'
calculus in Refs. \cite{valg,vsgg}), one proves that the Einstein equations (%
\ref{einstgrdef}) for a diagonal with respect to (\ref{ddif5}) and (\ref%
{dder5}) source,%
\begin{equation}
\mathbf{\Upsilon }_{\beta }^{\alpha }+\ ^{Z}\mathbf{\Upsilon }_{\beta
}^{\alpha }=[\Upsilon _{1}^{1}=\Upsilon _{2}(x^{i},v),\Upsilon
_{2}^{2}=\Upsilon _{2}(x^{i},v),\Upsilon _{3}^{3}=\Upsilon
_{4}(x^{i}),\Upsilon _{4}^{4}=\Upsilon _{4}(x^{i})]  \label{sdiag}
\end{equation}%
transform into this system of partial differential equations:
\begin{eqnarray}
\widehat{R}_{1}^{1} &=&\widehat{R}_{2}^{2}  \label{ep1a} \\
&=&\frac{1}{2g_{1}g_{2}}[\frac{g_{1}^{\bullet }g_{2}^{\bullet }}{2g_{1}}+%
\frac{(g_{2}^{\bullet })^{2}}{2g_{2}}-g_{2}^{\bullet \bullet }+\frac{%
g_{1}^{^{\prime }}g_{2}^{^{\prime }}}{2g_{2}}+\frac{(g_{1}^{^{\prime }})^{2}%
}{2g_{1}}-g_{1}^{^{\prime \prime }}]=-\Upsilon _{4}(x^{i}),  \notag \\
\widehat{S}_{3}^{3} &=&\widehat{S}_{4}^{4}=\frac{1}{2h_{3}h_{4}}\left[
h_{4}^{\ast }\left( \ln \sqrt{|h_{3}h_{4}|}\right) ^{\ast }-h_{4}^{\ast \ast
}\right] =-\Upsilon _{2}(x^{i},v),  \label{ep2a} \\
\widehat{R}_{3i} &=&-w_{i}\frac{\beta }{2h_{4}}-\frac{\alpha _{i}}{2h_{4}}=0,
\label{ep3a} \\
\widehat{R}_{4i} &=&-\frac{h_{3}}{2h_{4}}\left[ n_{i}^{\ast \ast }+\gamma
n_{i}^{\ast }\right] =0,  \label{ep4a}
\end{eqnarray}%
where, for $h_{3,4}^{\ast }\neq 0,$%
\begin{eqnarray}
\alpha _{i} &=&h_{4}^{\ast }\partial _{i}\phi ,\ \beta =h_{4}^{\ast }\ \phi
^{\ast },\ \gamma =\frac{3h_{4}^{\ast }}{2h_{4}}-\frac{h_{3}^{\ast }}{h_{3}},
\label{coef} \\
\phi &=&\ln |h_{3}^{\ast }/\sqrt{|h_{3}h_{4}|}|,  \label{coefa}
\end{eqnarray}%
when the necessary partial derivatives are written in the form \ $a^{\bullet
}=\partial a/\partial x^{1},$ $a^{\prime }=\partial a/\partial x^{2},$\ $%
a^{\ast }=\partial a/\partial v.$ In the vacuum case, we must consider $%
\Upsilon _{2,4}=0.$ We note that we use a source of type (\ref{sdiag}) in
order to show that the anholonomic frame method can be applied also for
non--vacuum solutions, for instance, when $\Upsilon _{2}=\lambda _{2}=const$
and $\Upsilon _{4}=\lambda _{4}=const,$ defining locally anisotropic
configurations generated by an anisotropic cosmological constant, which in
its turn, can be induced by certain ansatz for the so--called $H$--field
(absolutely antisymmetric third rank tensor field) in string theory \cite%
{vesnc,vsgg,valg}. Here we note that the off--diagonal gravitational
interactions can model locally anisotropic configurations even if $\lambda
_{2}=\lambda _{4},$ or both values vanish.

In string gravity, the nontrivial torsion components and source $\kappa
\mathbf{\Upsilon }_{\alpha \beta }$ can be related to certain effective
interactions with the strength (torsion)
\begin{equation*}
H_{\mu \nu \rho }=\mathbf{e}_{\mu }B_{\nu \rho }+\mathbf{e}_{\rho }B_{\mu
\nu }+\mathbf{e}_{\nu }B_{\rho \mu }
\end{equation*}%
of an antisymmetric field $B_{\nu \rho },$ when%
\begin{equation}
R_{\mu \nu }=-\frac{1}{4}H_{\mu }^{\ \nu \rho }H_{\nu \lambda \rho }
\label{c01}
\end{equation}%
and
\begin{equation}
D_{\lambda }H^{\lambda \mu \nu }=0,  \label{c02}
\end{equation}%
see details on string gravity, for instance, in Refs. \cite{string1,string2}%
. The conditions (\ref{c01}) and (\ref{c02}) are satisfied by the ansatz
\begin{equation}
H_{\mu \nu \rho }=\widehat{Z}_{\mu \nu \rho }+\widehat{H}_{\mu \nu \rho
}=\lambda _{\lbrack H]}\sqrt{\mid g_{\alpha \beta }\mid }\varepsilon _{\nu
\lambda \rho }  \label{ansh}
\end{equation}%
where $\varepsilon _{\nu \lambda \rho }$ is completely antisymmetric and the
distorsion (from the Levi--Civita connection) and
\begin{equation*}
\widehat{Z}_{\mu \alpha \beta }\mathbf{e}^{\mu }=\mathbf{e}_{\beta }\rfloor
\mathcal{T}_{\alpha }-\mathbf{e}_{\alpha }\rfloor \mathcal{T}_{\beta }+\frac{%
1}{2}\left( \mathbf{e}_{\alpha }\rfloor \mathbf{e}_{\beta }\rfloor \mathcal{T%
}_{\gamma }\right) \mathbf{e}^{\gamma }
\end{equation*}%
is defined by the torsion tensor (\ref{tors}). Our $H$--field ansatz is
different from those already used in string gravity when $\widehat{H}_{\mu
\nu \rho }=\lambda _{\lbrack H]}\sqrt{\mid g_{\alpha \beta }\mid }%
\varepsilon _{\nu \lambda \rho }.$ \ In our approach, we define $H_{\mu \nu
\rho }$ and $\widehat{Z}_{\mu \nu \rho }$ from the respective ansatz for the
$H$--field and nonholonomically deformed metric, compute the torsion tensor
for the canonical distinguished connection and, finally, define the
'deformed' H--field as $\widehat{H}_{\mu \nu \rho }=\lambda _{\lbrack H]}%
\sqrt{\mid g_{\alpha \beta }\mid }\varepsilon _{\nu \lambda \rho }-\widehat{Z%
}_{\mu \nu \rho }.$

Summarizing the results for an ansatz (\ref{ans5d}) with arbitrary
signatures $\epsilon _{\alpha }=\left( \epsilon _{1},\epsilon _{2},\epsilon
_{3},\epsilon _{4}\right) $ (where $\epsilon _{\alpha }=\pm 1)$ and $%
h_{3}^{\ast }\neq 0$ and $h_{4}^{\ast }\neq 0,$ one proves \cite%
{vesnc,valg,vsgg} that any off--diagonal metric
\begin{eqnarray}
\ ^{\circ }\mathbf{g} &=&e^{\psi (x^{i})}\left[ \epsilon _{1}\ dx^{1}\otimes
dx^{1}+\epsilon _{2}\ dx^{2}\otimes dx^{2}\right]  \notag \\
&&+\epsilon _{3}h_{0}^{2}(x^{i})\left[ f^{\ast }\left( x^{i},v\right) \right]
^{2}|\varsigma \left( x^{i},v\right) |\ \delta v\otimes \delta v  \notag \\
&&+\epsilon _{4}\left[ f\left( x^{i},v\right) -f_{0}(x^{i})\right] ^{2}\
\delta y^{4}\otimes \delta y^{4},  \notag \\
\delta v &=&dv+w_{k}\left( x^{i},v\right) dx^{k},\ \delta
y^{4}=dy^{4}+n_{k}\left( x^{i},v\right) dx^{k},  \label{gensol1}
\end{eqnarray}%
where $\psi (x^{i})$ is a solution of the 2D equation
\begin{equation*}
\epsilon _{1}\psi ^{\bullet \bullet }+\epsilon _{2}\psi ^{^{\prime \prime
}}=\Upsilon _{4},
\end{equation*}%
for a given source $\Upsilon _{4}\left( x^{i}\right) ,$%
\begin{equation*}
\varsigma \left( x^{i},v\right) =\varsigma _{\lbrack 0]}\left( x^{i}\right) -%
\frac{\epsilon _{3}}{8}h_{0}^{2}(x^{i})\int \Upsilon _{2}(x^{k},v)f^{\ast
}\left( x^{i},v\right) \left[ f\left( x^{i},v\right) -f_{0}(x^{i})\right] dv,
\end{equation*}%
and the N--connection coefficients $N_{i}^{3}=w_{i}(x^{k},v)$ and $%
N_{i}^{4}=n_{i}(x^{k},v)$ are computed following the formulas
\begin{equation}
w_{i}=-\frac{\partial _{i}\varsigma \left( x^{k},v\right) }{\varsigma ^{\ast
}\left( x^{k},v\right) }  \label{gensol1w}
\end{equation}%
and
\begin{equation}
n_{k}=\ ^{1}n_{k}\left( x^{i}\right) +\ ^{2}n_{k}\left( x^{i}\right) \int
\frac{\left[ f^{\ast }\left( x^{i},v\right) \right] ^{2}}{\left[ f\left(
x^{i},v\right) -f_{0}(x^{i})\right] ^{3}}\varsigma \left( x^{i},v\right) dv,
\label{gensol1n}
\end{equation}%
define an exact solution of the system of Einstein equations (\ref{ep1a})--(%
\ref{ep4a}). It should be emphasized that such solutions depend on arbitrary
nontrivial functions $f\left( x^{i},v\right) $ (with $f^{\ast }\neq 0),$ $%
f_{0}(x^{i}),$ $h_{0}^{2}(x^{i})$, $\ \varsigma _{\lbrack 0]}\left(
x^{i}\right) ,$ $\ ^{1}n_{k}\left( x^{i}\right) $ and $\ \ ^{2}n_{k}\left(
x^{i}\right) ,$ and sources $\Upsilon _{2}(x^{k},v),\Upsilon _{4}\left(
x^{i}\right) .$ Such values for the corresponding signatures $\epsilon
_{\alpha }=\pm 1$ have to be defined by certain boundary conditions and
physical considerations. These classes of solutions depending on integration
functions are more general than those for diagonal ansatz depending, for
instance, on one radial variable like in the case of the Schwarzschild
solution (when the Einstein equations are reduced to an effective nonlinear
ordinary differential equation, ODE). In the case of ODE, the integral
varieties depend on integration constants which can be defined from certain
boundary/ asymptotic and symmetry conditions, for instance, from the
constraint that far away from the horizon the Schwarzschild metric contains
corrections from the Newton potential. Because the ansatz (\ref{ans5d})
results in a system of nonlinear partial differential equations (\ref{ep1a}%
)--(\ref{ep4a}), the solutions depend not only on integration constants, but
on very general classes of integration functions.

The ansatz of type (\ref{ans5d}) with $h_{3}^{\ast }=0$ but $h_{4}^{\ast
}\neq 0$ (or, inversely, $h_{3}^{\ast }\neq 0$ but $h_{4}^{\ast }=0)$
consist more special cases and request a bit different method of
constructing exact solutions. Nevertheless, such type solutions are also
generic off--diagonal and they may be of substantial interest (the length of
paper does not allow to include an analysis of such particular cases).

A very general class of exact solutions of the Einstein equations with
nontrivial sources (\ref{sdiag}), in general relativity, is defined by the
ansatz
\begin{eqnarray}
\ _{\shortmid }^{\circ }\mathbf{g} &=&e^{\psi (x^{i})}\left[ \epsilon _{1}\
dx^{1}\otimes dx^{1}+\epsilon _{2}\ dx^{2}\otimes dx^{2}\right]
\label{eeqsol} \\
&&+h_{3}\left( x^{i},v\right) \ \delta v\otimes \delta v+h_{4}\left(
x^{i},v\right) \ \delta y^{4}\otimes \delta y^{4},  \notag \\
\delta v &=&dv+w_{1}\left( x^{i},v\right) dx^{1}+w_{2}\left( x^{i},v\right)
dx^{2},  \notag \\
\ \ \delta y^{4} &=&dy^{4}+n_{1}\left( x^{i}\right) dx^{1}+n_{2}\left(
x^{i}\right) dx^{2},  \notag
\end{eqnarray}%
with the coefficients restricted to satisfy the conditions
\begin{eqnarray}
\epsilon _{1}\psi ^{\bullet \bullet }+\epsilon _{2}\psi ^{^{\prime \prime }}
&=&\Upsilon _{4},  \notag \\
h_{4}^{\ast }\phi /h_{3}h_{4} &=&\Upsilon _{2},  \label{ep2b} \\
w_{1}^{\prime }-w_{2}^{\bullet }+w_{2}w_{1}^{\ast }-w_{1}w_{2}^{\ast } &=&0,
\notag \\
n_{1}^{\prime }-n_{2}^{\bullet } &=&0,  \notag
\end{eqnarray}%
for $w_{i}=\partial _{i}\phi /\phi ^{\ast },$ see (\ref{coefa}), for given
sources $\Upsilon _{4}(x^{k})$ and $\Upsilon _{2}(x^{k},v).$ We note that
the second equation in (\ref{ep2b}) relates two functions $h_{3}$ and $h_{4}$
and the third and forth equations satisfy the conditions (\ref{intcond}).

Even the ansatz (\ref{ans5d}) depends on three coordinates $(x^{k},v),$ it
allows us to construct more general classes of solutions for d--metrics,
depending on four coordinates: such solutions can be related by chains of
nonholonomic transforms (\ref{cgft}), (\ref{cngft}), or (\ref{htransm}) and (%
\ref{vtransm}). New classes of generic off--diagonal solutions will describe
nonhlonomic Einstein spaces related to string gravity, if one of the chain
metric is of type (\ref{gensol1}), or in Einstein gravity, if one of the
chain metric is of type (\ref{eeqsol}). The geometries of such spacetimes
are modelled equivalently by corresponding classes of N--anholonomic
algebroids provided with metric structures of type (\ref{adlegdm}).

\section{Einstein Gravity and Lagrange--K\"{a}h\-ler Spaces}

\label{sakeg} We show how nonholonomic Riemannian spaces and generalized
Lagrange algebroids can be transformed into almost Hermitian manifolds
enabled with nonintegrable almost complex structures.

\subsection{Generalized Lagrange--Hermitian algebroids}

\label{sglha}Let $\mathbf{\check{e}}_{\alpha }=(\mathbf{e}_{i},\check{e}%
_{b}) $ (\ref{ddera}) and $\mathbf{\check{e}}^{\alpha }=(e^{i},\ \mathbf{%
\check{e}}^{b})$ (\ref{ddifa}) be, respectively, some N--adapted frames and
co--frames on $\mathbf{V}^{n+n}.$ In a particular case, $\mathbf{V}%
^{n+n}=TM. $ We introduce a linear operator $\mathbf{\check{F}}$ acting on
the vectors on $\mathbf{V}^{n+n}$ following formulas
\begin{equation*}
\mathbf{\check{F}}(\mathbf{e}_{i})=-\check{e}_{i}\mbox{\ and \ }\mathbf{%
\check{F}}(\check{e}_{i})=\mathbf{e}_{i},
\end{equation*}%
where the superposition $\mathbf{\check{F}\circ \check{F}=-I,}$ for $\mathbf{%
I}$ being the unity matrix. On $TM,$ for vertically integrable
distributions, when $\mathbf{C}_{db}^{f}=0$ for (\ref{lie1d}), we get an
\textbf{almost Hermitian model }of generalized Lagrange space \cite%
{ma1987,ma}. The operator $\mathbf{\check{F}}$ reduces to a \textbf{complex
structure} if and only if both the h-- and v--distributions are integrable.

The metric $\mathbf{\check{g}}$ (\ref{adlegdm}) induces a 2--form associated
to $\mathbf{\check{F}}$ following formulas $\mathbf{\check{\theta}(X,Y)}%
\doteqdot \mathbf{\check{g}}\left( \mathbf{\check{F}X,Y}\right) $ for any
d--vectors $\mathbf{X}$ and $\mathbf{Y.}$ In local form, we have
\begin{equation}
\mathbf{\check{\theta}}=g_{ij}(x,y)\check{e}^{i}\wedge dx^{j}.  \label{asstr}
\end{equation}%
For v--integrable distributions, we shall write $\mathbf{F}$ and $\theta $
in order to emphasize that we model generalized Lagrange structures.

We compute
\begin{eqnarray}
d\mathbf{\check{\theta}} &=&\frac{1}{6}\sum\limits_{(ijk)}g_{is}\check{\Omega%
}_{jk}^{s}dx^{i}\wedge dx^{j}\wedge dx^{k}  \label{asstrsf} \\
&&+\frac{1}{2}\left( g_{ij\parallel k}-g_{ik\parallel j}\right) \check{e}%
^{i}\wedge dx^{j}\wedge dx^{k}+\frac{1}{2}\left( \check{e}_{k}g_{ij}-%
\check{e}_{i}g_{kj}\right) \check{e}^{k}\wedge \check{e}^{i}\wedge dx^{j},
\notag
\end{eqnarray}%
where
\begin{equation*}
g_{ij\parallel k}=\mathbf{e}_{k}g_{ij}-\check{B}_{ik}^{s}g_{sj}-\check{B}%
_{jk}^{s}g_{is},
\end{equation*}%
for $\check{B}_{ik}^{s}=\check{e}_{i}\check{N}_{k}^{s}$ and $\check{N}_{\
i}^{b}=\check{e}_{\ \underline{a}}^{b\ }\ N_{\ i}^{\underline{a}}.$

An \textbf{almost Hermitian model} of a generalized N--algebroid structure
is defined by a triple $\mathbf{\check{H}}^{2n}=(\mathbf{V}^{n+n},\mathbf{%
\check{g}},\mathbf{\check{F}}).$ A space $\mathbf{\check{H}}^{2n}$ is almost
K\" ahler if and only if $d\mathbf{\check{\theta}}=0,$ i.e.%
\begin{equation*}
\check{\Omega}_{jk}^{s}=0,g_{ij\parallel k}=g_{ik\parallel j},\check{e}%
_{k}g_{ij}=\check{e}_{i}g_{kj}.
\end{equation*}

A generalized Lagrange algebroid is not reducible to a Lagrange one if $%
\check{e}_{k}g_{ij}\neq \check{e}_{i}g_{kj},$ i.e. the almost sympletic
structure $\mathbf{\check{\theta}}$ (\ref{asstr}) is not integrable. One
considers also h-- (v--) Hermitian spaces $\mathbf{\check{H}}^{2n}$ if the
h-- (v--) distributions are integrable. For instance, the almost Hermitian
model of a Finsler (or Lagrange) space is an almost K\" ahler space \cite%
{mats} (or \cite{opr}).

An almost Hermitian connection $\mathbf{\check{D}}$ is of Lagrange type if
it preserve by parallelism the vertical distribution and is compatible with
the almost Hermitian structure $(\mathbf{\check{g}},\mathbf{\check{F}}),$
i.e. $\mathbf{\check{D}}_{\mathbf{X}}\mathbf{\check{g}}=0$ and $\mathbf{%
\check{D}}_{\mathbf{X}}\mathbf{\check{F}}=0$ for any d--vector $\mathbf{X}.$
Considering the canonical metrical d--connections for $\mathbf{\check{g}},$
we construct a canonical almost Hermitian connection $\ \mathbf{\check{D}},$
for $\mathbf{\check{H}}^{2n},$ with the coefficients $\ \mathbf{\check{\Gamma%
}}_{\ \beta \gamma }^{\alpha }=\left( \ \widehat{L}_{\ jk}^{i},\ \check{C}%
_{\ bc}^{a}\right) ,$ when
\begin{eqnarray}
\ \widehat{L}_{\ jk}^{i} &=&\frac{1}{2}g^{ih}(\mathbf{e}_{k}g_{jh}+\mathbf{e}%
_{j}g_{kh}-\mathbf{e}_{h}g_{jk}),  \label{cahcon} \\
\ \check{C}_{\ bc}^{a} &=&\check{e}_{\ \underline{a}}^{a}\check{e}_{b}^{\
\underline{b}}\check{e}_{c}^{\ \underline{c}}\widehat{C}_{\ \underline{b}%
\underline{c}}^{\underline{a}}+\check{e}_{\ \underline{a}}^{a}\partial _{%
\underline{c}}\check{e}_{b}^{\ \underline{a}},  \notag
\end{eqnarray}%
where $\check{e}_{b}=\check{e}_{b}^{\ \underline{b}}\partial _{\underline{b}%
} $ (\ref{vnhb}) and $\widehat{C}_{\ \underline{b}\underline{c}}^{\underline{%
a}}=\frac{1}{2}g^{\underline{a}\underline{e}}(e_{\underline{b}}g_{\underline{%
e}\underline{c}}+e_{\underline{c}}g_{\underline{e}\underline{b}}-e_{%
\underline{e}}g_{\underline{b}\underline{c}})$ computed as in (\ref{3cdctb})
but with respect to the ''underlined'' basis, when $g^{\underline{e}%
\underline{c}}$ with respect to $\partial _{\underline{e}}\otimes \partial _{%
\underline{c}}$ has the same values as $g^{ij}$ with respect to $\mathbf{e}%
_{i}\otimes \mathbf{e}_{j}.$

The curvature of (\ref{cahcon}) is
\begin{eqnarray}
\ \check{R}_{\ hjk}^{i} &=&\mathbf{e}_{k}\ \widehat{L}_{\ hj}^{i}-\mathbf{e}%
_{j}\ \widehat{L}_{\ hk}^{i}+\ \widehat{L}_{\ hj}^{m}\ \widehat{L}_{\
mk}^{i}-\ \widehat{L}_{\ hk}^{m}\ \widehat{L}_{\ mj}^{i}-\ \check{C}_{\
ha}^{i}\ \check{\Omega}_{\ kj}^{a},\ \   \notag \\
\check{P}_{\ jka}^{i} &=&\check{e}_{a}\ \widehat{L}_{\ jk}^{i}-\ \mathbf{%
\check{D}}_{k}\ \check{C}_{\ ja}^{i},\ \check{S}_{\ bcd}^{a}=\check{e}_{\
\underline{a}}^{a}\check{e}_{b}^{\ \underline{b}}\check{e}_{c}^{\ \underline{%
c}}\check{e}_{d}^{\ \underline{d}}\widehat{S}_{\ \underline{b}\underline{c}%
\underline{d}}^{\underline{a}},  \label{curvahcon}
\end{eqnarray}%
for $\widehat{S}_{\ \underline{b}\underline{c}\underline{d}}^{\underline{a}%
}=e_{\underline{d}}\widehat{C}_{\ \underline{b}\underline{c}}^{\underline{a}%
}-e_{\underline{c}}\widehat{C}_{\ \underline{b}\underline{d}}^{\underline{a}%
}+\widehat{C}_{\ \underline{b}\underline{c}}^{\underline{e}}\widehat{C}_{\
\underline{e}\underline{d}}^{\underline{a}}-\widehat{C}_{\ \underline{b}%
\underline{d}}^{\underline{e}}\widehat{C}_{\ \underline{e}\underline{c}}^{%
\underline{a}}$ computed as in (\ref{3dcurvtb}) but with respect to the
''underlined'' base.

\subsection{Almost Hermitian connections and general relativity}

We prove that the Einstein gravity on a (pseudo) Riemannian manifold $%
V^{n+n} $ can be equivalently redefined as an almost Hermitian model for
N--anholonomic Lie algebroids if a nonintegrable N--connection splitting is
prescribed. The Einstein theory can be also modified by considering certain
canonical lifts on tangent bundles. The first class of Finsler--Lagrange
like models \cite{vqgr3} preserves the local Lorentz symmetry and can be
applied for constructing exact solutions in Einstein gravity or for
developing some approaches to quantum gravity following methods of
geometric/deformation quantization. The second class of such models \cite%
{vqgr2} can be considered for some extensions to canonical quantum theories
of gravity which can be elaborated in a renormalizable form, but, in
general, result in violation of local Lorentz symmetry by such quantum
effects.

\subsubsection{Nonholonomic deformations in Einstein gravity}

Let us consider a metric $\underline{g}_{\alpha \beta }$ (\ref{ansatz}),
which for a $\left( n+n\right) $--splitting by a set of prescribed
coefficients $N_{i}^{a}(x,y)$ can be represented as a d--metric $\mathbf{g}$
(\ref{m1}), or $\mathbf{\check{g}}$ (\ref{adlegdm}), in N--algebroid form.
Respectively, we can write the Einstein equations in the form (\ref{einstgr}%
), or, equivalently, in the form (\ref{einstgrdef}) with the source $\ ^{Z}%
\mathbf{\Upsilon }_{\alpha \beta }$ defined by the off--diagonal metric
coefficients of $\underline{g}_{\alpha \beta },$ depending linearly on $%
N_{i}^{a},$ and generating the distorsion tensor $\ _{\shortmid }Z_{\ \alpha
\beta }^{\gamma }$ (\ref{cdeftc}).

Computing the Ricci and Einstein d--tensors by contracting the indices in (%
\ref{curvahcon}), we conclude that the Einstein equations written in terms
of the almost Hermitian d--connection can be also parametrized in the form (%
\ref{einstgrdef}). Such geometric structures are nonholonomic: working
respectively with $\underline{g},$ $\mathbf{g}$ or $\mathbf{\check{g},}$ we
elaborate equivalent geometric and physical models on $V^{n+n},\mathbf{V}%
^{n+n},$ or $^{N}\mathcal{A}\doteqdot (\mathbf{V}^{n+n},[\cdot ,\cdot ],%
\check{\rho})$ and $H^{2n}(\mathbf{V}^{n+n},\mathbf{\check{g},\check{F}}).$
Even for vacuum configurations, when $\mathbf{\Upsilon }_{\alpha \beta }=0,$
in the almost Hermitian model of the Einstein gravity, we have an effective
source $\ ^{Z}\mathbf{\Upsilon }_{\alpha \beta }$ induced by the
coefficients of generic off--diagonal metric. Nevertheless, there are
possible integrable configurations, when the conditions (\ref{intcond}) are
satisfied. In this case, $^{Z}\mathbf{\Upsilon }_{\alpha \beta }=0,$ and we
can construct effective Hermitian configurations defining vacuum Einstein
foliations.

One should be noted that the geometry of nonholonomic $2+2$ splitting in
general relativity, with nonholonomic frames and d--connections, or almost
Hermitian connections, is very different from the geometry of the well known
$3+1$ splitting ADM formalism, see \cite{mtw}, when only the Levi Civita
connection is used. Following the anholonomic frame method, we work with
different classes of connections and frames when some new symmetries and
invariants are distinguished and the field equations became exactly
integrable for some general metric ansatz. Constraining or redefining the
integral varieties and geometric objects, we can generate, for instance,
exact solutions in Einstein gravity and compute quantum corrections to such
solutions.

\subsubsection{Conformal lifts of Einstein structures to tangent bund\-les}

Let us consider a pseudo--Riemannian manifold $M$ enabled with a metric $\
_{\shortmid }g_{ij}(x)$ as a solution of the Einstein equations. We define a
procedure lifting $_{\shortmid }g_{ij}(x)$ conformally on $TM$ and inducing
a generalized Lagrange structure and a corresponding almost Hermitian
geometry. Let us introduce
\begin{equation*}
\ ^{\varpi }\mathcal{L}(x,y)\doteqdot \varpi ^{2}(x,y)g_{ab}(x)y^{a}y^{b}
\end{equation*}%
and use
\begin{equation}
\ ^{\varpi }g_{ab}\doteqdot \frac{1}{2}\frac{\partial ^{2}\ ^{\varpi }%
\mathcal{L}}{\partial y^{a}\partial y^{b}}  \label{cslm}
\end{equation}%
as a Lagrange Hessian for (\ref{slm}). A space $GL^{n}=(M,\ ^{\varpi
}g_{ij}(x,y))$ possess a weakly regular conformally deformed metric if $%
L^{n}=\left[ M,L=\sqrt{|\ ^{\varpi }\mathcal{L}|})\right] $ is a Lagrange
space. We can construct a canonical N--connection $\ ^{\varpi }N_{i}^{a}$
following formulas (\ref{glnc}), using $\ ^{\varpi }\mathcal{L}$ instead of $%
\mathcal{L}$ and $\ ^{\varpi }g_{ab}$ instead of $\ ^{\mathcal{L}}g_{ab}$ (%
\ref{glhf}), and define a d--metric on $TM,$%
\begin{equation}
\ ^{\varpi }\mathbf{g}=\ ^{\varpi }g_{ij}(x,y)\ dx^{i}\otimes dx^{j}+\
^{\varpi }g_{ij}(x,y)\ ^{\varpi }\mathbf{e}^{i}\otimes \ ^{\varpi }\mathbf{e}%
^{j},  \label{eslm}
\end{equation}%
where $^{\varpi }\mathbf{e}^{i}=dy^{i}+$ $^{\varpi }N_{i}^{j}dx^{i}.$ The
canonical d--connection and corresponding curvatures are constructed as in
generalized Lagrange geometry but using $\ ^{\varpi }\mathbf{g.}$

It is possible to reformulate the model with d--metric (\ref{eslm}) for
almost Hermitian spaces as we considered in section \ref{sglha}. For the
d--metric (\ref{eslm}), the model is elaborated for tangent bundles with
holonomic vertical frame structure. The linear operator $\mathbf{F}$
defining the almost complex structure acts on $\mathbf{TM}$ following
formulas
\begin{equation*}
\mathbf{F}(\ ^{\varpi }\mathbf{e}_{i})=-\partial _{i}\mbox{\ and \ }\mathbf{F%
}(\partial _{i})=\ ^{\varpi }\mathbf{e}_{i},
\end{equation*}%
when $\mathbf{F\circ F=-I,}$ for $\mathbf{I}$ being the unity matrix. The
operator $\mathbf{F}$ reduces to a complex structure if and only if the
h--distribution is integrable.

The metric $\ ^{\varpi }\mathbf{g}$ (\ref{eslm}) induces a 2--form
associated to $\mathbf{F}$ following formulas $\ ^{\varpi }\mathbf{\theta
(X,Y)}\doteqdot \ ^{\varpi }\mathbf{g}\left( \mathbf{FX,Y}\right) $ for any
d--vectors $\mathbf{X}$ and $\mathbf{Y.}$ In local form, we have
\begin{equation*}
\ ^{\varpi }\mathbf{\theta }=\ ^{\varpi }g_{ij}(x,y)dy^{i}\wedge dx^{j},
\end{equation*}%
similarly to (\ref{asstr}). The canonical d--connection $\ ^{\varpi }%
\widehat{\mathbf{D}},$ with N--adapted coefficients $^{\varpi }\mathbf{%
\Gamma }_{\ \beta \gamma }^{\alpha }=\left( \ ^{\varpi }\ \widehat{L}_{\
jk}^{i},\ \ ^{\varpi }\widehat{C}_{\ bc}^{a}\right) ,$ and corresponding
d--curvature can be computed respectively by the formulas (\ref{cahcon}) and
(\ref{curvahcon}) where $\check{e}_{b}^{\ \underline{b}}=\delta _{b}^{\
\underline{b}}$ and $\ ^{\varpi }g_{ij}$ are used instead of $g_{ij}.$

The model of almost Hermitian gravity $H^{2n}(\mathbf{TM},\ ^{\varpi }%
\mathbf{g,F})$ can be applied in order to construct different extensions of
general relativity to geometric quantum models on tangent bundle \cite{vqgr2}%
. Such models will result positively in violation of local Lorentz symmetry,
because the geometric objects depend on fiber variables $y^{a}.$ The
quasi--classical corrections can be obtained in the approximation $\varpi
\sim 1.$ We omit in this work consideration of quantum models, but note that
Finsler methods and almost K\"{a}hler geometry seem to be very useful for
such generalizations of Einstein gravity.

\section{Finsler--Lagrange Metrics in Einstein \& String Gravity}

\label{sflexso}We consider certain general conditions when Lagrange and
Finsler structures can be modelled as exact solutions in string and Einstein
gravity. Then, we analyze two explicit examples of exact solutions of the
Einstein equations modelling generalized Lagrange--Finsler geometries and
nonholonomic deformations of physically valuable equations in Einstein
gravity to such locally anisotropic configurations.

\subsection{Einstein spaces modelling generalized Finsler structures}

In this section, we outline the calculation leading from generalized
Lagrange and Finsler structures to exact solutions in gravity.

Let us consider a d--metric of type (\ref{adlegdm}), which is also
nonholonomically deformed in the h--part, when
\begin{equation}
\ ^{\varepsilon }\mathbf{\check{g}}=\ ^{\varepsilon }g_{i^{\prime }j^{\prime
}}(x^{k^{\prime }},y^{l^{\prime }})\left( e^{i^{\prime }}\otimes
e^{j^{\prime }}+\mathbf{\check{e}}^{i^{\prime }}\otimes \mathbf{\check{e}}%
^{j^{\prime }}\right) ,  \label{m2}
\end{equation}%
where $\ ^{\varepsilon }g_{i^{\prime }j^{\prime }}$ can be any metric
defined by nonholonomic transforms (\ref{ftgfs}) or a v--metric $\ ^{%
\mathcal{L}}g_{ij}$ (\ref{glhf}), $\ ^{L}g_{ij}$ (\ref{lqf}), or $\
^{F}g_{ij}$ (\ref{fhes}). The co--frame h-- and v--bases are
\begin{eqnarray*}
e^{i^{\prime }} &=&\ e_{\ i}^{i^{\prime }}(x,y)\ dx^{i}, \\
\mathbf{\check{e}}^{a^{\prime }} &=&\check{e}_{\ a}^{a^{\prime }}(x,y)\
\delta y^{a}=\check{e}_{\ a}^{a^{\prime }}\left( dy^{a}+\ _{\shortmid
}N_{i}^{a}dx^{i}\right) =e^{a^{\prime }}+\ _{\shortmid }\check{N}_{i^{\prime
}}^{a}e^{i^{\prime }},
\end{eqnarray*}%
for $e^{a^{\prime }}=\check{e}_{\ a}^{a^{\prime }}dy^{a}$ and $\ _{\shortmid
}\check{N}_{i^{\prime }}^{a}=\check{e}_{\ a}^{a^{\prime }}\ _{\shortmid
}N_{i^{\prime }}^{a}e_{\ i}^{i^{\prime }},$ when there are considered
nonholonomic transforms of type (\ref{htransm}), (\ref{vtransm}) and (\ref%
{pfnc}),
\begin{equation}
^{\varepsilon }g_{i^{\prime }j^{\prime }}=e_{i^{\prime }}^{\ i}e_{i^{\prime
}}^{\ i}\ \ _{\shortmid }g_{ij},\ \ _{\shortmid }h_{ab}=\ ^{\varepsilon
}g_{a^{\prime }b^{\prime }}\check{e}_{\ a}^{a^{\prime }}\check{e}_{\
b}^{b^{\prime }},\ \ _{\shortmid }N_{i}^{a}=\eta _{i}^{a}(x,y)\
^{\varepsilon }N_{i}^{a},  \label{nhdfes}
\end{equation}%
where we do not consider summation on indices for ''polarization'' functions
$\eta _{i}^{a}$ and $^{\varepsilon }N_{i}^{a}$ is a canonical connection
corresponding to $^{\varepsilon }g_{i^{\prime }j^{\prime }}.$

The d--metric (\ref{m2}) is equivalently transformed into the d--metric
\begin{eqnarray}
\ ^{\varepsilon }\mathbf{\check{g}} &=&\ _{\shortmid }g_{ij}(x)dx^{i}\otimes
dx^{j}+\ _{\shortmid }h_{ab}(x,y)\delta y^{a}\otimes \delta y^{a},
\label{m2a} \\
\delta y^{a} &=&dy^{a}+\ _{\shortmid }N_{i}^{a}(x,y)dx^{i},  \notag
\end{eqnarray}%
where the coefficients $\ _{\shortmid }g_{ij}(x),\ \ _{\shortmid
}h_{ab}(x,y) $ and $\ _{\shortmid }N_{i}^{a}(x,y)$ are constrained to be
defined by a class of exact solutions (\ref{gensol1}), in string gravity, or
(\ref{eeqsol}), in Einstein gravity. If it is possible to get the limit $%
\eta _{i}^{a}\rightarrow 1,$ we can say that an exact solution (\ref{m2a})
models exactly a respective (generalized) Lagrange, or Finsler,
configuration. We argue that we define a nonholonomic deformation of a
Finsler (Lagrange) space given by data $^{\varepsilon }g_{i^{\prime
}j^{\prime }}$ and $\ ^{\varepsilon }N_{i}^{a}$ as a class of exact
solutions of the Einstein equations given by data $\ _{\shortmid }g_{ij},\
_{\shortmid }h_{ab}$ and $\ _{\shortmid }N_{i}^{a},$ for any $\eta
_{i}^{a}(x,y)\neq 1.$ Such constructions are possible, if certain nontrivial
values of $e_{i^{\prime }}^{\ i},\check{e}_{\ a}^{a^{\prime }}$ and $\eta
_{i}^{a}$ can be algebraically defined from relations (\ref{nhdfes}) for any
defined sets of coefficients of the d--metric (\ref{m2}) and (\ref{m2a}).

Expressing a solution in the form (\ref{m2}), we can define the
corresponding almost Hermitian 1--form%
\begin{equation*}
\mathbf{\check{\theta}}=g_{i^{\prime }j^{\prime }}(x,y)\check{e}^{j^{\prime
}}\wedge e^{i^{\prime }},
\end{equation*}%
see (\ref{asstr}), and construct an almost Hermitian geometry characterizing
this solution, like we considered in section \ref{sglha}, but for
\begin{equation*}
\mathbf{\check{F}}(\mathbf{e}_{i^{\prime }})=-\check{e}_{i^{\prime }}%
\mbox{\
and \ }\mathbf{\check{F}}(\check{e}_{i^{\prime }})=\mathbf{e}_{i^{\prime }},
\end{equation*}%
when $\mathbf{e}_{i^{\prime }}=e_{i^{\prime }}^{\ i}\left( \frac{\partial }{%
\partial x^{i}}-\ _{\shortmid }N_{i}^{a}\frac{\partial }{\partial y^{a}}%
\right) =e_{i^{\prime }}-\ _{\shortmid }\check{N}_{i^{\prime }}^{a^{\prime }}%
\check{e}_{a^{\prime }}.$ This is convenient for further applications to
certain models of quantum gravity and geometry. For explicit constructions
of the solutions, it is more convenient to work with parametrizations of
type (\ref{m2a}).

Finally, in this section, we note that the general properties of integral
varieties of such classes of solutions are discussed in Refs. \cite%
{vparsol,vsgg}.

\subsection{Deformation of Einstein exact solutions into Lagrange--Finsler
metrics}

Let us consider a metric ansatz $\ _{\shortmid }g_{\alpha \beta }$ (\ref{m1}%
) with quadratic metric interval%
\begin{eqnarray}
ds^{2} &=&\ _{\shortmid }g_{1}(x^{1},x^{2})\left( dx^{1}\right) ^{2}+\
_{\shortmid }g_{2}(x^{1},x^{2})\left( dx^{2}\right) ^{2}  \label{exsolpr} \\
&&+\ _{\shortmid }h_{3}(x^{1},x^{2},v)\left[ dv+\ _{\shortmid
}w_{1}(x^{1},x^{2},v)dx^{1}+\ _{\shortmid }w_{2}(x^{1},x^{2},v)dx^{2}\right]
^{2}  \notag \\
&&+\ _{\shortmid }h_{4}(x^{1},x^{2},v)\left[ dy^{4}+\ _{\shortmid
}n_{1}(x^{1},x^{2},v)dx^{1}+\ _{\shortmid }n_{2}(x^{1},x^{2},v)dx^{2}\right]
^{2}  \notag
\end{eqnarray}%
defining an exact solution of the Einstein equations (\ref{einstgr}), for
the Levi--Civita connection, when the source $\Upsilon _{\alpha \beta }$ is
zero or defined by a cosmological constant. We parametrize the coordinates
in the form $u^{\alpha }=(x^{1},x^{2},y^{3}=v,y^{4})$ and the N--connection
coefficients as $\ _{\shortmid }N_{i}^{3}=\ _{\shortmid }w_{i}$ and $%
_{\shortmid }N_{i}^{4}=\ _{\shortmid }n_{i}.$

We nonholonomically deform the coefficients of the \textbf{primary}
d--metric (\ref{exsolpr}), similarly to (\ref{nhdfes}), when the \textbf{%
target} quadratic interval
\begin{eqnarray}
ds_{\eta }^{2} &=&g_{i}\left( dx^{i}\right) ^{2}+h_{a}\left(
dy^{a}+N_{i}^{a}dx^{i}\right) ^{2}  \label{targ1} \\
&=&e_{\ i}^{i^{\prime }}\ e_{\ j}^{j^{\prime }}\ ^{\varepsilon }g_{i^{\prime
}j^{\prime }}dx^{i}dx^{j}  \notag \\
&&+\check{e}_{\ a}^{a^{\prime }}\ \check{e}_{\ b}^{b^{\prime }}\
^{\varepsilon }g_{a^{\prime }b^{\prime }}\left( dy^{a}+\eta _{i}^{a}\
^{\varepsilon }N_{i}^{a}dx^{i}\right) \left( dy^{b}+\eta _{j}^{b}\
^{\varepsilon }N_{j}^{b}dx^{j}\right)  \notag
\end{eqnarray}%
can be equivalently parametrized in the form%
\begin{eqnarray}
ds_{\eta }^{2} &=&\eta _{j}\ _{\shortmid }g_{j}(x^{i})\left( dx^{j}\right)
^{2}  \label{targ1a} \\
&&+\eta _{3}(x^{i},v)\ _{\shortmid }h_{3}(x^{i},v)\left[ dv+\ ^{w}\eta
_{i}(x^{k},v)\ ^{\varepsilon }w_{i}(x^{k},v)dx^{i}\right] ^{2}  \notag \\
&&+\eta _{4}(x^{i},v)\ _{\shortmid }h_{4}(x^{i},v)\left[ dy^{4}+\ ^{n}\eta
_{i}(x^{k},v)\ \ ^{\varepsilon }n_{i}(x^{k},v)dx^{i}\right] ^{2},  \notag
\end{eqnarray}%
similarly to ansatz (\ref{ans5d}), and defines a solution of type (\ref%
{gensol1}) (with N--connecti\-on coefficients (\ref{gensol1w}) and (\ref%
{gensol1n})), for the canonical d--connection, or a solution of type (\ref%
{eeqsol}) with the coefficients subjected to solve the conditions (\ref{ep2b}%
).

The class of target metrics (\ref{targ1}) and (\ref{targ1a}) defining the
result of a nonholonomic deformation of the primary data $[\ _{\shortmid
}g_{i},\ _{\shortmid }h_{a},\ _{\shortmid }N_{i}^{b}]$ to a
Finsler--Lagrange configuration $[^{\varepsilon }g_{i^{\prime }j^{\prime
}},\ ^{\varepsilon }N_{j}^{b}]$ are parametrized by vales $e_{\
i}^{i^{\prime }},\check{e}_{\ a}^{a^{\prime }}$ and $\eta _{i}^{a}.$ These
values can be expressed in terms of some generation and integration
functions and the coefficients of the primary and Finsler like d--metrics
and N--connections in such a manner when a primary class of exact solutions
is transformed into a ''more general'' class of exact solutions. In a
particular case, we can search for solutions when the target metrics
transform into primary metrics under certain infinitesimal limits of the
nonholonomic deforms.

In general form, the solutions of equations (\ref{nhesp}) transformed into
the system of partial differential equations (\ref{ep1a})--(\ref{ep4a}), for
the d--metrics (\ref{targ1}), equivalently (\ref{targ1a}), are given by
corresponding sets of frame, N--connections and d--metric coefficients which
state for the h--part

\begin{eqnarray}
e_{\ 1}^{1^{\prime }} &=&\sqrt{|\eta _{1}|}\ \sqrt{|\ _{\shortmid }g_{1}|}%
\times \sqrt{|\ ^{\varepsilon }g_{1^{\prime }1^{\prime }}\ ^{\varepsilon
}g_{2^{\prime }2^{\prime }}\left[ \left( \ ^{\varepsilon }g_{1^{\prime
}1^{\prime }}\right) ^{2}\ ^{\varepsilon }g_{2^{\prime }2^{\prime }}+\left(
\ ^{\varepsilon }g_{1^{\prime }2^{\prime }}\right) ^{3}\right] ^{-1}|},
\notag \\
e_{\ 1}^{2^{\prime }} &=&\sqrt{|\eta _{2}|}\ \sqrt{|\ _{\shortmid }g_{1}|}\ /%
\sqrt{|\ ^{\varepsilon }g_{1^{\prime }1^{\prime }}\ ^{\varepsilon
}g_{2^{\prime }2^{\prime }}\left[ \left( \ ^{\varepsilon }g_{1^{\prime
}1^{\prime }}\right) ^{2}\ ^{\varepsilon }g_{2^{\prime }2^{\prime }}+\left(
\ ^{\varepsilon }g_{1^{\prime }2^{\prime }}\right) ^{3}\right] |},  \notag \\
e_{\ 2}^{1^{\prime }} &=&-\sqrt{|\eta _{2}|}\ \sqrt{|\ _{\shortmid }g_{2}|}%
\times g_{1^{\prime }2^{\prime }}/\sqrt{|\ ^{\varepsilon }g_{1^{\prime
}1^{\prime }}\left[ \left( \ ^{\varepsilon }g_{1^{\prime }1^{\prime
}}\right) \ ^{\varepsilon }g_{2^{\prime }2^{\prime }}-\left( \ ^{\varepsilon
}g_{1^{\prime }2^{\prime }}\right) ^{2}\right] |},  \notag \\
e_{\ 2}^{2^{\prime }} &=&\sqrt{|\eta _{2}|}\ \sqrt{|\ _{\shortmid }g_{2}|}%
\times \sqrt{|\ ^{\varepsilon }g_{1^{\prime }1^{\prime }}/\left[ \left( \
^{\varepsilon }g_{1^{\prime }1^{\prime }}\right) \ ^{\varepsilon
}g_{2^{\prime }2^{\prime }}-\left( \ ^{\varepsilon }g_{1^{\prime }2^{\prime
}}\right) ^{2}\right] |},  \label{hfr}
\end{eqnarray}%
and for the v--part

\begin{eqnarray}
e_{\ 3}^{3^{\prime }} &=&\sqrt{|\eta _{3}|}\ \sqrt{|\ _{\shortmid }h_{3}|}%
\times \sqrt{|\ ^{\varepsilon }g_{1^{\prime }1^{\prime }}\ ^{\varepsilon
}g_{2^{\prime }2^{\prime }}\left[ \left( \ ^{\varepsilon }g_{1^{\prime
}1^{\prime }}\right) ^{2}\ ^{\varepsilon }g_{2^{\prime }2^{\prime }}+\left(
\ ^{\varepsilon }g_{1^{\prime }2^{\prime }}\right) ^{3}\right] ^{-1}|},
\notag \\
e_{\ 3}^{4^{\prime }} &=&\sqrt{|\eta _{3}|}\ \sqrt{|\ _{\shortmid }h_{3}|}\ /%
\sqrt{|\ ^{\varepsilon }g_{1^{\prime }1^{\prime }}\ ^{\varepsilon
}g_{2^{\prime }2^{\prime }}\left[ \left( \ ^{\varepsilon }g_{1^{\prime
}1^{\prime }}\right) ^{2}\ ^{\varepsilon }g_{2^{\prime }2^{\prime }}+\left(
\ ^{\varepsilon }g_{1^{\prime }2^{\prime }}\right) ^{3}\right] |},  \notag \\
e_{\ 4}^{3^{\prime }} &=&-\sqrt{|\eta _{4}|}\ \sqrt{|\ _{\shortmid }h_{4}|}%
\times g_{1^{\prime }2^{\prime }}/\sqrt{|\ ^{\varepsilon }g_{1^{\prime
}1^{\prime }}\left[ \left( \ ^{\varepsilon }g_{1^{\prime }1^{\prime
}}\right) \ ^{\varepsilon }g_{2^{\prime }2^{\prime }}-\left( \ ^{\varepsilon
}g_{1^{\prime }2^{\prime }}\right) ^{2}\right] |},  \notag \\
e_{\ 4}^{4^{\prime }} &=&\sqrt{|\eta _{4}|}\ \sqrt{|\ _{\shortmid }h_{4}|}%
\times \sqrt{|\ ^{\varepsilon }g_{1^{\prime }1^{\prime }}/\left[ \left( \
^{\varepsilon }g_{1^{\prime }1^{\prime }}\right) \ ^{\varepsilon
}g_{2^{\prime }2^{\prime }}-\left( \ ^{\varepsilon }g_{1^{\prime }2^{\prime
}}\right) ^{2}\right] |},  \label{vfr}
\end{eqnarray}%
where h--polarizations $\eta _{j}$ are defined from $g_{j}=\eta _{j}\
_{\shortmid }g_{j}(x^{i})=$ $\epsilon _{j}e^{\psi (x^{i})},$ with signatures
$\epsilon _{i}=\pm 1,$ for $\psi (x^{i})$ being a solution of the 2D
equation
\begin{equation}
\epsilon _{1}\psi ^{\bullet \bullet }+\epsilon _{2}\psi ^{^{\prime \prime
}}=\lambda ,  \label{ep2b1}
\end{equation}%
for a given source $\Upsilon _{4}\left( x^{i}\right) =\lambda ,$ and the
v--polarizations $\eta _{a}$ defined from the data $h_{a}=\eta _{a}\
_{\shortmid }h_{a},$ for
\begin{eqnarray}
h_{3} &=&\epsilon _{3}h_{0}^{2}(x^{i})\left[ f^{\ast }\left( x^{i},v\right) %
\right] ^{2}|\ ^{\lambda }\varsigma \left( x^{i},v\right) |,  \notag \\
h_{4} &=&\epsilon _{4}\left[ f\left( x^{i},v\right) -f_{0}(x^{i})\right]
^{2},  \label{ep2b2}
\end{eqnarray}%
where
\begin{equation*}
\ ^{\lambda }\varsigma \left( x^{i},v\right) =\varsigma _{\lbrack 0]}\left(
x^{i}\right) -\frac{\epsilon _{3}}{8}\lambda h_{0}^{2}(x^{i})\int f^{\ast
}\left( x^{i},v\right) \left[ f\left( x^{i},v\right) -f_{0}(x^{i})\right] dv,
\end{equation*}%
for $\Upsilon _{2}(x^{k},v)=\lambda .$ The polarizations $\eta _{i}^{a}$ of
N--connection coefficients
\begin{equation*}
N_{i}^{3}=w_{i}=\ ^{w}\eta _{i}(x^{k},v)\ ^{\varepsilon }w_{i}(x^{k},v),\
N_{i}^{4}=n_{i}=\ ^{n}\eta _{i}(x^{k},v)\ \ ^{\varepsilon }n_{i}(x^{k},v)
\end{equation*}%
are computed from the respective formulas
\begin{equation}
\ ^{w}\eta _{i}\ ^{\varepsilon }w_{i}=-\frac{\partial _{i}\ ^{\lambda
}\varsigma \left( x^{k},v\right) }{\ ^{\lambda }\varsigma ^{\ast }\left(
x^{k},v\right) }  \label{gensol1wl}
\end{equation}%
and
\begin{equation}
\ ^{n}\eta _{k}\ ^{\varepsilon }n_{k}=\ ^{1}n_{k}\left( x^{i}\right) +\
^{2}n_{k}\left( x^{i}\right) \int \frac{\left[ f^{\ast }\left(
x^{i},v\right) \right] ^{2}}{\left[ f\left( x^{i},v\right) -f_{0}(x^{i})%
\right] ^{3}}\ ^{\lambda }\varsigma \left( x^{i},v\right) dv.
\label{gensol1nl}
\end{equation}

We generate a class of exact solutions for Einstein spaces with $\Upsilon
_{2}=\Upsilon _{4}=\lambda $ if the integral varieties defined by $%
g_{j},h_{a},w_{i}$ and $n_{i}$ are subjected to constraints (\ref{ep2b}).

\subsection{Solitonic pp--waves and their effective Lagrange spaces}

Let us consider a d--metric of type (\ref{exsolpr}),
\begin{equation}
\delta s_{[pw]}^{2}=-dx^{2}-dy^{2}-2\kappa (x,y,v)\ dv^{2}+\ dp^{2}/8\kappa
(x,y,v),  \label{5aux5}
\end{equation}%
where the local coordinates are $\ x^{1}=x,\ x^{2}=y,\ y^{3}=v,\ y^{4}=p,$
and the nontrivial metric coefficients are parametrized%
\begin{equation*}
\ _{\shortmid }g_{1}=-1,\ \ _{\shortmid }g_{2}=-1,\ _{\shortmid
}h_{3}=-2\kappa (x,y,v),\ \ _{\shortmid }h_{4}=1/\ 8\ \kappa (x,y,v).
\end{equation*}%
This is vacuum solution of the Einstein equation defining pp--waves \cite%
{peres}: for any $\kappa (x,y,v)$ solving
\begin{equation*}
\kappa _{xx}+\kappa _{yy}=0,
\end{equation*}%
with $v=z+t$ and $p=z-t,$ where $(x,y,z)$ are usual Cartesian coordinates
and $t$ is the time like coordinate. Two explicit examples of such solutions
are given by
\begin{equation*}
\kappa =(x^{2}-y^{2})\sin v,
\end{equation*}%
defining a plane monochromatic wave, or by
\begin{eqnarray*}
\kappa &=&\frac{xy}{\left( x^{2}+y^{2}\right) ^{2}\exp \left[ v_{0}^{2}-v^{2}%
\right] },\mbox{ for }|v|<v_{0}; \\
&=&0,\mbox{ for }|v|\geq v_{0},
\end{eqnarray*}%
defining a wave packet travelling with unit velocity in the negative $z$
direction.

We nonholonomically deform the vacuum solution (\ref{5aux5}) to a d--metric
of type (\ref{targ1a})
\begin{eqnarray}
ds_{\eta }^{2} &=&-e^{\psi (x,y)}\left[ \left( dx\right) ^{2}+(dy)^{2}\right]
\label{5auxd} \\
&&-\eta _{3}(x,y,v)\ \cdot 2\kappa (x,y,v)\left[ dv+\ ^{w}\eta
_{i}(x,y,v,p)\ ^{\varepsilon }w_{i}(x,y,v,p)dx^{i}\right] ^{2}  \notag \\
&&+\eta _{4}(x,y,v)\ \cdot \frac{1}{8\kappa (x,y,v)}\left[ dy^{4}+\ ^{n}\eta
_{i}(x,y,v,p)\ \ ^{\varepsilon }n_{i}(x,y,v,p)dx^{i}\right] ^{2},  \notag
\end{eqnarray}%
where the polarization functions $\eta _{1}=\eta _{2}=e^{\psi (x,y)},\eta
_{3,4}(x,y,v),\ ^{w}\eta _{i}(x,y,v)$ \ and $\ ^{n}\eta _{i}(x,y,v)\ $\ have
to be defined as solutions in the form (\ref{ep2b1}), (\ref{ep2b2}), (\ref%
{gensol1wl}) and (\ref{gensol1nl}) for a string gravity ansatz (\ref{ansh}),
$\lambda =\lambda _{H}^{2}/2,$ \ and \ a prescribed (in this section)
analogous mechanical system with $\ $%
\begin{equation}
N_{i}^{a}=\{w_{i}(x,y,v)=\ ^{w}\eta _{i}\ ^{L}w_{i},n_{i}(x,y,v)=\ ^{n}\eta
_{i}\ ^{\varepsilon }n_{i}\}  \label{cnclnd}
\end{equation}%
for $\varepsilon =L(x,y,v,p)$ considered as regular Lagrangian modelled on a
N--anholonomic manifold with holonomic coordinates $(x,y)$ and nonholonomic
coordinates $(v,p).$

A class of 3D solitonic configurations can defined by taking a polarization
function $\eta _{4}(x,y,v)=\eta (x,y,v)$ as a solution of solitonic equation%
\footnote{%
as a matter of principle we can consider\ that $\eta $ is a solution of any
3D solitonic, or other, nonlinear wave equation.}
\begin{equation}
\eta ^{\bullet \bullet }+\epsilon (\eta ^{\prime }+6\eta \ \eta ^{\ast
}+\eta ^{\ast \ast \ast })^{\ast }=0,\ \epsilon =\pm 1,  \label{5solit1}
\end{equation}%
and $\eta _{1}=\eta _{2}=e^{\psi (x,y)}$ as a solution of (\ref{ep2b1})
written as
\begin{equation}
\psi ^{\bullet \bullet }+\psi ^{\prime \prime }=\frac{\lambda _{H}^{2}}{2}.
\label{5lapl}
\end{equation}%
Introducing the\ above stated data for the ansatz (\ref{5auxd}) into the
equation (\ref{ep2a}), we get two equations relating $h_{3}=\eta _{3}\
_{\shortmid }h_{3}$ and $h_{4}=\eta _{4}\ _{\shortmid }h_{4},$
\begin{equation}
\eta _{4}=8\ \kappa (x,y,v)\left[ h_{4}^{[0]}(x,y)+\frac{1}{2\lambda _{H}^{2}%
}e^{2\eta (x,y,v)}\right]  \label{5sol2h5}
\end{equation}%
and
\begin{equation}
|\eta _{3}(x,y,v)|=\frac{e^{-2\eta (x,y,v)}}{2\kappa ^{2}(x,y,v)}\left[
\left( \sqrt{|\eta _{4}(x,y,v)|}\right) ^{\ast }\right] ^{2},
\label{5sol2h4}
\end{equation}%
where $h_{4}^{[0]}(x,y)$ is an integration function.

Having defined the coefficients $h_{a},$ we can solve the equations (\ref%
{ep3a}) and (\ref{ep4a}) expressing the coefficients (\ref{coef}) and (\ref%
{coefa}) through $\eta _{3}$ and $\eta _{4}$ defined by pp-- and solitonic
waves as in (\ref{5sol2h4}) and (\ref{5sol2h5}). The corresponding solutions
are
\begin{equation}
w_{1}=\ ^{w}\eta _{1}\ ^{L}w_{1}=\left( \phi ^{\ast }\right) ^{-1}\partial
_{x}\phi ,\ w_{2}=\ ^{w}\eta _{1}\ ^{L}w_{1}=\left( \phi ^{\ast }\right)
^{-1}\partial _{y}\phi ,  \label{5sol2w}
\end{equation}%
for $\phi ^{\ast }=\partial \phi /\partial v,$ see formulas (\ref{coefa})
and
\begin{equation}
n_{i}(x,y,v)=n_{i}^{[0]}(x,y)+n_{i}^{[1]}(x,y)\int \left| \eta
_{3}(x,y,v)\eta _{4}^{-3/2}(x,y,v)\right| dv,  \label{5sol2na}
\end{equation}%
where $n_{i}^{[0]}(x,y)$ and $n_{i}^{[1]}(x,y)$ are integration functions.

The values $e^{\psi (x,y)},$ $\eta _{3}$ (\ref{5sol2h4}), $\eta _{4}$ (\ref%
{5sol2h5}), $w_{i}$ (\ref{5sol2w}) and $n_{i}$ (\ref{5sol2na}) for the
ansatz (\ref{5auxd}) completely define a nonlinear superpositions of
solitonic and pp--waves as an exact solution of the Einstein equations in
string gravity if there are prescribed some initial values for the nonlinear
waves under consideration. In general, such solutions depend on some classes
of generation and integration functions.

It is possible to give a regular Lagrange analogous interpretation of an
explicit exact solution (\ref{5auxd}) if we prescribe a regular Lagrangian $%
\varepsilon =L(x,y,v,p),$ with Hessian $\ ^{L}g_{i^{\prime }j^{\prime }}=%
\frac{1}{2}\frac{\partial ^{2}L}{\partial y^{i^{\prime }}\partial
y^{j^{\prime }}},$ for $x^{i^{\prime }}=(x,y)$ and $y^{a^{\prime }}=(v,p).$
Introducing the values $\ ^{L}g_{i^{\prime }j^{\prime }},$ $\eta _{1}=\eta
_{2}=e^{\psi },\eta _{3},\eta _{4}$ and $_{\shortmid }h_{3},\ \ _{\shortmid
}h_{4},$ all defined above, into (\ref{hfr}) and (\ref{vfr}), we compute the
vierbein coefficients $e_{\ i}^{i^{\prime }}$ and $\check{e}_{\
a}^{a^{\prime }}$ which allows us to redefine equivalently the quadratic
element in the form (\ref{targ1}) as for a Lagrange N--algebroid for which
the N--connection coefficients $\ ^{L}N_{i}^{a}$ (\ref{cncl}) are
nonholonomically deformed to $N_{i}^{a}$ (\ref{cnclnd}). With respect to
such nonholonomic frames of references, an observer ''swimming in a string
gravitational ocean of interacting solitonic and pp--waves'' will see his
world as an analogous mechanical model defined by a regular Lagrangian $L.$

\subsection{Finsler--solitonic pp--waves in Schwarzschild spaces}

We consider a primary quadratic element
\begin{equation}
\delta s_{0}^{2}=-d\xi ^{2}-r^{2}(\xi )\ d\vartheta ^{2}-r^{2}(\xi )\sin
^{2}\vartheta \ d\varphi ^{2}+\varpi ^{2}(\xi )\ dt^{2},  \label{5aux1}
\end{equation}%
where the local coordinates and nontrivial metric coefficients are
parametriz\-ed in the form%
\begin{eqnarray}
x^{1} &=&\xi ,x^{2}=\vartheta ,y^{3}=\varphi ,y^{4}=t,  \label{5aux1p} \\
\ _{\shortmid }g_{1} &=&-1,\ \ _{\shortmid }g_{2}=-r^{2}(\xi ),\ \
_{\shortmid }h_{3}=-r^{2}(\xi )\sin ^{2}\vartheta ,\ \ _{\shortmid
}h_{4}=\varpi ^{2}(\xi ),  \notag
\end{eqnarray}%
for
\begin{equation*}
\xi =\int dr\ \left| 1-\frac{2\mu }{r}\right| ^{1/2}\mbox{\ and\ }\varpi
^{2}(r)=1-\frac{2\mu }{r}.
\end{equation*}%
For $\mu $ being a point mass, the element (\ref{5aux1}) defines the
Schwarzschild solution written in spacetime spherical coordinates $%
(r,\vartheta ,\varphi ,t).$

Our aim, is to find a nonholonomic deformation of metric (\ref{5aux1}) to a
class of new vacuum solutions modelled by certain types of Finsler
geometries.

The target stationary metrics are parametrized in the form similar to (\ref%
{targ1a}), see also (\ref{eeqsol}),
\begin{eqnarray}
ds_{\eta }^{2} &=&-e^{\psi (\xi ,\vartheta )}\left[ \left( d\xi \right)
^{2}+r^{2}(\xi )(d\vartheta )^{2}\right]  \label{5aux1pta} \\
&&-\eta _{3}(\xi ,\vartheta ,\varphi )\ \cdot r^{2}(\xi )\sin ^{2}\vartheta
\ \left[ d\varphi +\ ^{w}\eta _{i}(\xi ,\vartheta ,\varphi ,t)\
^{F}w_{i}(\xi ,\vartheta ,\varphi ,t)dx^{i}\right] ^{2}  \notag \\
&&+\eta _{4}(\xi ,\vartheta ,\varphi )\ \cdot \varpi ^{2}(\xi )\ \left[ dt+\
^{n}\eta _{i}(\xi ,\vartheta ,\varphi ,t)\ ^{F}n_{i}(\xi ,\vartheta ,\varphi
,t)dx^{i}\right] ^{2}.  \notag
\end{eqnarray}%
The polarization functions $\eta _{1}=\eta _{2}=e^{\psi (\xi ,\vartheta
)},\eta _{a}(\xi ,\vartheta ,\varphi ),\ ^{w}\eta _{i}(\xi ,\vartheta
,\varphi )$ \ and $\ ^{n}\eta _{i}(\xi ,\vartheta ,\varphi )\ $\ have to be
defined as solutions of (\ref{ep2b}) for $\Upsilon _{2}=\Upsilon _{4}=0$ and
a prescribed (in this section) locally anisotropic, on $\varphi ,$ geometry
with
\begin{equation*}
N_{i}^{a}=\{w_{i}(\xi ,\vartheta ,\varphi )=\ ^{w}\eta _{i}\ ^{F}w_{i},\
n_{i}(\xi ,\vartheta ,\varphi )=\ ^{n}\eta _{i}\ ^{F}n_{i}\},
\end{equation*}%
for $\varepsilon =F^{2}(\xi ,\vartheta ,\varphi ,t)$ considered as a
fundamental Finsler function for a Finsler geometry modelled on a
N--anholonomic manifold with holonomic coordinates $(r,\vartheta )$ and
nonholonomic coordinates $(\varphi ,t).$ We note that even the values $\
^{w}\eta _{i},\ ^{F}w_{i},$ $\ ^{n}\eta _{i},$ and $\ ^{F}n_{i}$ can depend
on time like variable $t,$ such dependencies must result in N--connection
coefficients of type $N_{i}^{a}(\xi ,\vartheta ,\varphi ).$

Putting together the coefficients solving the Einstein equations (\ref{ep2a}%
)--(\ref{ep4a}) and (\ref{ep2b}), the class of vacuum solutions in general
relativity related to (\ref{5aux1pta}) can be parametrized in the form
\begin{eqnarray}
ds_{\eta }^{2} &=&-e^{\psi (\xi ,\vartheta )}\left[ \left( d\xi \right)
^{2}+r^{2}(\xi )(d\vartheta )^{2}\right]  \label{aux6} \\
&&-h_{0}^{2}\ \left[ b^{\ast }(\xi ,\vartheta ,\varphi )\right] ^{2}\ \left[
d\varphi +w_{1}(\xi ,\vartheta )d\xi +w_{2}(\xi ,\vartheta )d\vartheta %
\right] ^{2}  \notag \\
&&+\left[ b(\xi ,\vartheta ,\varphi )-b_{0}(\xi ,\vartheta )\right] ^{2}\ %
\left[ dt+n_{1}(\xi ,\vartheta )d\xi +n_{2}(\xi ,\vartheta )d\vartheta %
\right] ^{2},  \notag
\end{eqnarray}%
where $h_{0}=const$ and the coefficients are constrained to solve the
equations
\begin{eqnarray}
\psi ^{\bullet \bullet }+\psi ^{^{\prime \prime }} &=&0,  \label{ep2bb} \\
w_{1}^{\prime }-w_{2}^{\bullet }+w_{2}w_{1}^{\ast }-w_{1}w_{2}^{\ast } &=&0,
\notag \\
n_{1}^{\prime }-n_{2}^{\bullet } &=&0,  \notag
\end{eqnarray}%
for instance, for $w_{1}=(b^{\ast })^{-1}(b+b_{0})^{\bullet },$ $%
w_{2}=(b^{\ast })^{-1}(b+b_{0})^{\prime },n_{2}^{\bullet }=\partial
n_{2}/\partial \xi $ and $n_{1}^{^{\prime }}=\partial n_{1}/\partial
\vartheta .$

The polarization functions relating (\ref{aux6}) to (\ref{5aux1pta}), are
computed in the form
\begin{eqnarray}
\eta _{1} &=&\eta _{2}=e^{\psi (\xi ,\vartheta )},\ \eta _{3}=\left[
h_{0}b^{\ast }/r(\xi )\sin \vartheta \right] ^{2},\eta _{4}=\left[
(b-b_{0})/\varpi \right] ^{2},  \label{polf6} \\
\ ^{w}\eta _{i} &=&w_{i}(\xi ,\vartheta )/\ ^{F}w_{i}(\xi ,\vartheta
,\varphi ,t),\ \ ^{n}\eta _{i}=n_{i}(\xi ,\vartheta )/\ ^{F}n_{i}(\xi
,\vartheta ,\varphi ,t).  \notag
\end{eqnarray}

The next step is to chose a Finsler geometry which will model (\ref{aux6}),
equivalently (\ref{5aux1pta}), as a Finsler like d--metric (\ref{targ1}).
For a fundamental Finsler function $F=F(\xi ,\vartheta ,\varphi ,t),$ where $%
x^{i^{\prime }}=(\xi ,\vartheta )$ are h--coordinates and $y^{a^{\prime
}}=\left( \varphi ,t\right) $ are v--coordinates, we compute $\
^{F}g_{a^{\prime }b^{\prime }}=(1/2)\partial ^{2}F/\partial y^{a^{\prime
}}\partial y^{b^{\prime }}$ following formulas (\ref{fhes}) and parametrize
the Cartan N--connection as $\ ^{C}N_{i}^{a}=\{\ ^{F}w_{i},\ ^{F}n_{i}\}.$
Introducing the values (\ref{5aux1p}),$\ ^{F}g_{i^{\prime }j^{\prime }}$ and
polarization functions (\ref{polf6}) into (\ref{hfr}) and (\ref{vfr}), we
compute the vierbein coefficients $e_{\ i}^{i^{\prime }}$ and $\check{e}_{\
a}^{a^{\prime }}$ which allows us to redefine equivalently the quadratic
element in the form (\ref{targ1}), in this case, for a Finsler N--algebroid
for which the N--connection coefficients $\ ^{C}N_{i}^{a}$ (\ref{cncl}) are
nonholonomically deformed to $N_{i}^{a}$ satisfying the last two conditions
in (\ref{ep2bb}). With respect to such nonholonomic frames of references, an
observer ''swimming in a locally anisotropic gravitational ocean'' will see
the nonholonomically deformed Schwarzschild geometry as an analogous Finsler
model defined by a fundamental Finser function $F.$

\section{Outlook and Conclusions}

In this review article, we gave a self--contained account of the core
developments on generalized Finsler--Lagrange geometries and their modelling
on (pseudo) Riemannian and Riemann--Cartan manifolds provided with preferred
nonholonomic frame structure. We have shown how the Einstein gravity and
certain string models of gravity with torsion can be equivalently
reformulated in the language of generalized Finsler and almost Hermitian/ K%
\"{a}hler geometries. It was also argued that former criticism and
conclusions on experimental constraints and theoretical difficulties of
Finsler like gravity theories were grounded only for certain classes of
theories with metric noncompatible connections on tangent bundles and/or
resulting in violation of local Lorentz symmetry. We emphasized that there
were omitted the results when for some well defined classes of nonholonomic
transforms of geometric structures we can model geometric structures with
local anisotropy, of Finsler--Lagrange type, and generalizations, on
(pseudo) Riemann spaces and Einstein manifolds.

Our idea was to consider not only some convenient coordinate and frame
transforms, which simplify the procedure of constructing exact solutions,
but also to define alternatively new classes of connections which can be
employed to generate new solutions in gravity. We proved that the solutions
for the so--called canonical distinguished connections can be equivalently
re--defined for the Levi Civita connection and/ or constrained to define
integral varieties of solutions in general relativity.

The main conclusion of this work is that we can avoid all existing
experimental restrictions and theoretical difficulties of Finsler physical
models if we work with metric compatible Finsler like structures on
nonholonomic (Riemann, or Riemann--Cartan) manifolds but not on tangent
bundles. In such cases, all nonholonomic constructions modelled as exact
solutions of the Einstein and matter field equations (with various string,
quantum field ... corrections) are compatible with the standard paradigm in
modern physics.

In other turn, we emphasize that in quantum gravity, statistical and
thermodynamical models with local anisotropy, gauge theories with
constraints and broken symmetry and in geometric mechanics, nonholonomic
configurations on (co) tangent bundles, of Finsler type and generalizations,
metric compatible or with nonmetricity, seem to be also very important.

Various directions in generalized Finsler geometry and applications has
matured enough so that some tenths of monographs have been written,
including some recent and updated: we cite here \cite%
{mats,ma1987,ma,mhl,mhf,mhss,mhh,bej,vsgg,vmon1,vstav,bcs,shen,ant,amaz,as1,as2,ap,bog}%
. These monographs approach and present the subjects from different
perspectives depending, of course, on the authors own taste, historical
period and interests both in geometry and physics. The monograph \cite{vsgg}
summarizes and develops the results oriented to application of Finsler
methods to standard theories of gravity (on nonholonomic manifolds, not only
on tangent bundles) and their noncommutative generalizations; it was also
provided a critical analysis of the constructions with nonmetricity and
violations of local Lorentz symmetry.

Finally, we suggest the reader to see a brief outline and comments on main
directions related to nonstandard applications in physics of Finsler
geometries and generalizations in Appendix.

\vskip5pt

\textbf{Acknowledgement: } The work is performed during a visit at the
Fields Institute. The author thanks M. Anastasiei, A. Bejancu, V. Ob\v{a}%
deanu and V. Oproiu for discussions and providing very important references
on the geometry of Finsler--Lagrange spaces, nonholonomic manifolds and
related almost K\"{a}hler geometry.

\appendix

\section{Historical and Bibliographical Comments}

One can be found by 3500 titles on key word ''Finsler'' in MathSciNet and
almost 150 titles in arxiv.org. It is not possible to review in an article
all known mathematical constructions and applications in various directions
in science related to Finsler geometry and generalizations. We shall sketch
only some very important lines of developments of such researches and
comment only a small part of results on nonstandard theories which, in our
subjective opinion, seem to have importance and certain perspectives to be
redefined for standard theories of gravity, mechanics and field interactions.

\setcounter{equation}{0} \renewcommand{\theequation}
{A.\arabic{equation}} \setcounter{subsection}{0}
\renewcommand{\thesubsection}
{A.\arabic{subsection}}

\subsection{ Moving frames, N--connections and nonholonomic (super) manifolds%
}

There are well known textbooks and monographs \cite{haw,mtw,wald,stw,sb}
where (pseudo--) Riemann geometry and general relativity are formulated in
arbitrary frame bases. The approach originates from the E. Cartan moving
frame method \cite{cartanfr1,cartanfr2} being developed both in abstract and
coordinate forms in modern gravity, see also supersymmetric generalizations
related to string gravity \cite{string3,string1,string2}.

The global definition of nonlinear connection (N--connection) is due to W.
Barthel (1963) \cite{barth}. In coefficient form, the N--connections can be
found in the E. Cartan's book on Finsler geometry \cite{cart} (1935) and in
A. Kawaguchi (1937,1952) works \cite{kaw1,kaw2}. The concept of
N--connection is also known as the Ehresmann connection \cite{ehr} (1955),
see also Grifone's works \cite{grifone}. The N--connection geometry was
developed in a series of works of the Romanian school of Finsler, Lagrange
and Hamilton geometries and higher order generalizations \cite%
{ma1987,ma,mhl,mhf,mhss,mhh,bej} (beginning the end of 50th of previous
century).

Then, the constructions with N--connections were generalized for
supersymmetric fiber variables \cite{bej} (1989,1990) following the DeWitt
approach to supermanifolds \cite{dewitt}. A definition of nonholonomic
supermanifolds was not possible in those works because it was not not yet
elaborated the concept of spinors for generalized Finsler spaces, see
details in section \ref{assscf}. Having accepted any global or local
constructions for supermanifolds and superbundles, for a well defined
nonholonomic spinor structure, it was possible to introduce N--connections
by a corresponding class of super--distributions and/or preferred systems of
superfields (i.e. super--vielbeinds). That allowed us to consider
nonholonomic generalizations of the geometry of supermanifolds, superstrings
and supergravity \cite{vncsup,vmon1,vcv}. Recently, the geometry of
semi--spray and N--connection structures on supermanifolds was developed in
Refs. \cite{aziz1,aziz2}. The geometry of nonholonomic supermanifolds has a
number of perspectives in such supergravity and superstring models when
non--compactified configurations and constraints on the superfield dynamics
are introduced into consideration.

The geometry on N--connections was extended and applied in gauge and
Einstein gravity with anholonomic/ non\-com\-mutative variables \cite%
{vgonch,vdgrg,dvgrg,vesnc}, for Clifford/ spinor bundles and algebroids
provided with N--connection structure \cite{cfs,vfs,vhs,vv,vstav,vclalg} and
on Fedosov--Lagrange manifolds \cite{esv}. Here, we emphasize that the
concept of N--connection has to be not confused with the linear connections
in gauge models with nonlinear realizations of some gauge groups related to
non--Abelian gauge potentials. Such constructions are completely different.

One should be noted that the idea of nonholonomic manifolds (as a geometric
background for geometric mechanics and generalized geometries) exists in
rigorous mathematical form due to the works of G. Vr\v{a}nceanu and Z. Horak %
\cite{vr1,vr1a,vr2,hor} (1926, 1931,1957,1927), see further developments in %
\cite{obad,mirnhv} and a modern approach and references in \cite{bejf}
(2006). For different classes of connections on nonholonomic manifolds,
there were computed the torsion and Riemannian tensors and investigated the
geometric properties of such spaces and considered certain applications in
geometric mechanics.

In Ref. \cite{leit}, see also references therein, the author argues that he
was able to define the Riemann tensor for a general nonholonomic manifold.
His homological considerations and analysis of former works on nonolonomic
(super) spaces was based on reviews \cite{leit1,versh,verg} on
supermanifolds, nonholonomic manifolds and mechanics. We note here that the
Riemannian tensors and Einstein equations were defined and computed
rigorously in various approaches to nonholonomic manifolds, Finsler and
Lagrange spaces and superbundles provided with nonlinear connections much
before mentioned publications and reviews by former Soviet mathematicians
(see, for instance, \cite{vr2,bw,cart,rund,ma1987,ma,bej,vmon1,vncsup}).

The approaches developed by the Romanian school on Finsler geometry and
generalizations have a number of connections to G. Vr\v{a}nceanu's results
(he published a four volume Course on Differential Geometry, in Romanian,
and some of them were translated in French, by 1957). In the G. Vr\v{a}%
nceanu, R. Miron, M. Anastasiei, A. Bejancu and other authors on the
geometry of nonholonomic manifolds and generalized Finsler--Lagrange spaces
there are not considered any methods of constructing exact solutions in
gravity theories (they were elaborated during the last decade, see below
section \ref{assfmes}). Nevertheless, the mentioned authors (and a number of
their co--authors) elaborated various applications in geometric mechanics,
Finsler generalizations of gravity, electro--gravitational fields, gauge
models and locally anisotropic supersymmetric variables. A brief summary of
R. Miron school's results on Finsler, Lagrange, Hamilton and higher order
generalizations and further perspectives in gravity and field theories is
given in Ref. \cite{vrmiron}.

\subsection{Finsler and Lagrange algebroid structures}

Lie algebroids were introduced as a generalization of the concepts of Lie
algebra and integrable distribution, see details in Ref. \cite{mcz}. An
extension of the theory of Lagrangians and Euler--Lagrange equations on Lie
algebroids was considered by A. Weinstein \cite{weinst}, see also Refs. \cite%
{liberm,mart}.

In our approach, we tried to model Lie algebroid structures as exact
solutions in gravity \cite{valgbh,valgtaub,valgdisk}. Such solutions were
defined by generic off--diagonal metrics and nonholonomic frames of
references. They were constructed following the anholonomic frame method,
outlined in Ref. \cite{vsgg} (see also the references from section \ref%
{assfmes}) when nonlinear connections are defined by splitting the
gravitational degrees of freedom into holonomic and anholonomic ones.

If in the usual approaches to Lie algebroids the geometric constructions are
related to Lie algebra generalizations and sections of vector bundles, in
order to define algebroid structures as solutions of the Einstein equations,
in general, it is necessary to work on nonholonomic manifolds modelling
certain types of Lie algebroid or Clifford--Lie algebroid structures. Some
explicit examples of such solutions were considered in Ref. \cite{valgtaub}
and the geometry of Clifford--Finsler algebroids and nonholonomic
Einstein--Dirac structures was elaborated in Ref. \cite{vclalg}. In a
general context, the theory of nonholonomic algebroids in relation to
nonholonomic manifolds, Finsler geometry and Lagrange--Hamilton spaces was
elaborated in Ref. \cite{valg}. Here we note that some topics on Lie
algebroids and Finsler and Lagrange geometry, without connections to modern
gravity but with certain orientation to application in mechanics, are
considered alternatively in Refs. \cite{hp1,hp2,anastalg}, see also
references therein.

\subsection{Higher order extensions of Lagrange and Hamilton spa\-ces}

The geometric approach to Lagrange and Hamilton mechanics elaborated as a
generalization of Finsler geometry was developed in a direction to include
higher order mechanics \cite{miat1,miat2}, see Ref. \cite{krup} as a summary
of alternative directions. The nonlinear connection formalism was developed
for higher order (co) tangent bundles which resulted in a series of
monographs on higher order Lagrange--Finsler and Hamilton--Cartan spaces %
\cite{mhl,mhf,mhss,mhh}, see also a recent work \cite{bmir} and a brief
review \cite{vrmiron}. Such higher order geometric mechanical constructions
are naturally adapted to corresponding (semi) spray configurations, from
which canonical nonlinear and distinguished connections can be derived.
Sure, this geometric formalism would be very important for developing
analogous models of gravity.

In parallel to the mentioned works on geometric mechanics, there were
elaborated certain new directions related to "higher order anisotropic"
configurations in high energy physics, gravity and string theory. The idea
was to consider higher order "shells" of extra dimensions which are not
completely compactified like in the Kaluza Klein theory and to take into
account certain possible correlations between spacetime dimensions and extra
dimensions (the higher dimension interactions being modelled by effective
higher order nonlinear connections). Such configurations can be derived as
exact solutions of the Einstein equations in extra dimension gravity, or in
certain low energy limits, but locally anisotropic, in string theory. It was
necessary to elaborate a corresponding (super) geometric formalism for the
higher order gauge theories, including gauge and Einstein gravity, see \cite%
{vggho}, higher order super spaces \cite{vncsup} and Clifford bundles and
spinor theory \cite{vhs,vv}, the bulk of results being summarized in
monographs \cite{vmon1,vstav}.

\subsection{Almost K\"{a}hler and nonholonomic structures}

\label{assakm} The constructions transforming nonholonomic Riemannian spaces
and generalized Lagrange algebroids into almost sympletic structures,
presented in section \ref{sakeg}, originate from a series of works on almost
K\"{a}hler models for Finsler \cite{mats} and Lagrange spaces and
generalizations \cite{opr1,opr2,opr,opp1,opp2,oppor}. We note that V. Oproiu
and co-authors performed the bulk of their lifts on (co) tangent bundles
working with linearized N--connections $N_{i}^{a}(u)=\Gamma
_{bj}^{a}(x)y^{b} $ (\ref{lincon}) but a number of formulas hold true for
more general nonholonomic structures with arbitrary $N_{i}^{a}(u).$

In order to model gravitational interactions by analogous
Finlser--La\-grange algebroid structures, we have to consider arbitrary
nonholonomic h-- and v--frames and N--connections. In such cases, the
effective geometric models are almost Hermitian ones with nonzero 2--form $d%
\mathbf{\check{\theta}}$ (\ref{asstrsf}). This is not a problem for the
anholonomic frame method of constructing exact solutions but result in more
sophisticate nonholonomic relations if we try to develop the approach, for
instance, for geometric quantization of Lagrange--Fedosov spaces, see \cite%
{esv} and references therein.

Finally, we note that from physical point of view there are two different
directions of elaborating analogous almost Hermitian / K\"{a}hler models of
gravitational interactions. If we we work with nonholonomic deformations of
geometric objects on the same class of manifolds, for instance, on a
semi--Riemannian one, we can preserve, in general, the local Lorentz
symmetry. Any lifts on tangent bundles and additional nonholonomic
transforms, positively result in violation of the local Lorentz symmetry and
modification of various types of possible gauge symmetries and conservation
laws. Both classes of such models present a substantial interest in modern
physics but only the first one can be related to the so--called standard
theories.

\subsection{Finsler methods and exact solutions}

\label{assfmes}

We applied the methods of Finsler--Lagrange geometry in order to elaborate
the so--called anholonomic frame method of constructing exact solutions in
Einstein, gauge, string and brane gravity and various locally anisotropic/
noncommutative generalizations \cite%
{vhep2,vtnut,vpdsw,vs,vswaw,vswsdpbh,vdgrg,dvgrg,vesnc,vsgg}. The idea was
to define such N--connection structures associated to nonholonomic frame
transforms, when the gravitational field equations are transformed into
general systems of partial differential equations which can be integrated in
general form. It was possible to construct, following geometric methods, new
classes of generic off--diagonal metrics, nonholonomic frames and linear
connections depending on 2-4 variables in 3-5 dimensional gravity. It became
obvious that the N--connection formalism can be applied also on Riemann,
Riemann--Cartan and metric--affine spaces and that an unified geometric
approach, with nonholonomic distributions, can be elaborated in order to
model generalized Finsler--Lagrange spaces both on nonholonomic manifolds or
on any (super, noncommutative, spinor...) bundle provided with N--connection
splitting. The main results are summarized in monograph \cite{vsgg} (2005)
and the basic constructions and some ansatz and examples are considered in
section \ref{sflexso}.

Here, we briefly outline some additional results giving a number of examples
of exact solutions with local anisotropy in standard and nonstandard models
of gravity:

The paper \cite{vesol1} was the first one containing the idea how the
N--connecti\-on formalism can be applied for generating generic
off--diagonal solutions in gravity. The work \cite{vhep2} presented a number
of examples with anholonom\-ic soliton--dilaton and black hole solutions in
general relativity and extensions to string or Finsler generalized theories.
Two papers \cite{vtnut,vpdsw} contain a research on nonholonomic
deformations of Taub NUT\ spinning metrics and locally anisotropic solitons
and Dirac spinor waves in such spaces.

A series of works \cite{vs,vswaw,vswsdpbh} is based on a collaboration of
authors on constructing exact solutions defining locally anisotropic
defining various types of locally anisotropic (ellipsoidal, toroidal, warped
...) wormhole / flux tubes and black holes moving in nontrivial solitonic
backgrounds in four and five dimensional gravity. The anholonomic frame
method was shown to be the most general one allowing to generate exact
solutions following geometric methods. There were elaborated a number of
examples which together with the the details on analytic computations were
summarized in Parts I and II of monograph \cite{vsgg}, see also review \cite%
{vesol2}.

There were constructed solutions defining static black ellipsoids \cite%
{vbel1,vbel2} which seem to be stable in Einstein gravity, with mater fields
and/ or geometric distorsions, and various noncommutative and metric--affine
generalizations of gravity, see \cite{vesnc} and Parts I and III in
monograph \cite{vsgg}. Solutions with generic local anisotropy, of Finsler
type and generalied ones, were constructed in gauge gravity \cite%
{vdgrg,dvgrg}, for various black hole and cosmological configurations, see
Part II in \cite{vsgg}, and by modelling explicit examplas of (disk, black
hole, solitonic and spinor waves ....) of solitonic spacetimes with Lie
algebroid symmetry \cite{valgbh,valgtaub,valgdisk}. The bulk of recent
results on parametric solutions and solitonic hierarchies are summarized in %
\cite{vparsol,vsh1,vsh2}. The anholonomic frame method provides also a
unique general geometric scheme for constructing exact solutions of Ricci
flow equations, see section \ref{ssrfs}.

\subsection{On standard and nonstandard models of Finsler geometry and
physics}

It is worth mentioned certain important directions additionally to those
mentioned in the points 1--5 from Introduction section and the previous
sections of the Attachment. Due to limits of space we have to leave out a
number of interesting developments: we discuss here briefly the
contributions of some authors and outline a few open issues, see also
Introduction to \cite{vsgg}.

\subsubsection{Finsler structures in gauge gravity and noncommutative gravity%
}

\label{assgt}The first original ideas to consider gauge group transforms and
additional gauge fields on Finsler spaces and generalizations belong to Y.
Takano \cite{takano,t1}, S. Ikeda \cite{ikeda} and G. Asanov and co--authors %
\cite{as1,ap,as2,aras}. Here one should be noted that the monograph \cite%
{ma1987} was the first one on Finsler and Lagrange spaces written rigorously
in the language of the geometry of vector and tangent bundles and related
geometric structures (fiber bundles and linear connections provide the
standard geometric formalism for the theory of gauge fields). Various types
of nonlinear connections (N--connections) on nonholonomic spaces (considered
to define a special class of nonlinear gauge fields) can be found in a
number of constructions in Finsler geometry, see \cite{cart,kaw1,kaw2}. The
monograph \cite{bej} contains a study of Yang--Mills theories in Finsler
spaces. In the same line of research, elaborating gauge and field theories
on bundle spaces, can be considered papers \cite{stav1,stav2}. Such models
can not be related to standard approaches in modern physics. Nevertheless,
they provide a number of geometric ideas which can be applied for
noncompactified higher dimension gravity models.

A series of works \cite{vggm1,vggm2,vggho,vgonch} is devoted to gauge models
of matter and gravity fields on generalized Lagrange and Finsler spaces in
relation to the Poincare, affine and de Sitter structure groups with actions
distinguished by the N--connection structure. The idea was to elaborate
gauge models which would embed Finsler like gravities into the class of
former theories on gauge gravity \cite{utiyama,pda1,pda2,tseyt}. Such
constructions, with nonholonomic frames and spinor--gauge transforms, were
related to the results on twistor--gauge gravity \cite%
{vtw,vtwl,vtwla,vvinity1,vvinity2,vvinity3}, and further supersymmetric
generalizations were summarized in the monograph \cite{vmon1}. This class of
locally anisotropic gauge models was elaborated following the frame bundle
and generalized connection formalism. They have strong connections to
standard gauge and gravity models in physics: for instance the nonholonomic
structures, metrics and connections can be defined for (pseudo) Riemannian
spaces and fiber bundles on sauch spaces.

Gauge models of higher order anisotropic gravity and nearly autoparallel
maps and their connections to Einstein and gauge models were analyzed in
Refs. \cite{vdgrg,dvgrg}. The approach was developed for noncommutative
versions of the Einstein and gauge gravity \cite%
{vggr,vcv,ncfggfg,ncfgsmt,vnccs} and provided with explicit examples of
exact solutions for nonholonomic noncommutative structures in gravity.

\subsubsection{Clifford structures, gerbes, and Lagrange--Finsler spinors}

\label{assscf}

The first who considered spinor variables in Finsler geometry was Y. Takano %
\cite{t1}. Perhaps, we can cite here the work \cite{ansh} on modelling
spinors on Hilbert manifolds, even explicit constructions related to Finsler
spaces are not given there (the author latter had fundamental contributions
in Finsler--Lagrange geometry and modelling such geometries on Hilbert
spaces). There were published a series of papers on geometries and physical
models with metrics depending on spinor variables, see \cite%
{ot1,ot2,ot3,ot4,ap,sm,sbmp,stav2}. It should be emphasized here that all
mentioned works do not contain a definition of spinor for Finsler like
spaces and generalizations. In the bulk, they provide certain constructions
when the existence of the bundle of two--spinors (i.e. two dimensional
spinor spaces) is supposed to exist on a Finsler like manifold for which the
metric and connections are considered to depend on two spinor variables.
Such models belong to ''nonstandard'' approaches to Finsler geometry and
generalizations.

A rigorous definition of spinors for Finsler spaces should contain a
fundamental relation between the Clifford structure and the metric structure
(like in Einstein gravity, one has to consider an anticommutator of "gamma"
matrices, generating the corresponding Clifford algebra, and the metric
quadratic form, all defined with respect to a local orthonormalized frame
basis). Such constructions for Finsler--Lagrange spaces were elaborated in
Refs. \cite{cfs,vfs} in the language of Clifford bundles provided with
N--connection structure. We note that the condition of metric compatibility
of Finsler--Lagrange connection plays a fundamental role in definition of
spinors on spaces provided with N--connection structure. For instance, it is
not possible to define directly the consept spinors in Finsler geometries
with metric noncompatible connections like \cite{bw,bcs}.\footnote{%
As a matter of principle, we can define Finsler--spinors using the metric
compatible Cartan connections and then to deform the geometric constructions
to those for the Chern connection, by using corresponding deformations
tensors for connections. So, following certain more sophisticate geometric
constructions, we can define nonholonomic Clifford bundles provided with
metric noncompatible linear connection structure. But there are not physical
arguments for such nonmetric structures.} Without fermions / spinors, it is
not possible to elaborate viable physical models. It was necessary to
construct the spinor geometry for Finsler spaces and generalizations. The
direction was analyzed in details, with a number of examples and exact
solutions in gravity models and generalizations to higher order Finsler,
Lagrange and Hamilton spaces, in Refs. \cite{vhs,vmon1,vv}. Such
constructions belong to the class of standard Clifford--Finsler structures.
Both standard and nonstandard approaches are analyzed in details in
monograph \cite{vstav}.

The standard constructions with Finsler--Lagrange spinors have further
developments for noncommutative Finsler geometry, see \cite{vnccs} and the
Part III in \cite{vsgg}, for explicit solutions see \cite{vesnc} and for
Clifford--Lagrange algebroid structures see \cite{vclalg,valgtaub}. One
could be topological restrictions in definition of spinor structures on
general Finsler or Lagrange spaces: certain generalized constructions for
such cases were performed following the gerbe formalism \cite{vgerb1,vgerb2}.

\subsubsection{Stochastic processes and locally anisotropic kinetics and
thermodynamics}

The famous P. Finsler's thesis \cite{fg} (1918) was written under
supervision of Prof. Caratheodory who had classical contributions in
thermodynamics. A theory of kinetics and thermodynamics based on
distribution functions on Finsler spaces was elaborated in monograph \cite%
{vlasov}.

Both the Riemann and Finsler geometry were applied to various problems in
information thermodynamics \cite{ing1,ing2,ingtam} and geometric
thermodynamics \cite{mrug1,janm,mnss,quev}. The approaches were summarized
in review \cite{ruppein}. There were proposed certain applications of
geometric thermodynamics with generalized Riemann--Finsler structures in
black hole physics \cite{vbht,vbhts,vbhtbh}.

A number of works were published on locally anisotropic diffusion in two
series of works elaborated in parallel by two collaborations of authors. The
first one is related to publications \cite%
{antz1,antz2,antz3,anth,antl,aim,amaz,ant} and the second one to \cite%
{vstoh1,vstoh3,vstoh4,vatsp} and Chapter 10 in monograph \cite{vmon1}. The
second approach was developed in relation to the theory of kinetic processes
and locally anisotropic thermodynamics \cite{vapkp,vatsp} (the works propose
certain applications in modern astrophysics and cosmological models, see
also such locally anisotropic cosmological solutions in Refs. \cite%
{vdgrg,stavc}). Chapter 5 in \cite{vmon1} contains the results on diffusion
theory on locally anisotropic superspaces.

Generalized Riemann--Finsler structures in thermodynamics and kinetic and
stochastic processes can be described in terms of both metric compatible and
noncompatible connections. For such constructions, there were not yet
elaborated criteria on standard and nonstandard models. The results cited in
this section can be elaborated following covariant calculus with different
generalized Finsler connections. Priorities should be given to certain more
"simple" of theoretical models with possible experimental verification and
applications, for instance, in modern thermodynamics, astrophysics and
cosmology.

\subsubsection{Nonholonomic curve flows and bi--Hamitonian structures}

A new direction of applications of the nonlinear connection formalism was
elaborated in Refs. \cite{vsh1,vsh2}. It was proposed to consider such
nonholonomic distributions and related frames on (semi) Riemannian and
generalized Finsler--Lagrange spaces when the curvature (\ref{dcurv}) of the
canonical d--connection (\ref{candcon}), with respect to certain N--adapted
bases, is parametrized by constant matrix coefficients. The geometric
information for such spaces can be encoded into bi--Hamilton structures and
solitonic hierarchies characterized by corresponding invariants and
conservation laws. The approach was generalized for nonholonomic Ricci flows %
\cite{vrf03}.

\subsubsection{Nonholonomic Ricci flows and Lagrange--Finsler spaces}

\label{ssrfs}Recently, it was elaborated a new direction in Riemann--Finsler
geometry, gravity with nonholonomic distributions and geometric mechanics,
the theory of nonholonomic Ricci flows and generalizations which is
positively related to standard theories of physics and can be generalized
similarly for evolution models on tangent and vector bundles, i.e. for
nonstandard theories. The idea was not just to extend the R. Hamilton \cite%
{ham1,ham2} and Grisha Perelman \cite{per1,per2,per3} results, on Ricci
flows of Riemannian metrics (see comprehensive reviews of results in Refs. %
\cite{caozhu,cao,kleiner,rbook}), to more sophisticated classes of
geometries, like Finsler and Lagrange geometry. A series of works \cite%
{vrf01,vrf02,vrf03,vrf04,vrf05} was written as a research of the Ricci flows
when (semi) Riemannian metrics are subjected to certain classes of
nonholonomic constraints. We proved that under well defined conditions, for
instance, a Finsler like metric can evolve into a Riemannian one, and
inversely. In general, the Ricci flows of such nonholonomic manifolds
contain nontrivial torsion structures but such flows can be alternatively,
and equivalently, described by flows of metrics and Levi Civita connections.
Nonholonomic distributions on Riemannian manifolds result in some geometric
multi--connection structures with, in general, nonzero torsion coefficients
induced by some off--diagonal metric coefficients transformed into the
N--connection coefficients. This is typical for metric compatible Finsler
like connections; as a matter of principle, the constructions can be
generalized for nonsymmetric metrics or Finsler connections with
nonmetricity.

The theory of nonholonomic Ricci flows and Perelman's functionals adap\-ted
to the N--connection structure allowed to formulate a new statistical
interpretation of Finsler--Lagrange spaces and related nonholonomic and/or
mechanical systems, see Ref. \cite{velfsrf}.

It should be noted that the anholonomic frame method works effectively not
only for generating new classes of exact solutions in gravity, as we
discussed in section \ref{assfmes}, but also for constructing exact
solutions for Ricci flow evolution of valuable physical metrics in gravity
(like solitonic and pp-waves, Taub NUT spaces, nonholonomically deformed
Schwarzschild metrics...), see Refs. \cite{vrfsol1,vvisrf1,vvisrf2}.
Perhaps, this is still the unique method which allows us to construct exact
solutions of Ricci flow equations in general form, by using geometric
methods and ideas from Finsler geometry.

\subsubsection{Quantum gravity and Finsler methods}

There are a few "standard" works related to nonholonomic Lagrange--Fedos\-ov
spaces \cite{esv} and geometric quantization of the Einstein gravity
transformed equivalently as an almost Hermitian / K\"{a}hler model \cite%
{vqgr1,vqgr2,vqgr3}, see discussion in section \ref{assakm}. This direction
is under elaboration for gravitational gauge models and certain nonholonomic
generalizations of gravity on manifolds and tangent bundles. Some results on
quantum models connected to Finsler geometry and gravity belonging to the
class of nonstandard theories will be analyzed in the next section.

\subsubsection{On some nonstandard but important contributions to Fins\-ler
geometry and physics}

We shall cite and briefly comment here some series of works concerning
nonstandard Finsler geometric and physical models for which a number of
important results can be re--defined on nonholonomic manifolds and may
present a substantial interest in the so--called standard theories, or are
related to certain recent developments in modern physics.

\paragraph{{\ }Gauge transforms on tangent bundle and Kaluza--Klein theory:}

{\ }\newline
Additionally to the discussion on Finsler geometry and generalized gauge
theories in section \ref{assgt}, we refer to a series of works by R. G. Beil %
\cite{beil1,beil2,beil3,beil4}.\ The author considered a Kaluza--Klein like
theory following the idea that the nature of spacetime is Finslerian and the
extra dimension is time like, i.e. not compactified. The geometry of moving
frame transports on Finsler spacetimes was related to local Poincare and
Lorentz transforms. The corresponding gauge transforms on tangent bundle
resulted in a Kaluza--Klein type theory, but not in a Yang--Mills one, which
was connected to a new type of quantum field theory. It should be noted that
A. Bejancu also elaborated gauge models in tangent bundle summarized in his
monograph \cite{bej}. Such gauge Finsler--Kaluza--Klein theories can be
considered in the usual gauge gravity or string gravity models with
nonholonomic distributions if the constructions are performed for
nonholonomic manifolds, see for instance, \cite{vsgg,vmon1,vesnc}.

\paragraph{{\quad } Generalized/ broken local Lorentz symmetries:}

{\ \qquad } \newline
We mention two directions on Finsler spacetime field theories with
generalized Lorentz and gauge symmetries. The first one, recently, is
connected to the so--called Finsleroid structures \cite{as3}, with
anisotropic kinematics and Finsler like generalizations of the local
Minkowski metric. A number of former investigations on locally anisotropic
gauge models, jet models, and Finsler like corrections to the Einstein
gravity can found in \cite{aras,as1,as2,ap} and references therein.

Group transforms defining a generalized local Lorentz invariance on a
Finsler like spacetime, instead a local Minkowski space are considered in %
\cite{bg1}, see also models with generalized Lorentz invariants and Dirac
equation \cite{bg2}, and the so--called relativistic theory of gravity
generalized to Finsler like spaces \cite{bog}. This class of theories by
definition belong to the class of nonstandard ones. It presents certain
interest for some approaches in modern physics related to violation of local
Lorentz invariance and violation of principle of equivalence.

\paragraph{{\quad } Classical \& quantum theories with local Lorentz
invariance:}

{\ \qquad } \newline
An "almost standard" approach with Finslerian fields and applications of
Finsler geometry methods is developed by H. E. Brandt. In work \cite{brandt0}%
, he proposed a Cartan like Finsler theory with Kaluza--Klein
generalizations on tangent bundle with K\"{a}hler geometrization starting
with Christoffel symbols on the base and considering local Lorentz
transforms on fibers. On total space, the theory possesses a nontrivial
almost complex structure and nontrivial torsion.

Further constructions are with maximal acceleration invariant quantum
fields, formulated in terms of the differential geometric structure of the
spacetime tangent bundle \cite{brandt1}. It was proposed a physically based
Planck scale effective regularization with a spectral cutoff at the Planck
mass. There were also considered Finslerian fields, strings and p--branes on
tangent bundle with local Lorentz invariance and maximal invariance, quantum
fields. There are Kaluza--Klein like fields but in general not compactified.

It was also attempted to elaborate a quantum field theory with Finslerian
quantum fields (scalar fields on tangent bundle/Finsler spacetime) and
microcausality with corresponding commutators of scalar fields \cite{brandt3}%
. The general idea was to work with Lorentz--invariant quantum fields and
maximal--acceleration invariants (for such quantum fields) in the spacetime
tangent bundle by using a Plank scale, causal domains and microcausal
constructions \cite{brandt4,brandt5}.

We also cite a series of works published recently in a Russian journal
"Hypercomplex Numbers in Physics and Geometry", see \cite{gar1,garpav1} and
references therein, with the aim to construct of pseudo--Riemannian geometry
on the basis of Berwald--Moore geometry, by using certain classes of
relativistic invariant Finsler geometry generating functions.

This class of theories is related to standard models of gravity and strings
but proposes new Finsler alternatives for quantum theories.

\paragraph{{\quad } Finslerian teleparallel and K\"{a}hler--Clifford
structures:}

{\ }\newline
In this section, we shall comment on a series of works by J. Vargas \ and D.
Torr \cite{varg1,varg2,varg3,varg4,varg5}. A very important idea for
applications to standard theories of physics (even, in general, the authors
work with locally anisotropic geometric models on tangent bundle) is that a
Riemann structure can be reconfigured on the Finsler bundle without loss of
information but with increased structural richness. This allows us to
consider canonical connections of Finsler metrics and Finslerian connections
on Riemannian metrics \cite{varg1}. Some constructions are similar, but
inverse, to our constructions \cite{vsgg} when Finsler geometries and
generalizations are modelled on (pseudo) Riemannian spaces.

Perhaps a source for such investigations can be found in a rigorous study by
E. Cartan and A. Einstein (1929) when a theory was elaborated in which the
electromagnetic field constitutes the time--like 2-form part of the torsion
of Finslerian teleparallel connections on pseudo--Riemannian metrics \cite%
{varg3}. The research by Vargas and Torr were performed following a
comparative analysis of fundamental geometric constructions elaborated by
Riemann, Cartan, Weyl, Klein, Clifford and K\"{a}hler and their explicit
realization in Finsler geometry and generalizations.

For B. Riemann, at a time when the theory of continuous groups had not yet
been founded, the fundamental geometric notion is that of distance (see
Riemann's famous inaugural dissertation 'On the hypothesis...', 1854, \cite%
{riemann}). Then, a geometric approach of a completely different nature has
developed between 1867 and 1914 by F. Klein, see an analysis oriented to
Finsler geometry in \cite{varg2}. For Klein, the fundamental geometric
notion is contained in the axiom of geometric equality, interpreted in the
light of the notion of group. We note here that the concept of geometric
equality vary from geometry to geometry and is contained in the axioms of
each geometry.

Weyl geometry is considered as the first type of Yang --Mills theories,
ahead of the times. It is directly related to the geometry of base manifolds
endowed with metric--compatible affine connection of the particular type
that Cartan called metric connections. One thus has to leave open the
possibility that some Yang--Mills theories may eventually become part of
classical differential geometry of some more general type, with some more
general (for instance, Finsler like) structure.

Various Finsler geometry models were elaborated following different
relations between metrics and connections (connections are in general
nonlinear but the authors tried to introduce certain effective linear ones).
The Vargas--Torr idea is to derive Finsler spaces from certain spacetime
structure moving the constructions on tangent bundles. This can be
considered as an ''almost standard'' modelling of physical interactions on
spaces with more rich (than Riemann geometry) structures. They formulated
corresponding K\"{a}hler and Clifford calculus for Finsler like connections,
analyzed Clifford structure of Kaluza--Klein spaces and tried to generate
''metrics without metric tensors''. Such constructions present a substantial
interest also for standard models of physics because we can model them by
nonholonomic distributions on (semi) Riemannian manifolds and
Riemann--Cartan spaces.

In \cite{varg4}, one works with Finsler bundles using the language of
differential forms but with nonmetricity, like Chern, see \cite{bcs}, which
is not compatible with the standard models of physics. The authors develop
the concept of affine Finsler connection involving bundles. Such Finsler
connections are defined even in the absence of a metric function and/or
metric. For the corresponding tangent bases and special bases (frames) and
affine Finsler connection, it is elaborated a teleparallel and K\"{a}hler
calculus. They compare their results with other approaches to
teleparallelism and Kaluza--Klein reformulation of Finslerian
teleparallelism. It is also considered the K\"{a}hler--Dirac equation for
such Finsler spaces.

It should be emphasized that the K\"{a}hler equations for forms are
equivalent for the Dirac equations only on flat spaces. For general curved
spaces, this result is not true. In general, it is not clear how to
formulate the Dirac equation for metric noncompatible connections if there
are not used certain auxiliary constructions for metric compatible ones. The
original contributions by Vargas--Torr is that they elaborated a K\"{a}hler
calculus with Clifford--valued multiforms for certain classes of Finsler
connections and found an analogous (K\"{a}hler equation) to the Dirac
equation for such cotangent spaces provided with Finsler like structure.
Nevertheless, this is not a theory of spinors for Finsler--Lagrange spaces
which was elaborated in Refs. \cite{cfs,vfs}, see discussion in section \ref%
{assscf}.

For applications of Finsler geometry methods to standard models of physics,
the most important is the idea that the classical and quantum geometric
structure of the spacetime and field/string dynamics may be richer than it
presently appears to be on Riemann, or Riemann--Cartan spaces. If in \cite%
{varg5} it is suggested to consider Clifford algebras for nonsymmetric
quadratic forms and generalized Finsler spaces, we consider that by
nonholonomic distributions on (semi) Riemann geometries we can also model
very rich geometric structures but preserving compatibility with the present
days paradigm of standard physics.

\paragraph{{\quad } Clifford manifolds and Finsler spaces:}

{\ } \newline
A model with maximal speed of light and maximal--acceleration relativity
principle in the spacetime tangent bundle and in the phase spaces (cotangent
bundle) was elaborated in Ref. \cite{castro4} following the approach on
relativity in C--spaces (Clifford manifolds), see \cite{castrop} and
references to the cited papers. The idea of maximal--acceleration is similar
to that from \cite{brandt4,brandt5} but it is developed on C--spaces.

The constructions on C--spaces are related to Finsler geometry by
considering dependencies on speed and accelerations following a ''new
program'' in physical theories \cite{castro1a,castro2}. We also cite the
work \cite{castro1} on $W$ geometry from Fedosov's deformation quantization.
It should be noted that C. Castro's works conventionally belong to the class
of nonstandard models of physics because they use constructions for the
tangent and cotangent bundles, or jets, even such spaces are modelled by
C--space structures. In other turn, they can be re--defined for nonholonomic
manifolds by using nonholonomic Clifford bundles \cite{cfs,vfs,vmon1,vstav}
and Fedosov--Lagrange manifolds \cite{esv}. Such results can be positively
related to standard models of physics.

Finally, in this section, we would mention the Hull's formulation \cite%
{hull1,hull2} of $W_{\infty }$ gravity as a gauge theory of the group of
sympletic diffeomorphisms of the cotangent bundle of two--dimensional
surface considering a generalized (Finsler like) scalar line element
\begin{equation*}
ds=\left( g_{\mu _{1}...\mu _{n}}dx^{\mu _{1}}...dx^{\mu _{n}}\right) ^{1/n}.
\end{equation*}%
If for this element one considers a canonical d--connection and extracts the
Levi--Civita connection, we generate a nonholonomic effective Riemannian
space with more rich structure. In this case, we can establish further
relations to noncommutative geometry and M(atrix) theory \cite{vesnc,vsgg}.
We conclude that such models can be elaborated both in standard and
nonstandard fashions.

\paragraph{{\quad } Deterministic quantum Finsler models:}

{\ } \newline
A series of works on Finsler geometry and applications \cite%
{gallego1,gallego2,gallego3,gallego4,gallego5} has the aim to solve certain
problems of quantum mechanics with dissipation and loss of information. The
constructions are based on 't Hooft's proposal \cite{thooft1,thooft2} to use
deterministic models in order to describe physical systems at the Planck
scale through a Hilbert space formulation of these models. R. Gallego uses
Finsler geometry with Chern connection in order to elaborate such
deterministic quantum models and relate dissipation to average of Chern
connection, with nonmetricity, in order to get the Levi Civita connection %
\cite{gallego2} and define a corresponding Hamiltonian following a model of
dissipative dynamics.

It was elaborated a corresponding formalism when translation of results is
considered from Riemannian to Finsler spaces and proposed the notions of
complete Finsler manifold with the ''average'' to a Riemannian manifold \cite%
{gallego3}. The dynamics at the quantum Planck scale with loss of
information is constructed supposing that a Finsler structure on tangent
bundle $TM$ evolves to a Riemannian structure also in $TM$ \cite{gallego4}.
The canonical quantization is considered on the dual tangent - tangent
bundle $T^{\ast }TM$ which is supposed to be the arena for deterministic
Finslerian models and dynamical systems with corresponding Poisson
structures. Finally, a nonstandard approach to quantum gravity with maximal
acceleration is considered in \cite{gallego5}.

It is not obligatory to use in researches on deterministic quantum model
only the Chern connection \cite{bcs} (similarly, and in a more simple form
one can be elaborated quantum Cartan--Finsler models with further
developments on Finsler, Lagrange and Hamilton geometry with metric
compatible connections \cite{ma1987,ma,bej,vmon1,vncsup}). We note that
having a canonical d--connection, we can always define exactly the
Levi--Civita connection, because both such linear connections are uniquely
defined by a generic off--diagonal metric tensor of type (\ref{ansatz}). It
is also not obligatory to average the Chern connection with nonmetricity in
order to get a metric compatible Levi Civita connection: we can do this by
subtracting from the Cartan, Chern, or other canonical connection the
corresponding deformation tensor (this topic is discussed in details in
Refs. \cite{vrf01,vrf02}). The experimental data does not constrain us to
use at Planck scale nonmetric connections (even there are quantum
uncertainty relations, there are not proofs that they are related strongly
to nonmetricity). A deterministic quantum dynamics can be modelled
alternatively with metric compatible Finsler like connections. Working with
nonholonmic manifolds and metric compatible connections, such constructions
will belong to the class of standard models, contrary to those performed on
tangent spaces and for nonmetric connections.

\paragraph{{\quad } Other directions:}

{\ } \newline
There were also proposed a number of other possible applications of Finsler
geometry in order to solve important geometric and physical problems. In
this final section, we mention a few of them.

The works by P. Stavrinos and co--authors \cite{stav3,stavrdiak} elaborate
certain Finsler like generalizations (nonstandard ones) in modern cosmology.

The idea to consider the Fermat principle on Finsler spaces was approached
by V. Perlick \cite{perl} \ following a variational principle with a
corresponding Lagrangian and Lagrangian, Euler--Lagrange equations. The
topic of nonlinear connections and distinguished connections is not
discussed in his work but it appears from the corresponding semi--spray
dynamics if the geometric constructions are adapted to the nonlinear
connection structure.

Possible hidden connections between general relativity and Finsler geometry
are considered in Ref. \cite{panahi}. An approach with homogeneous and
symmetric Finsler spaces, with Chern connection, as a generalization of
similar constructions for homogeneous/ symmetric Riemannian spaces, is
developed by authors \cite{lat1,lat2,lat3}. It is not clear if such results
hold true for nonholonomic Riemannian spaces and Finsler models defined by
metric compatible Finsler connections (for instance, with the Cartan
connection). Generalized Lagrange--Weyl structures and compatible
connections are analyzed in Ref. \cite{crasm}.

Geometric methods of Finsler--Lagrange geometry have been applied for a
study of systems of partial differential equations, multi--time Lagrange
spaces and dynamical systems and jet geometry \cite%
{udriste1,udriste1a,udriste2,neagudr,neagudr1}.

Finally, we note that the geometry of induced structures on submanifolds and
almost product Riemannian and/or Finsler manifolds \cite{hretc1,hretc2} has
certain applications in geometric quantization \cite{esv}.

\end{document}